\documentclass[longauth]{aa}  

\usepackage{graphicx}
\usepackage{txfonts}
\usepackage[labelfont=bf]{caption}
\bibpunct[; ]{(}{)}{;}{a}{}{;}

\usepackage{color}
\usepackage{xcolor}
\usepackage[colorlinks=true,
            linkcolor=blue,
            citecolor=blue,
            urlcolor=blue,
            breaklinks=True]{hyperref}
\newcommand{\rpicard}{\mbox{\texttt{rPICARD}}}
\newcommand{\meqsil}{\mbox{\texttt{MeqSilhouette}}}
\newcommand{\pipe}{\mbox{\texttt{SYMBA}}}
\newcommand{\casa}{\mbox{\texttt{CASA}}}
\newcommand{\ehtim}{\mbox{\texttt{eht-imaging}}}
\newcommand{\merra}{\mbox{\texttt{MERRA-2}}}

\begin{document}

   \title{SYMBA: An end-to-end VLBI synthetic data generation pipeline}
   \subtitle{Simulating Event Horizon Telescope observations of M87}

\author{F. Roelofs\inst{\ref{inst17}}\fnmsep\thanks{These authors contributed equally to this work.}\and
M. Janssen\inst{\ref{inst17}}\fnmsep\thanks{These authors contributed equally to this work.}\and
I. Natarajan\inst{\ref{inst39}}\and
R. Deane\inst{\ref{inst38},\ref{inst39}}\and
J. Davelaar\inst{\ref{inst17}}\and
H. Olivares\inst{\ref{inst17}}\and
O. Porth\inst{\ref{inst29},\ref{inst36}}\and
S. N. Paine \inst{\ref{inst11}}
K. L. Bouman\inst{\ref{inst4},\ref{inst11},\ref{inst14}}\and
R. P. J. Tilanus\inst{\ref{inst17},\ref{inst45},\ref{inst88}}\and
I. M. van Bemmel\inst{\ref{inst56}}\and
H. Falcke\inst{\ref{inst17}} \and
K. Akiyama\inst{\ref{inst1},\ref{inst2},\ref{inst3},\ref{inst4}}\and
A. Alberdi\inst{\ref{inst5}}\and
W. Alef\inst{\ref{inst6}}\and
K. Asada\inst{\ref{inst7}}\and
R. Azulay\inst{\ref{inst8},\ref{inst9},\ref{inst6}}\and
A. Baczko\inst{\ref{inst6}}\and
D. Ball\inst{\ref{inst10}}\and
M. Balokovi\'c\inst{\ref{inst4},\ref{inst11}}\and
J. Barrett\inst{\ref{inst2}}\and
D. Bintley\inst{\ref{inst12}}\and
L. Blackburn\inst{\ref{inst4},\ref{inst11}}\and
W. Boland\inst{\ref{inst13}}\and
G. C. Bower\inst{\ref{inst15}}\and
M. Bremer\inst{\ref{inst16}}\and
C. D. Brinkerink\inst{\ref{inst17}}\and
R. Brissenden\inst{\ref{inst4},\ref{inst11}}\and
S. Britzen\inst{\ref{inst6}}\and
A. E. Broderick\inst{\ref{inst18},\ref{inst19},\ref{inst20}}\and
D. Broguiere\inst{\ref{inst16}}\and
T. Bronzwaer\inst{\ref{inst17}}\and
D. Byun\inst{\ref{inst21},\ref{inst22}}\and
J. E. Carlstrom\inst{\ref{inst23},\ref{inst24},\ref{inst25},\ref{inst26}}\and
A. Chael\inst{\ref{inst100},\ref{inst101}}\and
C. Chan\inst{\ref{inst10},\ref{inst27}}\and
S. Chatterjee\inst{\ref{inst28}}\and
K. Chatterjee\inst{\ref{inst29}}\and
M. Chen\inst{\ref{inst15}}\and
Y. Chen \inst{\ref{inst30},\ref{inst31}}\and
I. Cho\inst{\ref{inst21},\ref{inst22}}\and
P. Christian\inst{\ref{inst10},\ref{inst11}}\and
J. E. Conway\inst{\ref{inst32}}\and
J. M. Cordes\inst{\ref{inst28}}\and
G. B. Crew\inst{\ref{inst2}}\and
Y. Cui\inst{\ref{inst33},\ref{inst34}}\and
M. De Laurentis\inst{\ref{inst35},\ref{inst36},\ref{inst37}}\and
J. Dempsey\inst{\ref{inst12}}\and
G. Desvignes\inst{\ref{inst6},\ref{inst102}}\and
J. Dexter\inst{\ref{inst40}}\and
S. S. Doeleman\inst{\ref{inst4},\ref{inst11}}\and
R. P. Eatough\inst{\ref{inst6}}\and
V. L. Fish\inst{\ref{inst2}}\and
E. Fomalont\inst{\ref{inst1}}\and
R. Fraga-Encinas\inst{\ref{inst17}}\and
P. Friberg\inst{\ref{inst12}}\and
C. M. Fromm\inst{\ref{inst36}}\and
J. L. G\'omez\inst{\ref{inst5}}\and
P. Galison\inst{\ref{inst4},\ref{inst41},\ref{inst42}}\and
C. F. Gammie\inst{\ref{inst43},\ref{inst44}}\and
R. Garc\'{\i}a\inst{\ref{inst16}}\and
O. Gentaz\inst{\ref{inst16}}\and
B. Georgiev\inst{\ref{inst19},\ref{inst20}}\and
C. Goddi\inst{\ref{inst17},\ref{inst45}}\and
R. Gold\inst{\ref{inst103},\ref{inst36},\ref{inst18}}\and
M. Gu \inst{\ref{inst30},\ref{inst46}}\and
M. Gurwell\inst{\ref{inst11}}\and
K. Hada\inst{\ref{inst33},\ref{inst34}}\and
M. H. Hecht\inst{\ref{inst2}}\and
R. Hesper\inst{\ref{inst47}}\and
L. C. Ho \inst{\ref{inst48},\ref{inst49}}\and
P. Ho\inst{\ref{inst7}}\and
M. Honma\inst{\ref{inst33},\ref{inst34}}\and
C. L. Huang\inst{\ref{inst7}}\and
L. Huang \inst{\ref{inst30},\ref{inst46}}\and
D. H. Hughes\inst{\ref{inst50}}\and
S. Ikeda\inst{\ref{inst3},\ref{inst51},\ref{inst52},\ref{inst53}}\and
M. Inoue\inst{\ref{inst7}}\and
S. Issaoun\inst{\ref{inst17}}\and
D. J. James\inst{\ref{inst4},\ref{inst11}}\and
B. T. Jannuzi\inst{\ref{inst10}}\and
B. Jeter\inst{\ref{inst19},\ref{inst20}}\and
W. Jiang \inst{\ref{inst30}}\and
M. D. Johnson\inst{\ref{inst4},\ref{inst11}}\and
S. Jorstad\inst{\ref{inst54},\ref{inst55}}\and
T. Jung\inst{\ref{inst21},\ref{inst22}}\and
M. Karami\inst{\ref{inst18},\ref{inst19}}\and
R. Karuppusamy\inst{\ref{inst6}}\and
T. Kawashima\inst{\ref{inst3}}\and
G. K. Keating\inst{\ref{inst11}}\and
M. Kettenis\inst{\ref{inst56}}\and
J. Kim\inst{\ref{inst6}}\and
J. Kim\inst{\ref{inst10}}\and
J. Kim\inst{\ref{inst21}}\and
M. Kino\inst{\ref{inst3},\ref{inst57}}\and
J. Y. Koay\inst{\ref{inst7}}\and
P. M. Koch\inst{\ref{inst7}}\and
S. Koyama\inst{\ref{inst7}}\and
M. Kramer\inst{\ref{inst6}}\and
C. Kramer\inst{\ref{inst16}}\and
T. P. Krichbaum\inst{\ref{inst6}}\and
C. Kuo\inst{\ref{inst58}}\and
T. R. Lauer\inst{\ref{inst59}}\and
S. Lee\inst{\ref{inst21}}\and
Y. Li \inst{\ref{inst60}}\and
Z. Li \inst{\ref{inst61},\ref{inst62}}\and
M. Lindqvist\inst{\ref{inst32}}\and
R. Lico\inst{\ref{inst6}}\and
K. Liu\inst{\ref{inst6}}\and
E. Liuzzo\inst{\ref{inst63}}\and
W. Lo\inst{\ref{inst7},\ref{inst64}}\and
A. P. Lobanov\inst{\ref{inst6}}\and
L. Loinard\inst{\ref{inst65},\ref{inst66}}\and
C. Lonsdale\inst{\ref{inst2}}\and
R. Lu \inst{\ref{inst30},\ref{inst31},\ref{inst6}}\and
N. R. MacDonald\inst{\ref{inst6}}\and
J. Mao \inst{\ref{inst67},\ref{inst68},\ref{inst69}}\and
S. Markoff\inst{\ref{inst29},\ref{inst70}}\and
D. P. Marrone\inst{\ref{inst10}}\and
A. P. Marscher\inst{\ref{inst54}}\and
I. Mart\'{\i}-Vidal\inst{\ref{inst8}}\and
S. Matsushita\inst{\ref{inst7}}\and
L. D. Matthews\inst{\ref{inst2}}\and
L. Medeiros\inst{\ref{inst72},\ref{inst10},\ref{inst104}}\and
K. M. Menten\inst{\ref{inst6}}\and
Y. Mizuno\inst{\ref{inst36}}\and
I. Mizuno\inst{\ref{inst12}}\and
J. M. Moran\inst{\ref{inst4},\ref{inst11}}\and
K. Moriyama\inst{\ref{inst2},\ref{inst33}}\and
M. Moscibrodzka\inst{\ref{inst17}}\and
C. M\"uller\inst{\ref{inst6},\ref{inst17}}\and
H. Nagai\inst{\ref{inst3},\ref{inst34}}\and
N. M. Nagar\inst{\ref{inst73}}\and
M. Nakamura\inst{\ref{inst7}}\and
R. Narayan\inst{\ref{inst4},\ref{inst11}}\and
G. Narayanan\inst{\ref{inst74}}\and
R. Neri\inst{\ref{inst16}}\and
C. Ni\inst{\ref{inst19},\ref{inst20}}\and
A. Noutsos\inst{\ref{inst6}}\and
H. Okino\inst{\ref{inst33},\ref{inst75}}\and
H. Olivares\inst{\ref{inst36}}\and
G. N. Ortiz-Le\'on\inst{\ref{inst6}}\and
T. Oyama\inst{\ref{inst33}}\and
F. Özel\inst{\ref{inst10}}\and
D. C. M. Palumbo\inst{\ref{inst4},\ref{inst11}}\and
N. Patel\inst{\ref{inst11}}\and
U. Pen\inst{\ref{inst18},\ref{inst76},\ref{inst77},\ref{inst78}}\and
D. W. Pesce\inst{\ref{inst4},\ref{inst11}}\and
V. Pi\'etu\inst{\ref{inst16}}\and
R. Plambeck\inst{\ref{inst79}}\and
A. PopStefanija\inst{\ref{inst74}}\and
B. Prather\inst{\ref{inst43}}\and
J. A. Preciado-L\'opez\inst{\ref{inst18}}\and
D. Psaltis\inst{\ref{inst10}}\and
H. Pu\inst{\ref{inst18}}\and
V. Ramakrishnan\inst{\ref{inst73}}\and
R. Rao\inst{\ref{inst15}}\and
M. G. Rawlings\inst{\ref{inst12}}\and
A. W. Raymond\inst{\ref{inst4},\ref{inst11}}\and
L. Rezzolla\inst{\ref{inst36}}\and
B. Ripperda\inst{\ref{inst98},\ref{inst99}}\and
A. Rogers\inst{\ref{inst2}}\and
E. Ros\inst{\ref{inst6}}\and
M. Rose\inst{\ref{inst10}}\and
A. Roshanineshat\inst{\ref{inst10}}\and
H. Rottmann\inst{\ref{inst6}}\and
A. L. Roy\inst{\ref{inst6}}\and
C. Ruszczyk\inst{\ref{inst2}}\and
B. R. Ryan\inst{\ref{inst80},\ref{inst81}}\and
K. L. J. Rygl\inst{\ref{inst63}}\and
S. S\'anchez\inst{\ref{inst82}}\and
D. S\'anchez-Arguelles\inst{\ref{inst50},\ref{inst83}}\and
M. Sasada\inst{\ref{inst33},\ref{inst84}}\and
T. Savolainen\inst{\ref{inst6},\ref{inst85},\ref{inst86}}\and
F. P. Schloerb\inst{\ref{inst74}}\and
K. Schuster\inst{\ref{inst16}}\and
L. Shao\inst{\ref{inst6},\ref{inst49}}\and
Z. Shen \inst{\ref{inst30},\ref{inst31}}\and
D. Small\inst{\ref{inst56}}\and
B. Won Sohn\inst{\ref{inst21},\ref{inst22},\ref{inst87}}\and
J. SooHoo\inst{\ref{inst2}}\and
F. Tazaki\inst{\ref{inst33}}\and
P. Tiede\inst{\ref{inst19},\ref{inst20}}\and
M. Titus\inst{\ref{inst2}}\and
K. Toma\inst{\ref{inst89},\ref{inst90}}\and
P. Torne\inst{\ref{inst6},\ref{inst82}}\and
E. Traianou\inst{\ref{inst6}}\and
T. Trent\inst{\ref{inst10}}\and
S. Trippe\inst{\ref{inst91}}\and
S. Tsuda\inst{\ref{inst33}}\and
H. J. van Langevelde\inst{\ref{inst56},\ref{inst92}}\and
D. R. van Rossum\inst{\ref{inst17}}\and
J. Wagner\inst{\ref{inst6}}\and
J. Wardle\inst{\ref{inst93}}\and
J. Weintroub\inst{\ref{inst4},\ref{inst11}}\and
N. Wex\inst{\ref{inst6}}\and
R. Wharton\inst{\ref{inst6}}\and
M. Wielgus\inst{\ref{inst4},\ref{inst11}}\and
G. N. Wong\inst{\ref{inst43},\ref{inst80}}\and
Q. Wu\inst{\ref{inst94}}\and
A. Young\inst{\ref{inst17}}\and
K. Young\inst{\ref{inst11}}\and
Z. Younsi\inst{\ref{inst95},\ref{inst36}}\and
F. Yuan \inst{\ref{inst30},\ref{inst46},\ref{inst96}}\and
Y. Yuan \inst{\ref{inst97}}\and
J. A. Zensus\inst{\ref{inst6}}\and
G. Zhao\inst{\ref{inst21}}\and
S. Zhao\inst{\ref{inst17},\ref{inst61}}\and
Z. Zhu\inst{\ref{inst42}} \\ (The Event Horizon Telescope Collaboration)
}

\institute{
Department of Astrophysics, Institute for Mathematics, Astrophysics and Particle Physics (IMAPP), Radboud University, P.O. Box 9010, 6500 GL Nijmegen, The Netherlands\label{inst17} \\ \email{f.roelofs@astro.ru.nl}\and
Centre for Radio Astronomy Techniques and Technologies, Department of Physics and Electronics, Rhodes University, Grahamstown 6140, South Africa\label{inst39}\and
Department of Physics, University of Pretoria, Lynnwood Road, Hatfield, Pretoria 0083, South Africa\label{inst38}\and
Institut f\"ur Theoretische Physik, Goethe-Universit\"at Frankfurt, Max-von-Laue-Stra{\ss}e 1, D-60438 Frankfurt am Main, Germany\label{inst36}\and
Anton Pannekoek Institute for Astronomy, University of Amsterdam, Science Park 904, 1098 XH, Amsterdam, The Netherlands\label{inst29}\and
Center for Astrophysics | Harvard \& Smithsonian, 60 Garden Street, Cambridge, MA 02138, USA\label{inst11}\and
Black Hole Initiative at Harvard University, 20 Garden Street, Cambridge, MA 02138, USA\label{inst4}\and
California Institute of Technology, 1200 East California Boulevard, Pasadena, CA 91125, USA\label{inst14}\and
Leiden Observatory---Allegro, Leiden University, P.O. Box 9513, 2300 RA Leiden, The Netherlands\label{inst45}\and
Netherlands Organisation for Scientific Research (NWO), Postbus 93138, 2509 AC Den Haag, The Netherlands\label{inst88}\and
Joint Institute for VLBI ERIC (JIVE), Oude Hoogeveensedijk 4, 7991 PD Dwingeloo, The Netherlands\label{inst56}\and
National Radio Astronomy Observatory, 520 Edgemont Rd, Charlottesville, VA 22903, USA\label{inst1}\and
Massachusetts Institute of Technology Haystack Observatory, 99 Millstone Road, Westford, MA 01886, USA\label{inst2}\and
National Astronomical Observatory of Japan, 2-21-1 Osawa, Mitaka, Tokyo 181-8588, Japan\label{inst3}\and
Instituto de Astrof\'{\i}sica de Andaluc\'{\i}a-CSIC, Glorieta de la Astronom\'{\i}a s/n, E-18008 Granada, Spain\label{inst5}\and
Max-Planck-Institut f\"ur Radioastronomie, Auf dem H\"ugel 69, D-53121 Bonn, Germany\label{inst6}\and
Institute of Astronomy and Astrophysics, Academia Sinica, 11F of Astronomy-Mathematics Building, AS/NTU No. 1, Sec. 4, Roosevelt Rd, Taipei 10617, Taiwan, R.O.C.\label{inst7}\and
Departament d'Astronomia i Astrof\'{\i}sica, Universitat de Val\`encia, C. Dr. Moliner 50, E-46100 Burjassot, Val\`encia, Spain\label{inst8}\and
Observatori Astronòmic, Universitat de Val\`encia, C. Catedr\'atico Jos\'e Beltr\'an 2, E-46980 Paterna, Val\`encia, Spain\label{inst9}\and
Steward Observatory and Department of Astronomy, University of Arizona, 933 N. Cherry Ave., Tucson, AZ 85721, USA\label{inst10}\and
East Asian Observatory, 660 N. A'ohoku Place, Hilo, HI 96720, USA\label{inst12}\and
Nederlandse Onderzoekschool voor Astronomie (NOVA), PO Box 9513, 2300 RA Leiden, The Netherlands\label{inst13}\and
Institute of Astronomy and Astrophysics, Academia Sinica, 645 N. A'ohoku Place, Hilo, HI 96720, USA\label{inst15}\and
Institut de Radioastronomie Millim\'etrique, 300 rue de la Piscine, F-38406 Saint Martin d'H\`eres, France\label{inst16}\and
Perimeter Institute for Theoretical Physics, 31 Caroline Street North, Waterloo, ON, N2L 2Y5, Canada\label{inst18}\and
Department of Physics and Astronomy, University of Waterloo, 200 University Avenue West, Waterloo, ON, N2L 3G1, Canada\label{inst19}\and
Waterloo Centre for Astrophysics, University of Waterloo, Waterloo, ON N2L 3G1 Canada\label{inst20}\and
Korea Astronomy and Space Science Institute, Daedeok-daero 776, Yuseong-gu, Daejeon 34055, Republic of Korea\label{inst21}\and
University of Science and Technology, Gajeong-ro 217, Yuseong-gu, Daejeon 34113, Republic of Korea\label{inst22}\and
Kavli Institute for Cosmological Physics, University of Chicago, 5640 South Ellis Avenue, Chicago, IL 60637, USA\label{inst23}\and
Department of Astronomy and Astrophysics, University of Chicago, 5640 South Ellis Avenue, Chicago, IL 60637, USA\label{inst24}\and
Department of Physics, University of Chicago, 5640 South Ellis Avenue, Chicago, IL 60637, USA\label{inst25}\and
Enrico Fermi Institute, University of Chicago, 5640 South Ellis Avenue, Chicago, IL 60637, USA\label{inst26}\and
Princeton Center for Theoretical Science, Jadwin Hall, Princeton University, Princeton, NJ 08544, USA\label{inst100}\and
NASA Hubble Fellowship Program, Einstein Fellow\label{inst101}\and
Data Science Institute, University of Arizona, 1230 N. Cherry Ave., Tucson, AZ 85721, USA\label{inst27}\and
Cornell Center for Astrophysics and Planetary Science, Cornell University, Ithaca, NY 14853, USA\label{inst28}\and
Shanghai Astronomical Observatory, Chinese Academy of Sciences, 80 Nandan Road, Shanghai 200030, People's Republic of China\label{inst30}\newline\and
Key Laboratory of Radio Astronomy, Chinese Academy of Sciences, Nanjing 210008, People's Republic of China\label{inst31}\and
Department of Space, Earth and Environment, Chalmers University of Technology, Onsala Space Observatory, SE-43992 Onsala, Sweden\label{inst32}\and
Mizusawa VLBI Observatory, National Astronomical Observatory of Japan, 2-12 Hoshigaoka, Mizusawa, Oshu, Iwate 023-0861, Japan\label{inst33}\and
Department of Astronomical Science, The Graduate University for Advanced Studies (SOKENDAI), 2-21-1 Osawa, Mitaka, Tokyo 181-8588, Japan\label{inst34}\and
Dipartimento di Fisica ``E. Pancini'', Universit\'a di Napoli ``Federico II'', Compl. Univ. di Monte S. Angelo, Edificio G, Via Cinthia, I-80126, Napoli, Italy\label{inst35}\and
INFN Sez. di Napoli, Compl. Univ. di Monte S. Angelo, Edificio G, Via Cinthia, I-80126, Napoli, Italy\label{inst37}\and
LESIA, Observatoire de Paris, Universit\'e PSL, CNRS, Sorbonne Universit\'e, Universit\'e de Paris, 5 place Jules Janssen, 92195 Meudon, France\label{inst102}\and
Max-Planck-Institut f\"ur Extraterrestrische Physik, Giessenbachstr. 1, D-85748 Garching, Germany\label{inst40}\and
Department of History of Science, Harvard University, Cambridge, MA 02138, USA\label{inst41}\and
Department of Physics, Harvard University, Cambridge, MA 02138, USA\label{inst42}\and
Department of Physics, University of Illinois, 1110 West Green St, Urbana, IL 61801, USA\label{inst43}\and
Department of Astronomy, University of Illinois at Urbana-Champaign, 1002 West Green Street, Urbana, IL 61801, USA\label{inst44}\and
CP3-Origins, University of Southern Denmark, Campusvej 55, DK-5230 Odense M, Denmark\label{inst103}
Key Laboratory for Research in Galaxies and Cosmology, Chinese Academy of Sciences, Shanghai 200030, People's Republic of China\label{inst46}\and
NOVA Sub-mm Instrumentation Group, Kapteyn Astronomical Institute, University of Groningen, Landleven 12, 9747 AD Groningen, The Netherlands\label{inst47}\and
Department of Astronomy, School of Physics, Peking University, Beijing 100871, People's Republic of China\label{inst48}\and
Kavli Institute for Astronomy and Astrophysics, Peking University, Beijing 100871, People's Republic of China\label{inst49}\and
Instituto Nacional de Astrof\'{\i}sica, \'Optica y Electr\'onica. Apartado Postal 51 y 216, 72000. Puebla Pue., M\'exico\label{inst50}\and
The Institute of Statistical Mathematics, 10-3 Midori-cho, Tachikawa, Tokyo, 190-8562, Japan\label{inst51}\and
Department of Statistical Science, The Graduate University for Advanced Studies (SOKENDAI), 10-3 Midori-cho, Tachikawa, Tokyo 190-8562, Japan\label{inst52}\and
Kavli Institute for the Physics and Mathematics of the Universe, The University of Tokyo, 5-1-5 Kashiwanoha, Kashiwa, 277-8583, Japan\label{inst53}\and
Institute for Astrophysical Research, Boston University, 725 Commonwealth Ave., Boston, MA 02215, USA\label{inst54}\and
Astronomical Institute, St. Petersburg University, Universitetskij pr., 28, Petrodvorets,198504 St.Petersburg, Russia\label{inst55}\and
Kogakuin University of Technology \& Engineering, Academic Support Center, 2665-1 Nakano, Hachioji, Tokyo 192-0015, Japan\label{inst57}\and
Physics Department, National Sun Yat-Sen University, No. 70, Lien-Hai Rd, Kaosiung City 80424, Taiwan, R.O.C\label{inst58}\and
National Optical Astronomy Observatory, 950 North Cherry Ave., Tucson, AZ 85719, USA\label{inst59}\and
Key Laboratory for Particle Astrophysics, Institute of High Energy Physics, Chinese Academy of Sciences, 19B Yuquan Road, Shijingshan District, Beijing, People's Republic of China\label{inst60}\and
School of Astronomy and Space Science, Nanjing University, Nanjing 210023, People's Republic of China\label{inst61}\and
Key Laboratory of Modern Astronomy and Astrophysics, Nanjing University, Nanjing 210023, People's Republic of China\label{inst62}\and
Italian ALMA Regional Centre, INAF-Istituto di Radioastronomia, Via P. Gobetti 101, I-40129 Bologna, Italy\label{inst63}\and
Department of Physics, National Taiwan University, No.1, Sect.4, Roosevelt Rd., Taipei 10617, Taiwan, R.O.C\label{inst64}\and
Instituto de Radioastronom\'{\i}a y Astrof\'{\i}sica, Universidad Nacional Aut\'onoma de M\'exico, Morelia 58089, M\'exico\label{inst65}\and
Instituto de Astronom\'{\i}a, Universidad Nacional Aut\'onoma de M\'exico, CdMx 04510, M\'exico\label{inst66}\and
Yunnan Observatories, Chinese Academy of Sciences, 650011 Kunming, Yunnan Province, People's Republic of China\label{inst67}\and
Center for Astronomical Mega-Science, Chinese Academy of Sciences, 20A Datun Road, Chaoyang District, Beijing, 100012, People's Republic of China\label{inst68}\and
Key Laboratory for the Structure and Evolution of Celestial Objects, Chinese Academy of Sciences, 650011 Kunming, People's Republic of China\label{inst69}\and
Gravitation Astroparticle Physics Amsterdam (GRAPPA) Institute, University of Amsterdam, Science Park 904, 1098 XH Amsterdam, The Netherlands\label{inst70}\and
Centro Astron\'omico de Yebes (IGN), Apartado 148, E-19180 Yebes, Spain\label{inst71}\and
School of Natural Sciences, Institute for Advanced Study, 1 Einstein Drive, Princeton, NJ 08540, USA\label{inst72}\and
NSF Astronomy and Astrophysics Postdoctoral Fellow under award no. AST-1903847\label{inst104}\and
Astronomy Department, Universidad de Concepci\'on, Casilla 160-C, Concepci\'on, Chile\label{inst73}\and
Department of Astronomy, University of Massachusetts, 01003, Amherst, MA, USA\label{inst74}\and
Department of Astronomy, Graduate School of Science, The University of Tokyo, 7-3-1 Hongo, Bunkyo-ku, Tokyo 113-0033, Japan\label{inst75}\and
Canadian Institute for Theoretical Astrophysics, University of Toronto, 60 St. George Street, Toronto, ON M5S 3H8, Canada\label{inst76}\and
Dunlap Institute for Astronomy and Astrophysics, University of Toronto, 50 St. George Street, Toronto, ON M5S 3H4, Canada\label{inst77}\and
Canadian Institute for Advanced Research, 180 Dundas St West, Toronto, ON M5G 1Z8, Canada\label{inst78}\and
Radio Astronomy Laboratory, University of California, Berkeley, CA 94720, USA\label{inst79}\and
Department of Astrophysical Sciences, Peyton Hall, Princeton University, Princeton, NJ 08544, USA\label{inst98}\newpage\and
Center for Computational Astrophysics, Flatiron Institute, 162 Fifth Avenue, New York, NY 10010, USA\label{inst99}\and
CCS-2, Los Alamos National Laboratory, P.O. Box 1663, Los Alamos, NM 87545, USA\label{inst80}\and
Center for Theoretical Astrophysics, Los Alamos National Laboratory, Los Alamos, NM, 87545, USA\label{inst81}\and
Instituto de Radioastronom\'{\i}a Milim\'etrica, IRAM, Avenida Divina Pastora 7, Local 20, E-18012, Granada, Spain\label{inst82}\and
Consejo Nacional de Ciencia y Tecnolog\'{\i}a, Av. Insurgentes Sur 1582, 03940, Ciudad de M\'exico, M\'exico\label{inst83}\and
Hiroshima Astrophysical Science Center, Hiroshima University, 1-3-1 Kagamiyama, Higashi-Hiroshima, Hiroshima 739-8526, Japan\label{inst84}\and
Aalto University Department of Electronics and Nanoengineering, PL 15500, FI-00076 Aalto, Finland\label{inst85}\and
Aalto University Mets\"ahovi Radio Observatory, Mets\"ahovintie 114, FI-02540 Kylm\"al\"a, Finland\label{inst86}\and
Department of Astronomy, Yonsei University, Yonsei-ro 50, Seodaemun-gu, 03722 Seoul, Republic of Korea\label{inst87}\and
Frontier Research Institute for Interdisciplinary Sciences, Tohoku University, Sendai 980-8578, Japan\label{inst89}\and
Astronomical Institute, Tohoku University, Sendai 980-8578, Japan\label{inst90}\and
Department of Physics and Astronomy, Seoul National University, Gwanak-gu, Seoul 08826, Republic of Korea\label{inst91}\and
Leiden Observatory, Leiden University, Postbus 2300, 9513 RA Leiden, The Netherlands\label{inst92}\and
Physics Department, Brandeis University, 415 South Street, Waltham, MA 02453, USA\label{inst93}\and
School of Physics, Huazhong University of Science and Technology, Wuhan, Hubei, 430074, People's Republic of China\label{inst94}\and
Mullard Space Science Laboratory, University College London, Holmbury St. Mary, Dorking, Surrey, RH5 6NT, UK\label{inst95}\and
School of Astronomy and Space Sciences, University of Chinese Academy of Sciences, No. 19A Yuquan Road, Beijing 100049, People's Republic of China\label{inst96}\and
Astronomy Department, University of Science and Technology of China, Hefei 230026, People's Republic of China\label{inst97}
}

   \date{Received 3 September 2019 / Accepted 18 October 2019}

 
  \abstract
   {
   Realistic synthetic observations of theoretical source models are essential for our understanding of real observational data. In using synthetic data, one can verify the extent to which source parameters can be recovered and evaluate how various data corruption effects can be calibrated. These studies are the most important when proposing observations of new sources, in the characterization of the capabilities of new or upgraded instruments, and when verifying model-based theoretical predictions in a direct comparison with observational data.
   }
   {
   We present the SYnthetic Measurement creator for long Baseline Arrays (\pipe{}), a novel synthetic data generation pipeline for Very Long Baseline Interferometry (VLBI) observations. \pipe{} takes into account several realistic atmospheric, instrumental, and calibration effects.
   }
   {
   We used \pipe{} to create synthetic observations for the Event Horizon Telescope (EHT), a millimetre VLBI array, which has recently captured the first image of a black hole shadow. After testing \pipe{} with simple source and corruption models, we study the importance of including all corruption and calibration effects, compared to the addition of thermal noise only. Using synthetic data based on two example general relativistic magnetohydrodynamics (GRMHD) model images of M87, we performed case studies to assess the image quality that can be obtained with the current and future EHT array for different weather conditions.
   }
   {
   Our synthetic observations show that the effects of atmospheric and instrumental corruptions on the measured visibilities are significant. Despite these effects, we demonstrate how the overall structure of our GRMHD source models can be recovered robustly with the EHT2017 array after performing calibration steps, which include fringe fitting, a priori amplitude and network calibration, and self-calibration. With the planned addition of new stations to the EHT array in the coming years, images could be reconstructed with higher angular resolution and dynamic range. In our case study, these improvements allowed for a distinction between a thermal and a non-thermal GRMHD model based on salient features in reconstructed images.
   }
   {}

   \titlerunning{SYMBA: An end-to-end VLBI synthetic data generation pipeline}
   \maketitle
%

\section{Introduction}

\nocite{eht-paperI}
\nocite{eht-paperII}
\nocite{eht-paperIII}
\nocite{eht-paperIV}
\nocite{eht-paperV}
\nocite{eht-paperVI}

The giant elliptical galaxy M87 hosts an active galactic nucleus (AGN) with a radio jet extending to kpc scales \citep[e.g.][]{Owen2000}. The radio core of M87 shifts inwards with increasing frequency as the jet becomes optically thin closer to the central black hole, resulting in a flat radio spectrum as predicted by analytical models \citep{Blandford1979, Falcke1995}. The radio core of M87 coincides with the central engine at 43 GHz \citep{Hada2011}. At millimetre wavelengths, emission near the event horizon becomes optically thin. Due to strong gravitational lensing, the black hole is predicted to cast a `shadow' on this emission \citep{Falcke2000, Dexter2012, Moscibrodzka2016}. The shadow is a region exhibiting an emission deficit produced by the capture of photons by the event horizon, with a size enhanced by strong gravitational lensing.

For a Schwarzschild (non-spinning) black hole, the apparent radius of the black hole shadow is $\sqrt{27}R_g$, with $R_g=GM/c^2$ the gravitational radius where $G$ is Newton's gravitational constant, $M$ is the black hole mass, and $c$ is the speed of light. The difference in shadow size between a rotating black hole \citep{Kerr1963} and the Schwarzschild solution is marginal ($\lesssim 4\%$), since the apparent size is nearly independent of the black hole spin \citep{Bardeen1973, Takahashi2004, Johannsen2010}. Estimates for the mass of the supermassive black hole at the centre of M87 have historically ranged between  $(3.5^{+0.9}_{-0.7}) \times 10^9 M_{\odot}$ from gas-dynamical measurements \citep{Walsh2013}, and $(6.6 \pm 0.4) \times 10^9 M_{\odot}$ from stellar-dynamical measurements \citep{Gebhardt2011}. At a distance of $(16.4 \pm 0.5)$ Mpc \citep{Bird2010}, the mass measurements correspond to an apparent diameter of the shadow between $\sim22$ $\mu$as and $42$ $\mu$as.

At 230 GHz, Earth-sized baselines give a nominal resolution of $\sim23$ $\mu$as, which is certainly sufficient to resolve the black hole shadow of M87 for the high-mass estimate. M87 is therefore one of the prime targets of the Event Horizon Telescope (EHT), the Earth-sized mm-Very Long Baseline Interferometry (VLBI) array aiming to image a black hole shadow \mbox{\citep{eht-paperII}}. 
The other prime candidate is Sagittarius A* (Sgr A*). With a better constrained shadow size of $\sim53$ $\mu$as, this is the black hole with the largest predicted angular size in the sky. Interstellar scattering effects and variability on short time scales (minutes) may make reconstructing the black hole shadow challenging for this source. On the other hand, it provides us with opportunities to study scattering effects \citep{Johnson2016so, Dexter2017, Johnson2018} and real-time dynamics of the accretion flow \citep[e.g.][]{Doeleman2009cl, Fish2009, Dexter2010, Medeiros2016, Roelofs2017, Johnson2017, Bouman2017}. In this paper, we focus on synthetic EHT observations of M87, where orbital timescales are much larger than those of the observations. 

With the EHT data sets and images, it is possible to test general relativity in a unique environment \citep[e.g.][]{Bambi2009, Johannsen2010, Psaltis2015, Goddi2016, eht-paperI}. Also, constraints can be put on models of the accretion flow around supermassive black holes  \citep[e.g.][]{Falcke2000b, Yuan2003, Dexter2010, Dexter2012, Moscibrodzka2014, Moscibrodzka2016, Chan2015, Broderick2016, Gold2017, eht-paperV}.

In 2017, the EHT consisted of the IRAM 30-metre telescope on Pico Veleta in Spain, the Large Millimeter Telescope (LMT) in Mexico, the Atacama Large Millemeter Array (ALMA), the Atacama Pathfinder Experiment (APEX) telescope in Chile, the Sub-Millimeter Telescope (SMT) in Arizona, the Sub-Millimeter Array and James Clerk Maxwell Telescope (JCMT) in Hawaii, and the South Pole Telescope (SPT). In the April 2017 observing run (hereafter EHT2017) and a subsequent two-year analysis period, the EHT imaged the M87 black hole shadow within a $42 \pm 3$ $\mu$as asymmetric emission ring \citep{eht-paperIV, eht-paperVI}. The measured ring size, when associated with a black hole shadow, leads to an angular size of one gravitational radius of $3.8\pm0.4$ $\mu$as \citep{eht-paperVI}. At the adopted distance of $16.8^{+0.8}_{-0.7}$ Mpc that was calculated from multiple measurements \citep{Bird2010, Blakeslee2009, Cantiello2018}, this angular size corresponds to a black hole mass of $(6.5\pm0.2|_{\text{stat}}\pm0.7|_{\text{sys}})\times 10^9M_{\odot}$, which is consistent with the stellar-dynamical mass measurement by \citet{Gebhardt2011}.

Over the years, synthetic data have proven to be of importance for demonstrating the capabilities of the EHT. They were also essential for developing new strategies to increase the scientific output of the rich, yet challenging, observations. 

\citet{Doeleman2009cl} and \citet{Fish2009} used the Astronomical Image Processing System (AIPS)\footnote{\url{http://www.aips.nrao.edu}} task UVCON to calculate model visibilities for the EHT array, showing that signatures of source variability could be detected in Sgr A* by using interferometric closure quantities and polarimetric ratios. The MIT Array Performance Simulator (MAPS)\footnote{\url{https://www.haystack.mit.edu/ast/arrays/maps}.} was used in several EHT synthetic imaging studies. \citet{Lu2014} used it to test the ability of the EHT to reconstruct images of the black hole shadow for several models of the accretion flow of M87. \citet{Fish2014} demonstrate that for Sgr A*, the blurring effect of interstellar scattering could be mitigated if the properties of the scattering kernel are known. \citet{Lu2016} showed that source variability could also be mitigated by observing the source for multiple epochs and applying visibility averaging, normalization, and smoothing to reconstruct an image of the average source structure.

Typically, the only data corruption included in these synthetic data sets is thermal noise, although \citet{Fish2009} also included instrumental polarization. More corruptions can be added with the \ehtim{} library\footnote{\url{https://github.com/achael/eht-imaging}.}. \citet{Chael2016, Chael2018} simulated polarimetric EHT images of Sgr A* and M87, and included randomly varying complex station gains and elevation-dependent atmospheric opacity terms. With the stochastic optics module in \ehtim{}, the input model images can be scattered using a variable refractive scattering screen, and the scattering can be mitigated by solving for the scattering screen and image simultaneously \citep{Johnson2016so}.
However, scattering effects are only relevant for observations of Sgr A*.
\ehtim{} can also simulate observations following a real observing schedule, and copy the $uv$-coverage and thermal noise directly from existing data sets. It also includes polarimetric leakage corruptions \citep{eht-paperIV}.

Despite these recent advances in synthetic data generation, there are still differences between synthetic and real mm-VLBI data sets. So far, synthetic EHT data sets have not been frequency-resolved, and gain offsets have only been included as random relative offsets drawn from a Gaussian with a fixed standard deviation, rather than being based on a physical model.

Moreover, no calibration effects are taken into account in the synthetic data products. It is essentially assumed that residual delays, phase decoherence due to atmospheric turbulence, and signal attenuation caused by the atmospheric opacity are perfectly calibrated. In \ehtim{}, atmospheric turbulence can be included by fully randomizing the phases (with the option of fixing them within a scan). In real mm-VLBI data, atmospheric turbulence results in rapid phase wraps. The correlated phases are not fully randomized, but evolve continuously over frequency and time, allowing to perform fringe fitting and average complex visibilities coherently on time scales set by the atmospheric coherence time. 

In this paper, we present the SYnthetic Measurement creator for long Baseline Arrays (\pipe{}) -- a new synthetic VLBI data generation and calibration pipeline.\footnote{\url{https://bitbucket.org/M_Janssen/symba}.}

We generate raw synthetic data with \meqsil{}\footnote{\url{https://github.com/rdeane/MeqSilhouette_public_v0.1}.} \citep{Blecher2017,Natarajan2019}, which includes a tropospheric module and physically motivated antenna pointing offsets (Section \ref{sec:meqsilhouette}). We then calibrate the raw data using the new \casa{} \citep{McMullin2007} VLBI data calibration pipeline \rpicard{}\footnote{\url{https://bitbucket.org/M_Janssen/picard}.} \citep{Janssen2019}, applying a fringe fit and a priori amplitude calibration (Section \ref{sec:rpicard}).  The overall computing workflow of \pipe{} is outlined in Section~\ref{sec:computingworkflow}. We describe our simulated observational setup (antenna and weather parameters and observing schedule) in Section~\ref{sec:obsparameters} and our input source models for the synthetic data generation in Section~\ref{sec:sourcemodels}. In Section~\ref{sec:study_corandcal}, we demonstrate the effects of simulated data corruptions and subsequent calibration. We illustrate the capabilities of \pipe{} in Section~\ref{sec:case_studies} based on three scientific case studies. In these studies we show 1) how well we can distinguish between two example general relativistic magnetohydrodynamics (GRMHD) models with different descriptions for the electron temperatures with the current and future EHT array, 2) how the EHT would perform under different weather conditions, and 3) how pre-2017 models of M87 compare to the observed image of the black hole shadow. In Section \ref{sec:summary}, we summarize our conclusions and discuss future work.

\section{Synthetic data generation with \meqsil{}}
\label{sec:meqsilhouette}
\meqsil{} \citep{Blecher2017, Natarajan2019} is a synthetic data generator designed to simulate high frequency VLBI observations. While visibilities of real radio interferometric observations are produced by correlating recorded voltage streams from pairs of telescopes, \meqsil{} predicts visibilities directly from the Fourier Transform of an input sky model. For simple ASCII input models (e.g. a set of Gaussian components, each with an independent spectral index), \texttt{MeqTrees} \mbox{\citep{Noordam2010}} is used for the visibility prediction. FITS-based\footnote{See \url{https://fits.gsfc.nasa.gov/fits_documentation.html} for a definition of the FITS standard.} sky models are converted with the \mbox{\texttt{wsclean}} \mbox{\citep{Offringa2014}} algorithm. The signal path is described by the Measurement Equation formalism \mbox{\citep{Hamaker1996}}, breaking down the various effects on the visibilities into a chain of complex $2\times 2$ Jones matrices \mbox{\citep{Jones1941, Smirnov2011a, Smirnov2011b, Smirnov2011c}}. \meqsil{} generates frequency-resolved visibilities, with a bandwidth and number of channels set by the user. Frequency-resolved visibilities are required for the calibration of signal path variations introduced by the troposphere. In particular, synthetic data from \meqsil{} has been used to validate the CASA-based data reduction path of the EHT. Moreover, channelized data allows for the introduction of frequency dependent leakage of polarized signals at telescopes’ receivers, the inclusion of wavelength dependent Faraday rotation and spectral indices in source models, and multi-frequency aperture synthesis, which can yield significant improvements to the $uv$-coverage.\footnote{For example, the EHT is currently able to observe with two sidebands separated by 18 GHz, which \meqsil{} can replicate.}
It is also possible to generate corrupted data sets from time-dependent polarized emission models in full Stokes and to follow an observed schedule from a VEX file.\footnote{See \url{https://vlbi.org/vlbi-standards/vex/} for a definition of the VEX file format.} A key design driver of \meqsil{} is to generate synthetic data (and associated meta-data) in a format that is seamlessly ingested by the \casa{} software package. The native format is the MeasurementSet (MS)\footnote{See \url{https://casa.nrao.edu/Memos/229.html} for the definition of the MeasurementSet format.}, but the visibilities can also be exported to UVFITS.\footnote{See \url{ftp://ftp.aoc.nrao.edu/pub/software/aips/TEXT/PUBL/AIPSMEM117.PS} for a description of the UVFITS file data format.} We briefly describe the added tropospheric and instrumental corruptions below, referring to \mbox{\citet{Blecher2017}} and \citet{Natarajan2019} for more details.

\subsection{Tropospheric corruptions}
\label{sec:trop}
The effects of the troposphere on the measured visibilities can be separated into those resulting from a mean atmospheric profile, and those resulting from atmospheric turbulence. 

\subsubsection{Mean troposphere}
\label{sec:meantrop}
The mean troposphere causes time delays, resulting in phase slopes versus frequency and an attenuation of the visibility amplitudes due to absorption of radiation in molecular transitions \citep{TMS2017}. In the mm-wave regime, absorption lines are mostly caused by rotational transitions of H$_2$O and O$_2$. Apart from the individual lines, there is a general increase of the opacity with frequency due to the cumulative effect of pressure-broadened H$_2$O lines peaking in the THz-regime \citep{Carilli1999}.

\meqsil{} calculates the attenuation and time delays using the Atmospheric Transmission at Microwaves (\texttt{ATM}) software \citep{Pardo2001}. It integrates the radiative transfer equation
\begin{equation}
\label{eq:radtransfer}
\frac{\mathrm{d}I_{\nu}(s)}{\mathrm{d}s} = \epsilon_{\nu}(s) - \kappa_{\nu}(s)I_{\nu}(s)\,,
\end{equation}
where $I_{\nu}(s)$ is the specific intensity at frequency $\nu$ at path length coordinate $s$, and $\epsilon_{\nu}$ and $\kappa_{\nu}$ are the emission and absorption coefficients, respectively. In thermodynamic equilibrium, the latter are related through Kirchhoff's law,
\begin{equation}
\label{eq:kirchhoff}
\frac{\epsilon_{\nu}}{\kappa_{\nu}} = B_{\nu}(T)\,,
\end{equation}
where $B_{\nu}(T)$ is the Planck spectrum at temperature $T$. In order to integrate Equation~\ref{eq:radtransfer}, \texttt{ATM} calculates $\kappa_{\nu}$ as a function of altitude. For a specific transition, $\kappa_{\nu}$ is proportional to the photon energy, the transition probability (Einstein coefficient), molecular densities of the lower and upper states, and the line shape including pressure and Doppler broadening. $\kappa_{\nu}$ is related to the refractive index of the medium via the Kramers-Kronig relations. The introduced time delay is then calculated from the refractive index. 

As is evident from Kirchhoff's law (Equation~\ref{eq:kirchhoff}), the atmosphere not only absorbs, but also emits radiation. This process leads to an increase in system temperature (sky noise), which also follows from the integration in \texttt{ATM} and is included in the noise budget with an elevation and therefore time-dependent contribution. 

\subsubsection{Turbulent troposphere}
Apart from the mean troposphere induced amplitude attenuation, signal delay, and sky noise, a major source of data corruptions in the mm regime is tropospheric turbulence. Rapid evolution of the spatial distribution of tropospheric water vapour causes the signal path delay to vary on short ($\sim10$ s) time scales. This then leads to rapid and unpredictable rotations of the visibility phase, posing challenges for fringe fitting. Because of atmospheric turbulence, uncalibrated visibilities can not be coherently averaged beyond the atmospheric coherence time. Absolute phase information can only be recovered with phase-referencing \citep{Beasley1995}. For imaging mm-VLBI data, one often needs to rely on closure phases \citep[e.g.][]{Chael2018}. Closure phase is the sum of visibility phases on a triangle of baselines, in which many station-based instrumental and atmospheric corruptions cancel out \citep{Jennison1958,Rogers1974}.

In \meqsil{}, turbulent phase errors are added to the visibilities assuming that the atmospheric turbulence can be represented by a thin phase-changing scattering screen. Similar to simulations of interstellar scattering \citep[e.g.][]{Johnson2015}, the turbulent substructure of the screen is assumed to be constant in time while the screen itself is moving with a constant transverse velocity $v$. The screen velocity sets the atmospheric coherence time together with the spatial phase turbulence scale on the screen. The introduced phase offsets are described by a phase structure function that takes a power law form,
\begin{equation}
D_{\phi}(\vec{x}, \vec{x'}) = \langle[\phi(\vec{x}+\vec{x'})-\phi(\vec{x})]^2\rangle \approx \mu (r/r_0)^{\beta}\,,
\end{equation}
where $\vec{x}$ and $\vec{x'}$ are spatial coordinates on the screen, $r^2=(\vec{x}-\vec{x'})^2$, $r_0$ is the phase coherence length such that $D_{\phi}(r_0) = 1$ rad, $\mu=\csc{(\mathrm{elevation})}$ is the airmass towards the horizon\footnote{The $\csc{(\mathrm{elevation})}$ dependence of the airmass is an approximation assuming a planar rather than a spherical atmosphere, which breaks down at elevations below $\sim$\,10 degrees \citep{Paine2019}. For the synthetic observations in this work, we set the elevation limit to 10 degrees as is typically done for real VLBI observations. Hence, the $\csc{(\mathrm{elevation})}$ approximation has a negligible effect on our results.}, and $\beta=5/3$ if one assumes Kolmogorov turbulence, which is supported by \citet{Carilli1999}. The nature of the scattering is set by the ratio of $r_0$ and the Fresnel scale $r_{\mathrm{F}}=\sqrt{\lambda D_{\mathrm{os}}/2\pi}$, where $D_{\mathrm{os}}$ is the distance between the observer and the scattering screen. With $r_0$ measured to be $\sim\,50-700$ m \citep{Masson1994, Radford1998} and a water vapour scale height of 2 km, we have $r_\mathrm{F}\approx 0.45$ m $<r_0$ and are in the weak scattering regime. This means that most of the received power on the ground originates from a screen area $A_\mathrm{weak}\approx\pi r_{\mathrm{F}}^2$, rather than from disjoint patches, as is the case for interstellar scattering. At a distance of 2 km, 1 mas corresponds to $\sim10$ $\mu$m, and the Field of View (FoV) of the array is much smaller than $r_0$. The phase error is therefore assumed to be constant across the FoV, and the structure function can be written as $D(t)=D(r)|_{r=vt}$, where $v$ is the bulk transverse velocity of the phase screen. From this, a phase error time sequence can be computed directly. Due to the long baselines, atmospheric corruptions can be modelled independently at each station \citep{Carilli1999}. For a given coherence time $t_{\mathrm{c}}=r_0/v$ \citep{Treuhaft1987} at a reference frequency $\nu_0$, \mbox{\citet{Blecher2017}} showed that the temporal variance of the phase for a power-law turbulence as a function of frequency $\nu$ can be modelled as
\begin{equation}
    \label{eq:coher_def}
    \sigma^2_\phi(t_{\mathrm{c}}, \nu) = \left[ \frac{\mu}{\beta^2 + 3\beta + 2} \right] \left( \frac{t_\mathrm{int}}{t_{\mathrm{c}}} \right)^\beta \left( \frac{\nu}{\nu_0} \right) \, \mathrm{rad}^2 \, ,
\end{equation}
where $t_\mathrm{int}$ is the data integration time and $\nu_0$ is taken as the lowest frequency in the data. \meqsil{} uses Equation~\ref{eq:coher_def} to compute the tropospheric phase turbulence using $\beta = 5/3$. A constant amount of precipitable water vapour at zenith ($\mathrm{PVW}_0$) is assumed, mixed evenly into the atmosphere. An increase in the phase variance due to the PWV therefore enters through the amount of airmass towards the horizon in Equation~\ref{eq:coher_def}. The specified coherence time $t_{\mathrm{c}}=t_{\mathrm{c}}(\mathrm{PWV}_0)$ should decrease with increasing precipitable water vapour content in the atmosphere, although other factors such as wind speed also affect $t_{\mathrm{c}}$. No sudden phase jumps due to inhomogeneities in the atmosphere (e.g. clouds or airmass boundary kinks) along the line of sight are simulated. Phase turbulence and resulting decorrelation within an integration time $t_\mathrm{int}$ are not simulated by \meqsil{}. For realistic results, $t_{\mathrm{int}}$ should therefore preferably be set to well within $t_{\mathrm{c}}$, as is the case for real observations. Delay-related decoherence effects within individual frequency channels are also not simulated. It is assumed that frequency resolution is sufficiently high to make this effect negligible, as it is done in modern correlators.

\subsection{Receiver noise}
The System Equivalent Flux Density (SEFD) of a station is a measure for its overall noise contribution. \meqsil{} reads $\mathcal{S}_\mathrm{rx}$, the contribution from the receiver noise to the SEFD, from input files. Receiver temperatures $T_\mathrm{rx}$ are typically determined from real data by extrapolating system temperatures to zero airmass and the receiver noise contribution in units of Jansky (Jy) follows as 
\begin{equation}
\mathcal{S}_\mathrm{rx} = \frac{T_\mathrm{rx}}{\mathrm{DPFU}}\,.
\end{equation}
Here, the DPFU is the telescope's `degree per flux unit' gain, defined as $\mathrm{DPFU} = \eta_\mathrm{ap} A_\mathrm{dish} / \left( 2 k_\mathrm{B} \right)$, with $\eta_\mathrm{ap}$ the aperture efficiency (taken to be constant during observations), $A_\mathrm{dish}$ the geometric area of the dish, and $k_\mathrm{B}$ the Boltzmann constant.

\subsection{The full noise budget}
Visibilities on all baselines are corrupted by the addition of noise as a complex Gaussian variable with standard deviation
\begin{equation}
\sigma_{mn}=\frac{1}{\eta_\mathrm{Q}}\sqrt{\frac{\mathrm{SEFD}_m{\mathrm{SEFD}_n}}{2\Delta\nu t_{\mathrm{int}}}}\,,
\end{equation} 
where $\mathrm{SEFD}_\mathrm{m}$ is the system equivalent flux density from station $m$ with combined contributions from the atmosphere and receiver, $\Delta\nu$ is the channel bandwidth, $t_{\mathrm{int}}$ is the correlator integration time, and $\eta_\mathrm{Q}$ is a quantization efficiency factor, set to 0.88 for standard 2-bit quantization. We assume perfect quantization thresholds when simulating the cross-correlation data. Therefore, we do not need to simulate the auto-correlations to correct for erroneous sampler thresholds.
All noise sources along the signal chain (sky noise, turbulence, and thermal noise from the instrument) enter into $\sigma_{mn}$.
\meqsil{} produces visibilities in a circular polarization basis, that is LL, RR, LR, and RL. The noise on, for example, the Stokes I data is a factor $\sqrt{2}$ smaller.

\subsection{Antenna pointing errors}
\label{sec:pointing}
Pointing offsets of individual antennas manifest as a time and station dependent amplitude error. They cause a drop of the visibility amplitudes $Z_{mn}$ on a $m$-$n$ baseline as the maximum of the antenna primary beam is not pointed on the source. The primary beam profile of a station $m$ is modelled as a Gaussian with a full width at half maximum $\mathcal{P}_{\mathrm{FWHM},\,m}$, which is related to the Gaussian's standard deviation by a factor of $2\sqrt{2\ln{2}} \approx 2.35$. A Gaussian beam is justified since the pointing offsets are not large enough that a Gaussian and Bessel function deviate (i.e. near the first null), see \citet{Middelberg2013}. No further systematic point effects, such as refraction, are considered here. Pointing offsets $\rho_m$ are drawn from a normal distribution $\mathcal{N}$ centred around zero, with a standard deviation given by a specified rms pointing offset $\mathcal{P}_{\mathrm{rms},\,m}$. The resulting visibility amplitude loss
\begin{eqnarray}
\frac{\Delta Z_{mn}}{Z_{mn}} & = &  \mathrm{exp}\left(-8\ln^2{2}\left[\frac{\rho_m^2}{\mathcal{P}_{\mathrm{FWHM},\,m}^2} + \frac{\rho_n^2}{\mathcal{P}_{\mathrm{FWHM},\,n}^2}\right]\right)\,, \\
\rho_m & = & \mathcal{N}\left(\mu=0, \sigma=\mathcal{P}_{\mathrm{rms},\,m}\right)\,, \nonumber
\end{eqnarray}
describes a data corruption effect caused by an erroneous source tracking of the telescopes.

In \pipe{}, we employ two types of pointing offsets, which occur on short and long timescales, respectively. 
The short timescale variations are caused by the atmospheric seeing and wind shaking the telescope, resulting in a displacement of the sky source with respect to an otherwise perfectly pointed telescope beam. Here, \pipe{} draws values of $\rho_m$ from $\mathcal{P}_{\mathrm{rms},\,m}$ on timescales set by the atmospheric coherence time.
The long timescale variations are caused by sub-optimal pointing solutions adopted by a telescope. \pipe{} simulates these by adopting a new value of $\rho_m$ every $N\sim5$ scans and letting these pointing offsets deteriorate by $\xi \sim 0.1$ in every scan until a new offset is determined. For simplicity, the $\rho_m$ are drawn from the same $\mathcal{P}_{\mathrm{rms},\,m}$, multiplied by a factor $\alpha \sim 1.5$. For a scan number $M$, the effect of an incorrect pointing model is thus given as
\begin{equation}
    \label{longTpointing}
    \rho_m  = \left( 1 + \xi \right)^{M \bmod{} N} \mathcal{N}\left(\mu=0, \sigma=\alpha \mathcal{P}_{\mathrm{rms},\,m}\right)\,.
\end{equation}

\subsection{Leakage and gain errors}
Complex gain errors $\mathcal{G}$, that would translate to errors in the \mbox{DPFUs} and phase gains in real observations, and complex leakage effects ($\mathcal{D}$-terms) can be added as well. For observed/corrupted (obs) visibilities from a baseline of stations $m$ and $n$,
$\mathcal{D}$-terms cause artificial instrumental polarization as a rotation of the cross-hand visibilities in the complex plane by twice the station's feed rotation angles $\chi$ \citep{Conway1969}:
\begin{eqnarray}
\label{dterms}
\mathrm{RL}_{mn}^\mathrm{obs} & = & \mathrm{RL}_{mn}^\mathrm{true} + \left[ \mathcal{D}^R_m e^{2 i \chi_m} + \left(\mathcal{D}^L_n\right)^* e^{2 i \chi_n} \right] \mathrm{I} \; , \\
\mathrm{LR}_{mn}^\mathrm{obs} & = & \mathrm{LR}_{mn}^\mathrm{true} + \left[ \mathcal{D}^L_m e^{-2 i \chi_m} + \left(\mathcal{D}^R_n\right)^* e^{-2 i \chi_n} \right]  \mathrm{I} \; .
\end{eqnarray}
Here, $\mathcal{D}$ are the leakage terms, with a superscript indicating the polarization, and $i=\sqrt{-1}$. The star denotes complex conjugation. More complex and realistic polarimetric effects are available in the forthcoming release of {\tt MeqSilhouette v2} \citep{Natarajan2019}.

\section{Synthetic data calibration with \rpicard{}}
\label{sec:rpicard}

The goal of \pipe{} is to create synthetic observations which match real data as closely as possible. After the simulation of physically motivated data corruptions by \meqsil{}, the synthetic data are passed through the \rpicard{} calibration pipeline \citep{Janssen2019}. The data are treated in the same way as actual correlated visibilities and a model-agnostic calibration \citep{Smirnov2011a} of phases and amplitudes is performed based on information typically available for real observations.

The atmospheric signal attenuation introduced by \meqsil{} is corrected by recording opacity values for each station at the start of each scan. This is the equivalent of measuring opacity-corrected system temperatures with a hot-load calibration scan in real VLBI observations \citep{Ulich1976}, which leaves intra-scan opacity variations unaccounted for. As \meqsil{} does not simulate the digitization when radio telescopes record data, nor the correlation process, the simulated visibilities are already scaled to units of flux density, as derived from the input source model. Therefore, unity amplitude gains are used and the system temperatures are set to $\exp{(\tau)}$ for the amplitude calibration, with $\tau$ describing the atmospheric opacity (see Sect.~4.2 in \citet{Janssen2019}). Amplitude losses due to pointing offsets can not be corrected with this standard VLBI amplitude calibration method.

   \begin{figure}[b]
   \centering
   \includegraphics[width=0.5\textwidth]{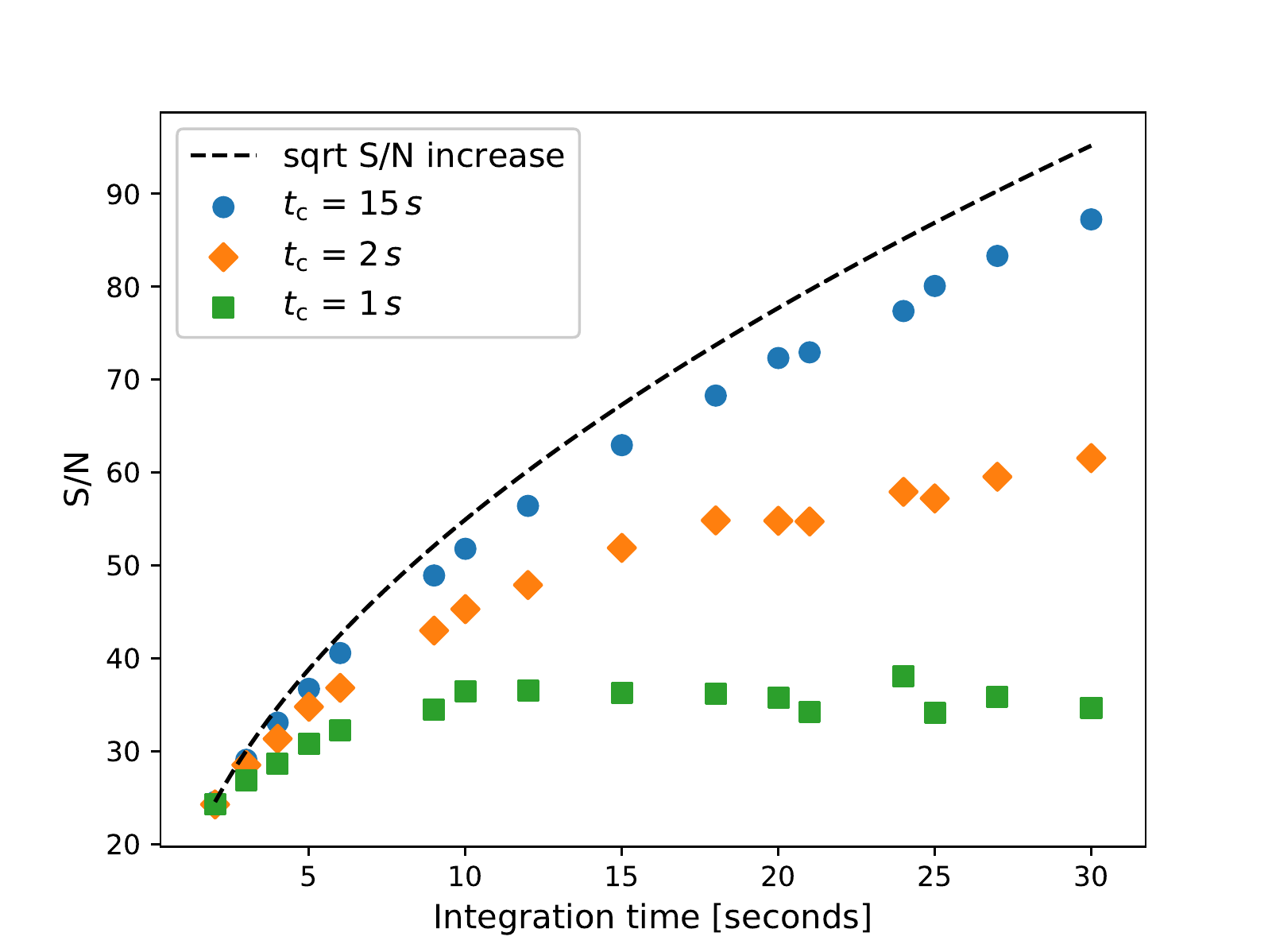}
      \caption{S/N estimates for \rpicard{} fringe solutions.
               The plotted points indicate the estimated average FFT S/N values by the \casa{} \textit{fringefit} code for different integration times (solution intervals), segmenting a 15 minute long scan of a \meqsil{} observation of a 4 Jy point source on the ALMA-APEX baseline.
               Different symbols correspond to different coherence times (Equation~\ref{eq:coher_def}) used for the simulation of atmospheric turbulence. The dashed line shows the expected
               increase in S/N for an infinite coherence time without added noise corruptions.
              }
         \label{cohersnr}
   \end{figure}

   \begin{figure}[h]
   \centering
   \includegraphics[width=0.46\textwidth]{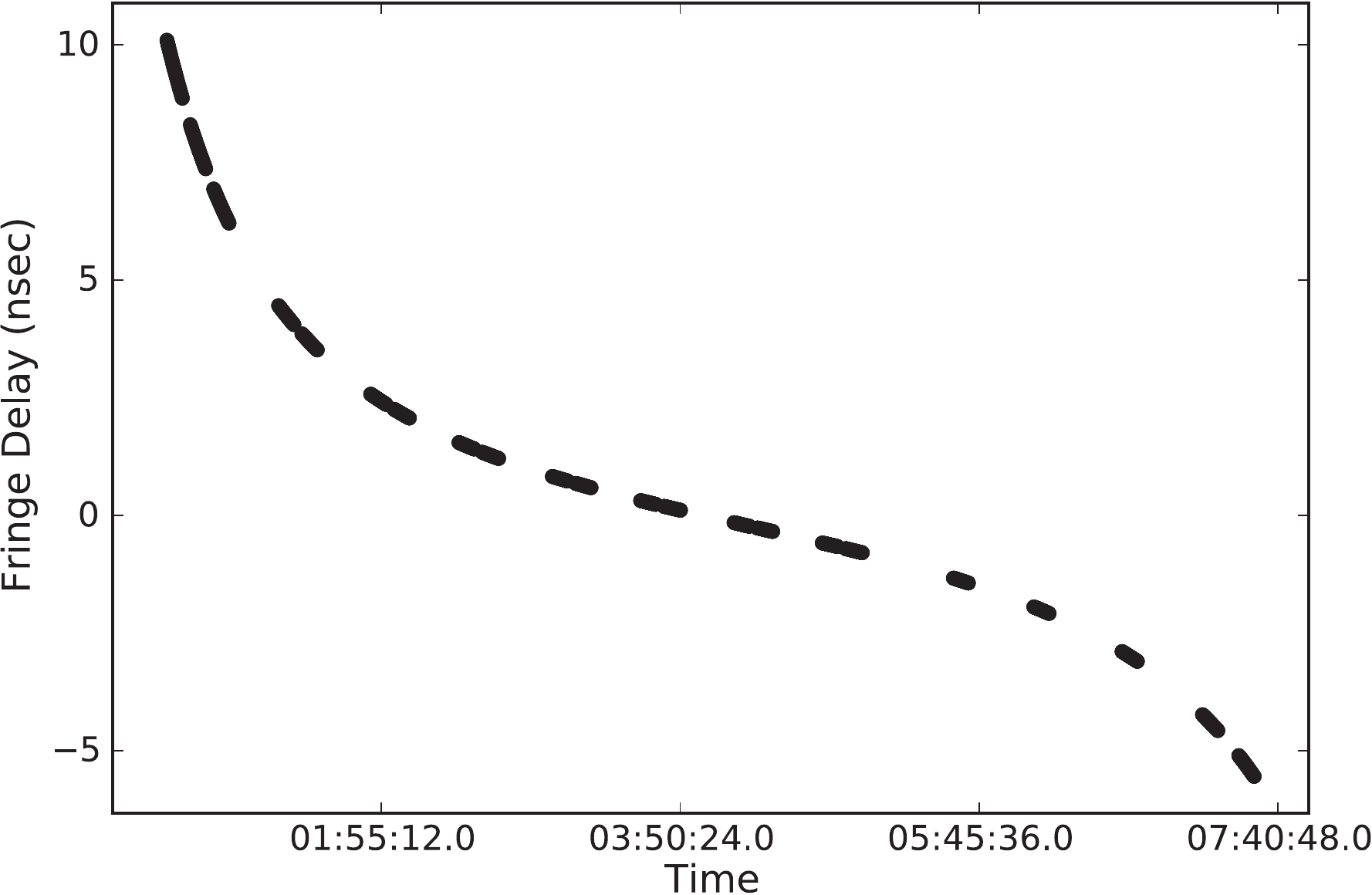}
      \caption{Delay between ALMA and LMT. The delay is solved a function of time by the fringe fitting calibration step. The input source model is a 4 Jy point source.
              }
         \label{fig:delay}
   \end{figure}
   
      \begin{figure*}[h!]
   \centering
   \includegraphics[width=0.75\textwidth]{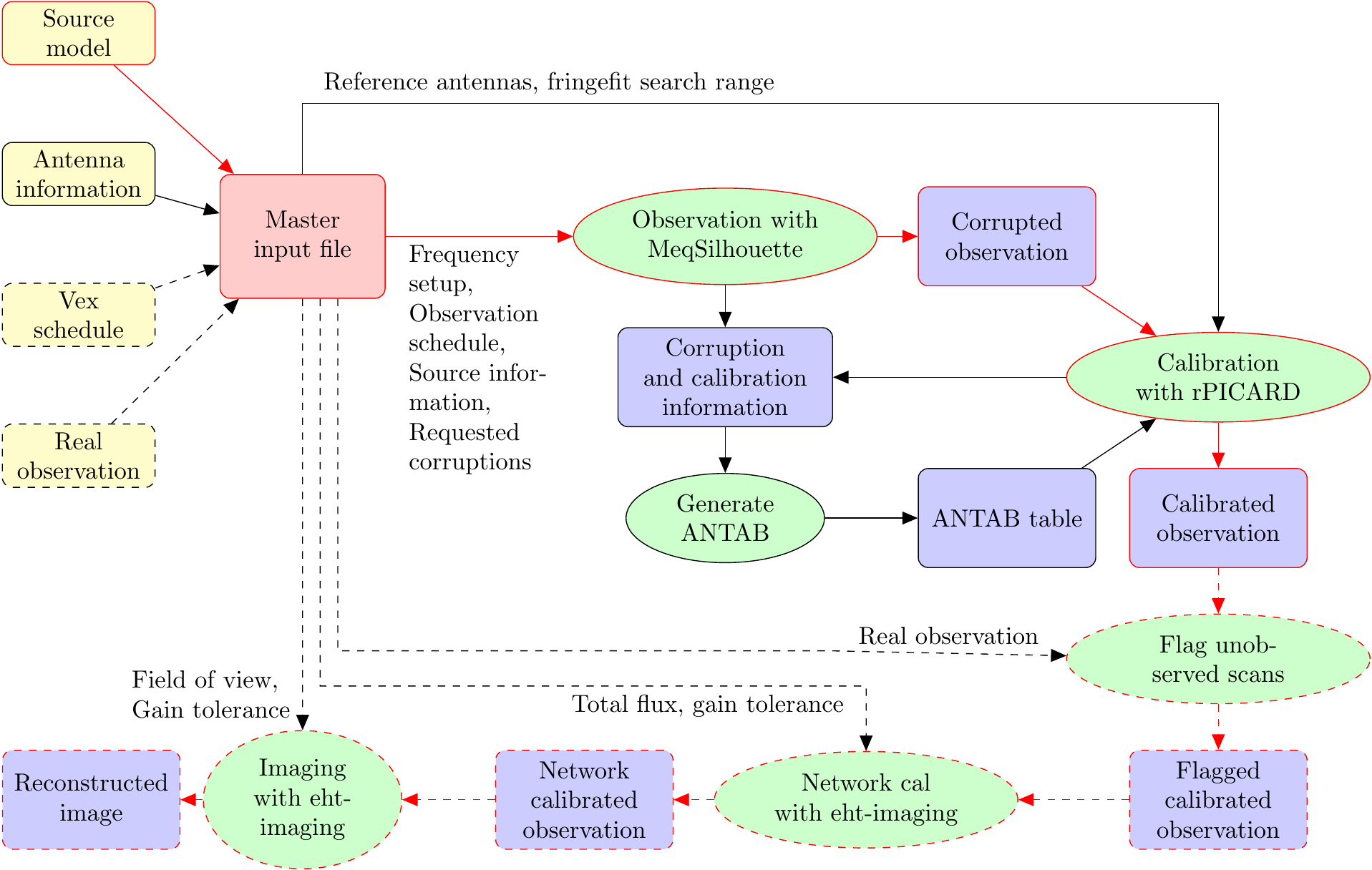}
      \caption{Computing workflow flowchart of \pipe{}. Red borders and arrows indicate the main data path. Dashed borders and arrows indicate optional steps that may be skipped (for example, imaging could be done without network calibration). Yellow boxes are auxiliary input files; the master input file is indicated by the red box. Green ellipses are actions, and blue boxes are data products. Text next to arrows lists the information from the master input file that is used for a specific action.}
         \label{fig:flowchart}
   \end{figure*}

The phases are calibrated with the \casa{} Schwab-Cotton \citep{Schwab1983} fringe fitter implementation. With this method, station gains for phases, rates, and delays are solved with respect to a chosen reference station. \rpicard{} uses a prioritized list of reference stations (based on availability). For the EHT, these are ALMA $\rightarrow$ LMT $\rightarrow$ APEX $\rightarrow$ SMT $\rightarrow$ PV.
All solutions are re-referenced to a single common station in the end. Optimal fringe fit solution intervals are found based on the signal-to-noise ratio (S/N) of the data in each scan. The search intervals range from twice the data integration time (typically $\sim$\,0.5-1 s) to 60 s. Within this interval, the smallest timescale which yields fringe detections with S/N>5.5 on all baselines for which the source can be detected, is chosen \citep{Janssen2019b}. Figure~\ref{cohersnr} shows estimated S/N values for a range of fringe fit solution intervals and different simulated coherence times.
The presence of (frequency independent) atmospheric delays and absence of instrumental delays in the synthetic data warrants a combined fringe fit solution over the whole frequency band for a maximum S/N.
Usually, \rpicard{} would smooth solved delays within scans to remove potential outliers. This is done under the assumption that an a priori delay model like Calc/Solve\footnote{\url{http://astrogeo.org/psolve/}.} has been applied at the correlation stage, which already takes out the bulk of the delay offsets. For the synthetic data generation, no atmospheric delay model is applied and \rpicard{} has to solve for steep residual delay gradients caused by the wet and dry atmospheric components within scans (Figure~\ref{fig:delay}). Smoothing of solved delays is therefore disabled here.

The last step of the calibration pipeline is the application of the amplitude and phase calibration tables, and averaging of the data in frequency within each spectral window. The calibrated and averaged data are then exported in the UVFITS file format.
Optionally, an additional UVFITS file can be provided as input. \pipe{} then uses \ehtim{} to reproduce the $uv$-coverage from that file. For a UVFITS file from a real observation, this means taking into account time periods where telescopes drop out of the observed schedule and all non-detections. Thereby, a comparison of synthetic and real data is unaffected by $uv$-coverage.

Finally, the synthetic UVFITS data are averaged in 10 second intervals and a `network calibration' \citep{Fish2011, Johnson2015, Blackburn2019, eht-paperIII} is performed with the \ehtim{} software.
The gains of non-isolated (redundant) stations, which have a very short baseline to another nearby station can be calibrated if the model of the observed source is known at large scales. For the 2017 EHT observations, ALMA was able to provide accurate large-scale source models, allowing for a network calibration of the co-located ALMA/APEX and SMA/JCMT sites. 
For our synthetic observations, we use the known total flux density of the input model.

\section{Computing workflow}
\label{sec:computingworkflow}

\pipe{} is controlled by a single input ASCII file. The observed schedule can either follow a VEX file or explicitly set start time, duration, number of scans, and gaps between scans. If the VEX file has been used for a real observation, a UVFITS file can be provided to match the $uv$-coverage. All antenna and weather parameters are also set in ASCII files. The input source model can be provided as FITS or ASCII file, as a single model or multiple frames from a time-variable source, and contain only Stokes I or full polarization information.
The input model is Fourier Transformed and corrupted by \meqsil{}. The resultant visibilities are calibrated by \rpicard{}, and optionally network calibrated and imaged by \ehtim{}.
\pipe{} outputs a FITS file of the final reconstructed source model, the calibrated and self-calibrated visibilities in UVFITS and ASCII format, and diagnostic plots of the calibration process.
The pipeline is fully dockerized.\footnote{\url{https://www.docker.com/}}. An overview of the workflow is shown in Figure~\ref{fig:flowchart}.

\section{Simulated observation setup}
\label{sec:obsparameters}
\begin{table*}[h]
\centering
\caption{Antenna parameters adopted in our synthetic observations.}
\label{tab:antennas}
\begin{tabular}{cc|cccccccccc}
Year        & Antenna & X (m)  & Y (m) & Z (m) & $D$ (m) & $\eta_{\textrm{ap}}$ & $\mathcal{S}_\mathrm{rx}$ (Jy) & $\mathcal{G}_\mathrm{err}$ & $\mathcal{D}$ & $\mathcal{P}_\mathrm{rms}$ (") &  $\mathcal{P}_\mathrm{FWHM}$ (")\\ \hline
2017 & ALMA &  2225061 & -5440057 & -2481681 & 70 & 0.73 & 60   & 1.02 & 0.05 & 1.0 & 27\\
     & APEX &  2225040 & -5441198 & -2479303 & 12 & 0.63 & 3300 & 0.97 & 0.05 & 1.0 & 27\\
     & JCMT & -5464585 & -2493001 &  2150654 & 15 & 0.52 & 6500 & 1.05 & 0.05 & 1.0 & 20\\
     & LMT  &  -768716 & -5988507 &  2063355 & 32 & 0.31 & 2400 & 0.85 & 0.05 & 1.0 & 10\\
     & PV   &  5088968 &   -301681 &  3825012 & 30 & 0.43 & 1000 & 1.03 & 0.05 & 0.5 & 11\\
     & SMA  & -5464555 & -2492928 &  2150797 & 16 & 0.73 & 3300 & 0.96 & 0.05 & 1.5 & 55\\
     & SMT  & -1828796 & -5054407 &  3427865 & 10 & 0.57 & 7700 & 0.93 & 0.05 & 1.0 & 32\\ \hline
2018 & GLT  &   541647 & -1388536 &  6180829 & 12 & 0.63 & 3300 & 1.08 & 0.05 & 1.0 & 27\\
2020 & KP   & -1994314 & -5037909 &  3357619 & 12 & 0.63 & 3300 & 0.96 & 0.05 & 1.0 & 27\\
     & PDB  &  4523951 &   468037 &  4460264 & 47 & 0.52 & 750  & 0.95 & 0.05 & 1.0 & 20\\
2020+& AMT  &  5627890 &  1637767 & -2512493 & 15 & 0.52 & 1990 & 1.03 & 0.05 & 1.0 & 20\\ \hline
\end{tabular}
\end{table*}

\begin{table}[h]
\centering
\caption{Weather parameters adopted in our synthetic observations.}
\label{tab:weather}
\begin{tabular}{c|cccc}
Antenna & PWV (mm) &  $P_{\mathrm{g}}$ (mb) & $T_{\mathrm{g}}$ (K) & $t_{\mathrm{c}}$ (s) \\ \hline
ALMA & 1.5 & 555 & 271 & 10 \\
APEX & 1.5 & 555 & 271 & 10 \\
JCMT & 1.5 & 626 & 278 & 5 \\
LMT  & 5.7 & 604 & 275 & 6 \\
PV   & 2.9 & 723 & 270 & 4 \\
SMA  & 1.5 & 626 & 278 & 5 \\
SMT  & 4.4 & 695 & 276 & 3 \\ \hline
GLT  & 1.7 & 1000 & 254 & 5 \\ 
KP   & 2.5 & 793 & 282 & 3 \\
PDB  & 3.0 & 747 & 270 & 3 \\
AMT  & 6.2 & 772 & 287 & 3 \\ \hline
\end{tabular}
\end{table}

\pipe{} is able to create synthetic observations for any VLBI array. Here, we outline the antenna and weather parameters and observing schedules adopted for the creation of our synthetic data sets.

\subsection{EHT2017 array}
Our primary array consists of the 2017 EHT stations, excluding the SPT station for which M87 is always below the horizon. The antenna parameters are summarized in Table~\ref{tab:antennas}. The receiver SEFDs of the primary array have been estimated by extrapolating system temperature measurements to zero airmass, following \citet{Janssen2019}. Full width at half maximum 230 GHz beam sizes ($\mathcal{P}_\mathrm{FWHM}$) and dish diameters ($D$) were taken from the websites and documentation for each individual site.
Pointing rms offsets ($\mathcal{P}_\mathrm{rms}$) have been based on a priori station information and typical inter- and intra-scan amplitude variations seen in EHT data. All offsets lie within official telescope specifications.
Aperture efficiencies ($\eta_{\textrm{ap}}$) were estimated with $\sim10$\% accuracy from planet observations \citep{Janssen2019b, eht-paperIII}. 
In our synthetic observations, we have added gain errors ($\mathcal{G}_\mathrm{err}$) listed in Table~\ref{tab:antennas} in accordance with these uncertainties.
Additionally, a polarization leakage corruption has been added at a $\mathcal{D}=5\%$ level for all stations. This corruption has been left uncalibrated by \rpicard{}, to mimic the current capabilities of the EHT, which did not perform a polarization calibration for the first scientific data release \citep{eht-paperIII}.

The weather parameters are summarized in Table \ref{tab:weather}. For the ground temperature $T_{\mathrm{g}}$, pressure $P_{\mathrm{g}}$, and precipitable water vapour PWV, we used the median values measured during the EHT2017 campaign (5-11 April) at the individual primary sites, logged by the VLBI monitor \citep{eht-paperII}. No weather information was available from the VLBI monitor for ALMA. We adopted the values measured at the nearby station APEX. 

The radiometers at the sites measure the atmospheric opacity $\tau$, while \meqsil{} takes the PWV as input. The 225 GHz opacity can be converted to PWV in mm using
\begin{equation}
\label{eq:tau}
\mathrm{PWV} = \frac{\tau-\tau_{\mathrm{dry-air}}}{B},
\end{equation}
where $\tau_{\mathrm{dry-air}}$ is the dry air opacity and the slope $B$ is in mmH$_2$O$^{-1}$. $B$ and $\tau_{\mathrm{dry-air}}$ have been measured at some sites and both tend to decrease with site altitude, but the errors on these measurements are not well known \citep[][and references therein]{TMS2017,Thomas-Osip2007}: the calibration of $B$ needs an accurate independent measure of the water vapour column density at the same site as the radiometer, which is only available for a few EHT sites. Also, $\tau_{\mathrm{dry-air}}$ is generally small (order $10^{-2}$), making it challenging to measure. 

For these reasons, climatological modelling likely provides better estimates than empirical measurements here. To estimate $B$ and $\tau_{\mathrm{dry-air}}$, we use the Modern-Era Retrospective Analysis for Research and Applications, version 2 (\merra{}) from the NASA Goddard Earth Sciences Data and Information Services Center (GES DISC) \citep{Gelaro2017}. In a reanalysis model like \merra{}, variables such as the air temperature and mixing ratios of different molecules are computed based on ground- and space-based measurements. They depend on time, atmospheric pressure level, and latitude and longitude coordinates. We use 2006-2016 \merra{} data averaged over seasons (per three months) and latitude zones (antarctic and arctic, southern and northern mid-latitudes, and tropical)\footnote{As available on \url{https://www.cfa.harvard.edu/~spaine/am/cookbook/unix/zonal/}.}. For each pressure layer and latitude zone, we then perform radiative transfer at 225 GHz with the \texttt{am} atmospheric model software \citep{Paine2019} with and without water vapour included to calculate $B$ and $\tau_{\mathrm{dry-air}}$ in the March-April-May season (which is the usual EHT observing season). We then interpolate these to the pressure level of each EHT site and calculate the PWV from the measured $\tau$ using equation \ref{eq:tau}.

Atmospheric coherence times $t_{\mathrm{c}}$ were estimated based on the characteristics of the 2017 EHT measurements for the primary array. Precise station-based coherence times are difficult to obtain and will vary from day to day due to changes in the weather conditions. For this paper, estimates are taken that match well to decent to poor weather. The values are summarized in Table \ref{tab:weather}. A larger parameter space will be studied in future work to characterize the effect of varying weather conditions.

\subsection{Enhanced EHT array}
Apart from simulated observations with the stations that joined the 2017 EHT campaign, we also simulate observations with an enhanced EHT array including four additional stations. The Greenland Telescope \citep[GLT,][]{Raffin2014} is currently located at Thule air base (it will be relocated to Summit Station near the peak of the Greenland ice sheet) and joined the EHT in 2018. The 12-m telescope on Kitt Peak \citep[KP,][]{Freund2014} in Arizona and the IRAM NOEMA interferometer on Plateau de Bure \citep[PDB,][]{Guilloteau1992} in France were to join in the cancelled 2020 observations and will join in future campaigns. Finally, the Africa Millimetre Telescope \citep[AMT,][]{Backes2016}, is planned to be built on the Gamsberg in Namibia. 

For these sites, we estimated weather parameters using the \merra{} inst3\_3d\_asm\_Np data product, which has a time resolution of 3 hours, and is distributed on a grid having 0.625 degree longitude by 0.5 degree latitude with 42 vertical pressure levels between 0.1 and 1000 mbar. From this dataset, we took the 25th percentile (representing good weather) of the air temperature and specific humidity measured on 11 April in the last two decades (1999-2018).\footnote{It should be noted that the current EHT observing strategy is to trigger a few observing days in a March/April observing window, based on optimal weather conditions across all sites \mbox{\citep{eht-paperII}}.}
At each pressure level, these quantities were then linearly interpolated between the four grid points nearest to the observatory site. We then performed an integration of the humidity over the pressure levels using the \texttt{am} atmospheric model software \citep{Paine2019} to obtain the total PWV above the site. The starting point for the integration over the pressure levels was determined by interpolating the geopotential height (pressure as a function of altitude) to the altitude of the site. The geopotential height data were downloaded through NASA's Giovanni portal. The resulting weather parameters are listed in Table \ref{tab:weather}. The GLT site is close to sea level, but the closest \merra{} grid points are further inland at higher altitudes. Hence, the air temperature and specific humidity were extrapolated from a pressure level of 925 mbar to the GLT site pressure level of 1000 mbar before the integration was done in \texttt{am}.

The receiver temperatures and aperture efficiencies for the future stations were estimated from existing stations. The GLT and KP antennas are ALMA prototypes like APEX, so the values for APEX were adopted here. The NOEMA interferometer has ten 15-metre dishes, so the sensitivity was scaled accordingly from the JCMT, including a phasing efficiency of 87\%. The currently envisioned dish for the AMT is the now defunct Swedish-ESO Submillimetre Telescope \citep[SEST,][]{Booth1989} telescope in Chile. With a sideband separating receiver, the current estimate for the SEFD of the AMT is 1990 Jy (A. Young, priv. comm.).

Hereafter, the EHT2017 array plus GLT, KP, and PDB are referred to as EHT2020. When the AMT is also included, it is referred to as EHT2020+AMT.

\subsection{$uv$-coverage}

   \begin{figure}[t]
   \centering
   \includegraphics[width=.49\textwidth]{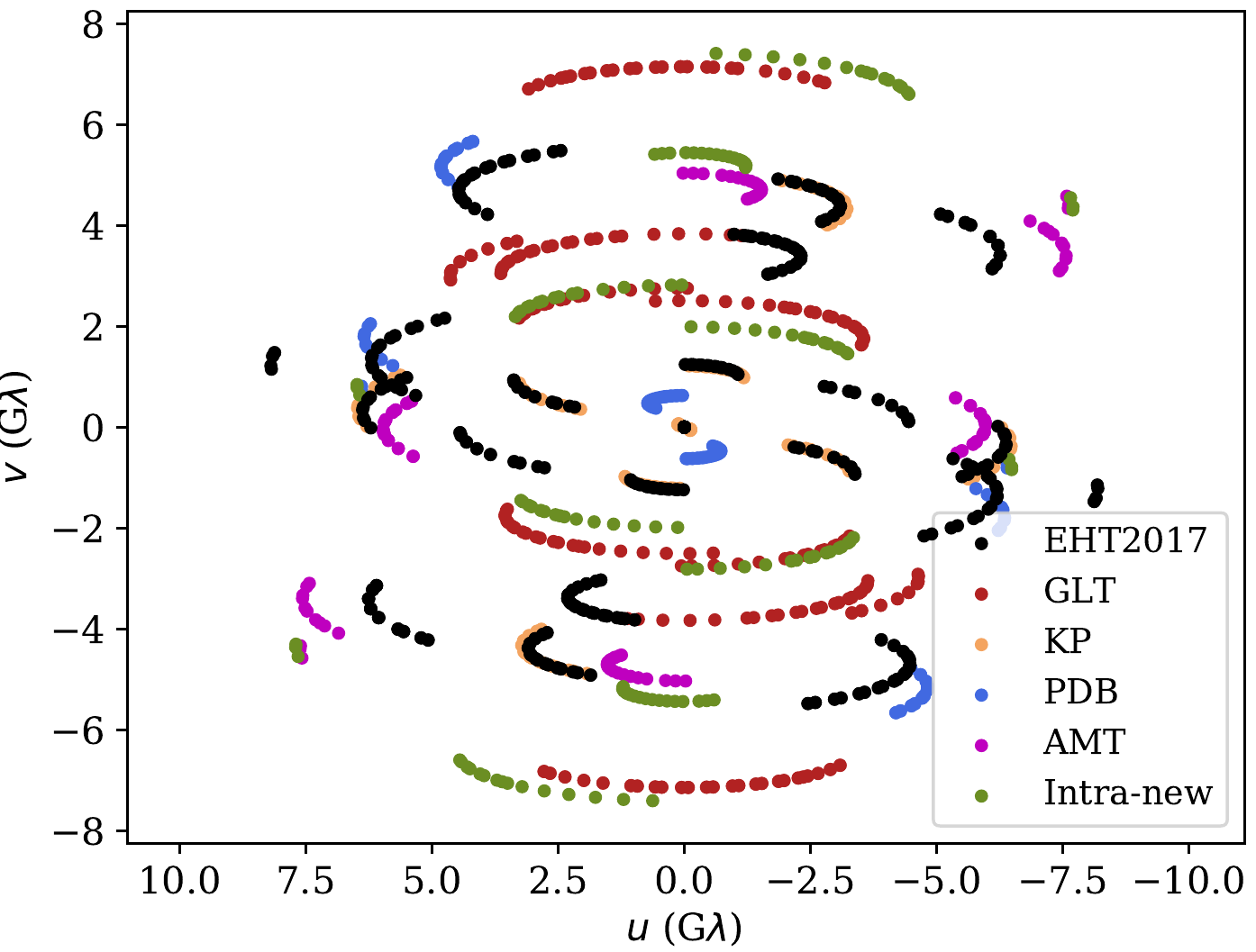}
      \caption{$uv$-coverage towards M87. Different colors show baselines within the EHT2017 array, baselines between the EHT2017 array and four (potential) new stations, and baselines between the new stations (labelled as `Intra-new').
      }
         \label{fig:uv}
   \end{figure}
   
Figure \ref{fig:uv} shows the $uv$-coverage towards M87 for the EHT2017 array and expansions with future stations. The EHT2017 schedule for 11 April was adopted. To accommodate the eastward expansion of the array with the AMT and PDB, ten-minute scans were prepended to the schedule at 30-minute intervals starting when the source is at an elevation of more than ten degrees at both the AMT and PDB. The GLT, strategically located between the European and American mainland, adds north-south baselines to all stations, significantly increasing the north-south resolution due to long baselines to ALMA/APEX. KP and PDB add short baselines to the SMT and PV, respectively, filling the $uv$-gaps between the intrasite baselines and the SMT-LMT baseline. These gaps on short $uv$-spacings pose challenges for image reconstruction with the EHT2017 array \citep{eht-paperIV}. Finally, the AMT adds north-south baselines to the European stations, east-west baselines to ALMA/APEX, and increases the north-east to south-west resolution by adding baselines to the LMT and SMT/KP. The AMT has a larger impact for observations of more southern sources like Sgr A*.
Unless noted otherwise, all synthetic data sets in this work are generated based on the 11 April observing schedule for a source in the direction of M87 for the EHT2017 and EHT2020 arrays, and the extended schedule described above is used for EHT2020+AMT array.
\newpage

\section{Source models}
\label{sec:sourcemodels}

This section describes the set of input source models we use to exercise the various aspects of the pipeline and perform scientific case studies.

\subsection{Geometrical models}

\subsubsection{Point source model}
\label{sec:model_pointsource}

We use a simple 4 Jy point source model to study signal corruption and calibration effects.

\subsubsection{Crescent model}
\label{sec:model_crescent}
\begin{figure}[t]
   \centering
   \includegraphics[width=0.35\textwidth]{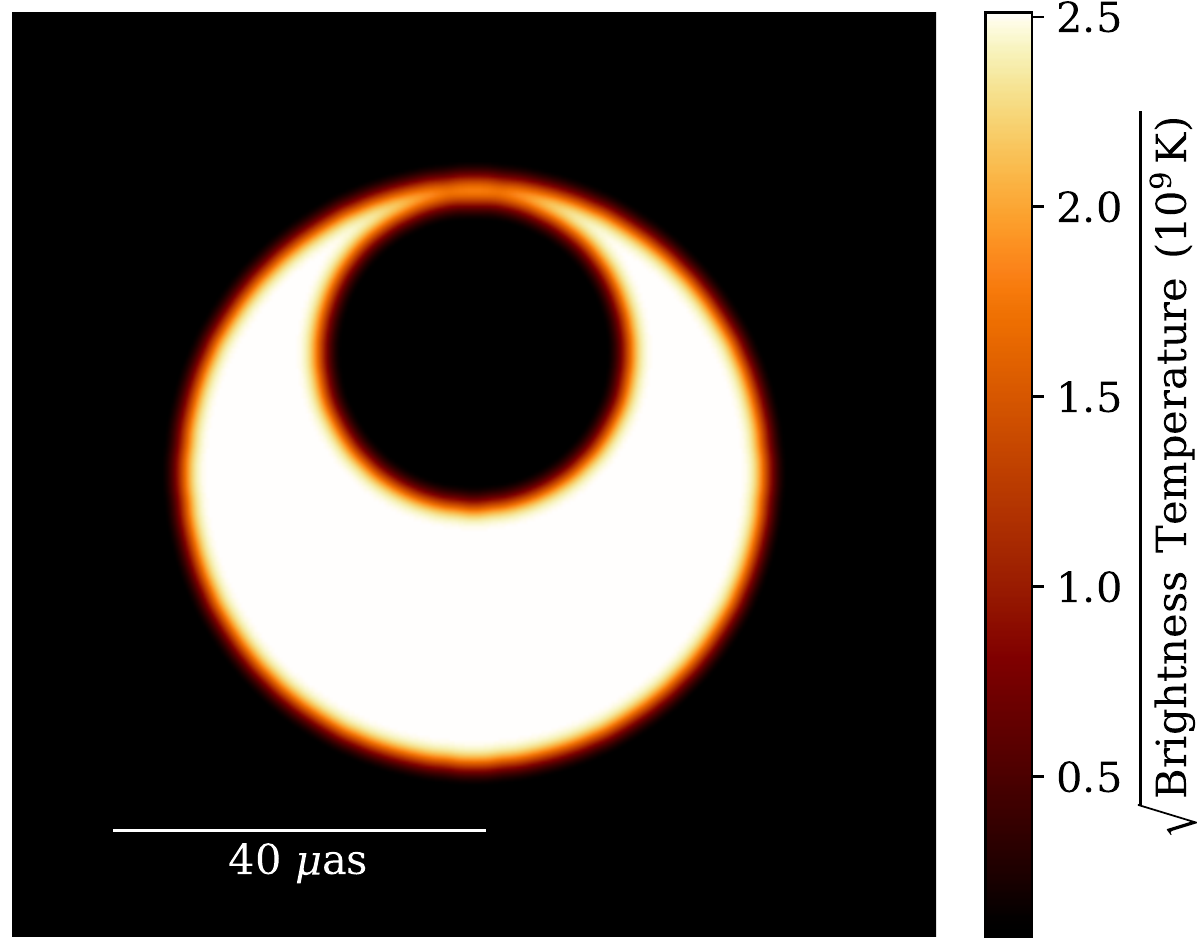}
    \caption{Crescent model from \citet{Kamruddin2013} used for our simulated observations. This images and the images elsewhere in this paper are displayed on a square root scale, unless indicated otherwise.}
     \label{fig:crescent}
  \end{figure}

As an intermediate step between a point source and GRMHD model, we use the geometric crescent model from \mbox{\citet{Kamruddin2013}}. This model consists of two disks with equal brightness that are subtracted from each other. We set the large disk radius to 31 $\mu$as and the small disk radius to 17 $\mu$as. The small disk was offset by 13 $\mu$as towards the north and subtracted from the large disk. The total flux was set to 0.5 Jy and the model was blurred with a 2 $\mu$as beam in order to smear out the sharp edges. The model is shown in Figure \ref{fig:crescent}.
  
\subsection{GRMHD models}

\subsubsection{Fiducial models}
\label{sec:models_davelaar}

  \begin{figure}[h]
   \centering
   \includegraphics[width=0.24\textwidth]{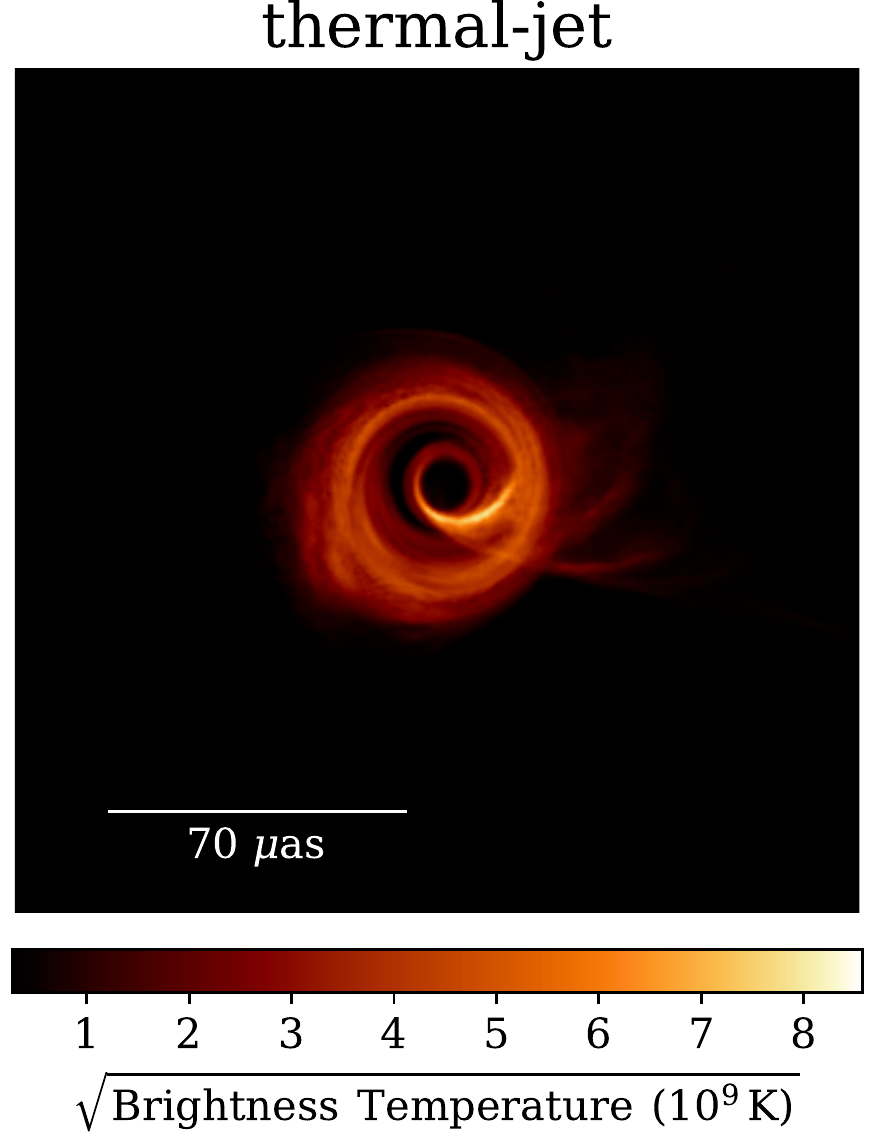}
   \includegraphics[width=0.24\textwidth]{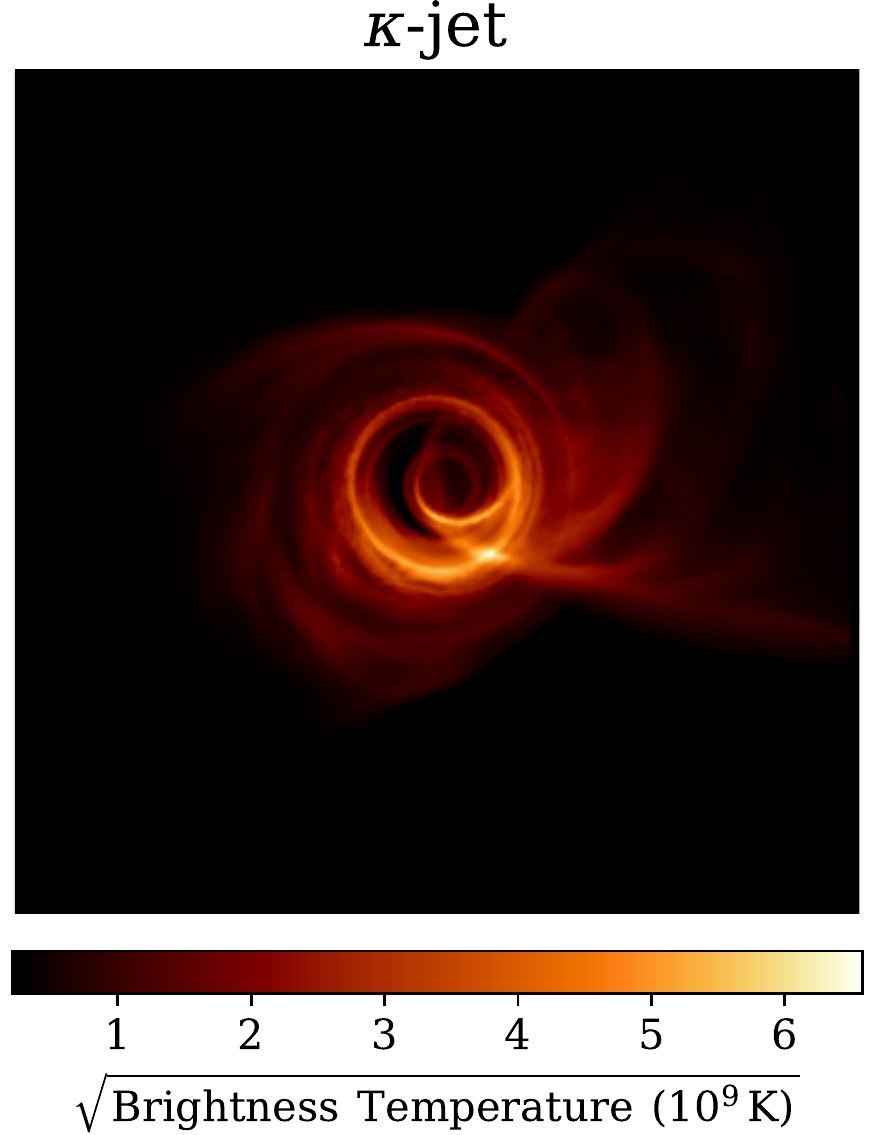} \\
   \includegraphics[width=0.24\textwidth]{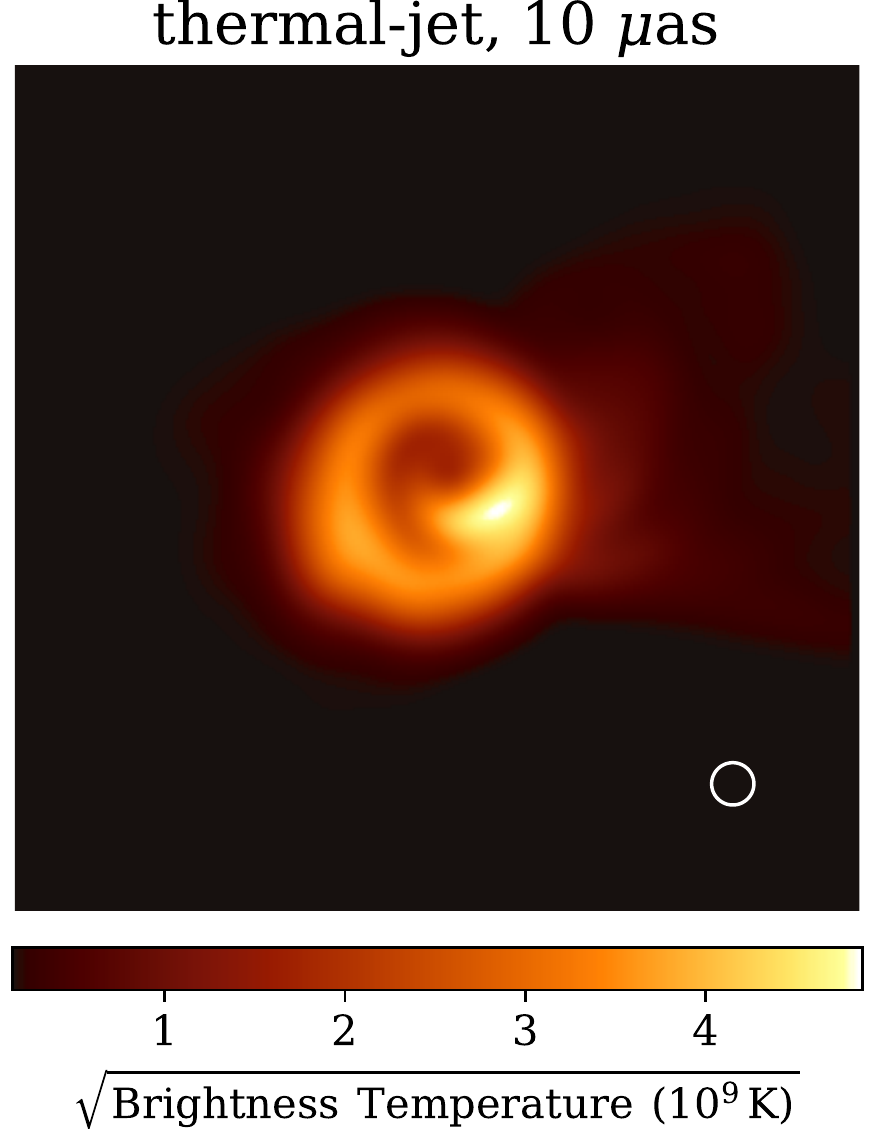}
   \includegraphics[width=0.24\textwidth]{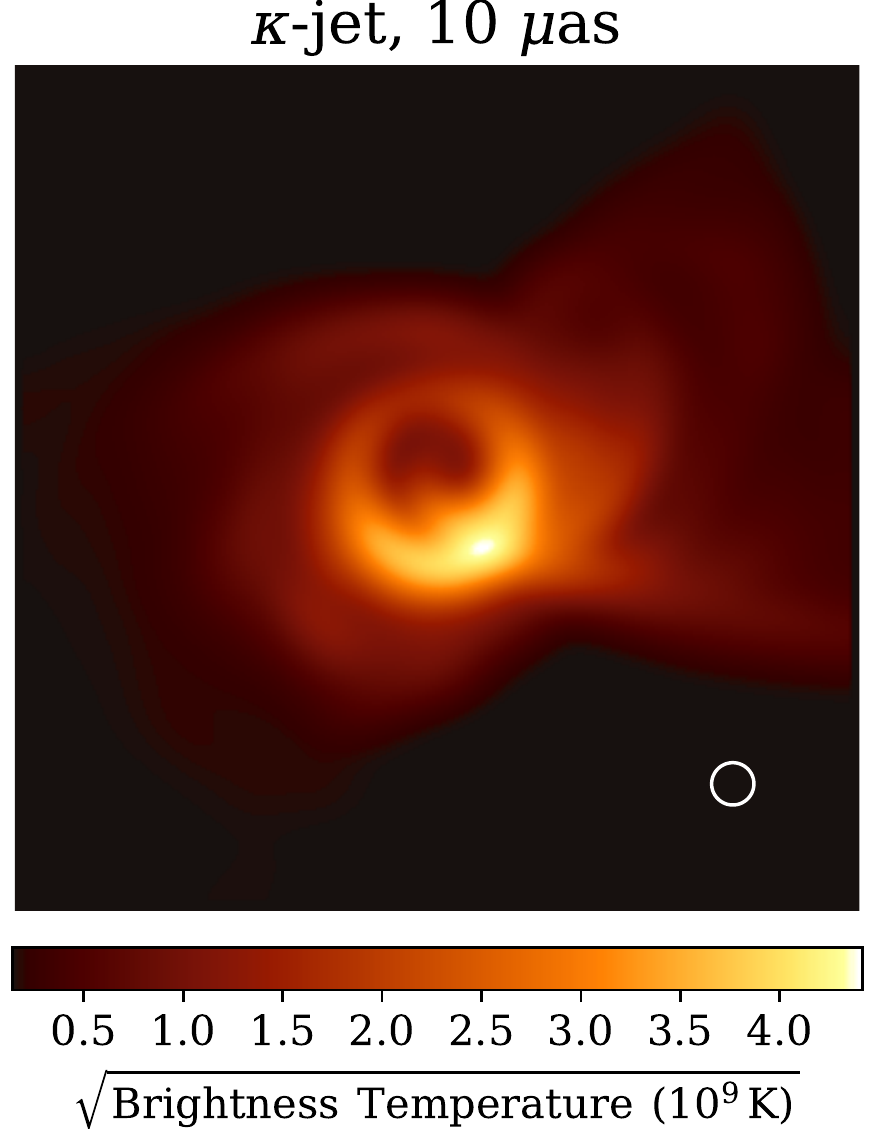} \\
   \includegraphics[width=0.24\textwidth]{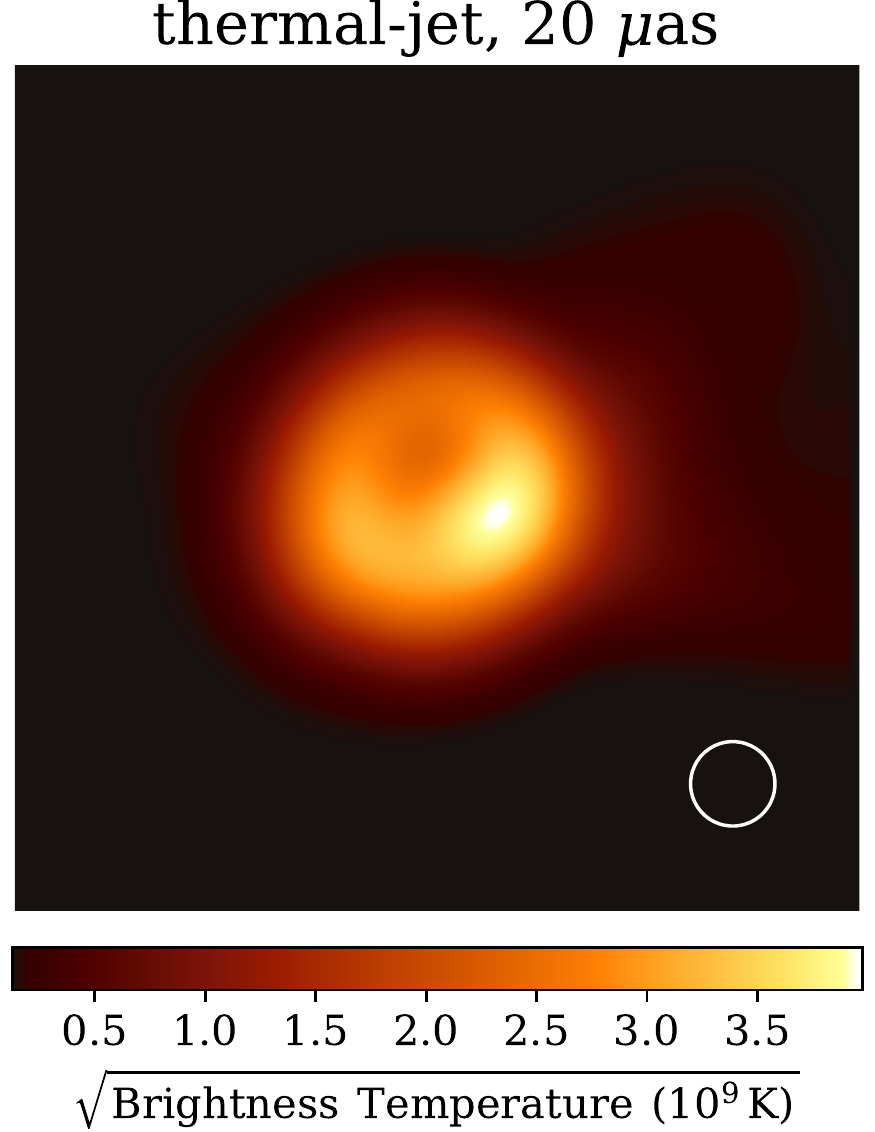} 
   \includegraphics[width=0.24\textwidth]{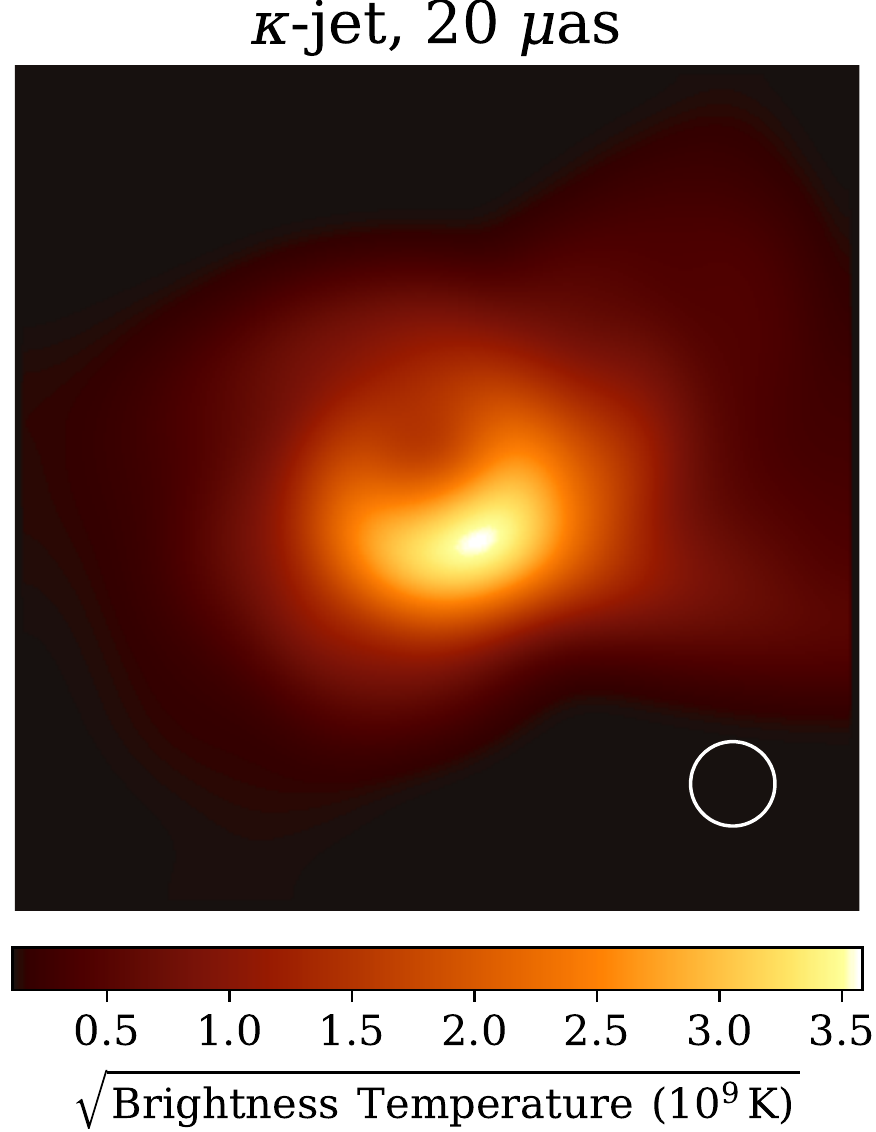}
   \caption{ Thermal-jet (left) and $\kappa$-jet (right) GRMHD models \citep{Davelaar2019} used as input for \pipe{}. The models were blurred with a circular Gaussian beam with FWHM of 10 (middle) and 20 (bottom) $\mu$as, showing the models at different resolutions without including any observation effects.}
   \label{fig:models}
  \end{figure}

We base our scientific studies primarily on a GRMHD simulation of the jet launching region of M87 from \citet{Davelaar2019}.
This GRMHD simulation is performed with the code {\tt BHAC} \citep{Porth2017} in Cartesian Kerr-Schild Coordinates with eight levels of Adaptive Mesh Refinement. The black hole is set to have an angular momentum of $a=\frac{Jc}{GM^2 } = 0.9375 $, where $J$ is the specific angular momentum, $G$ the gravitational constant, $M$ the mass of the black hole, and $c$ the speed of light. The black hole spin influences the appearance of the accretion flow, but the shadow size does not change by more than $\sim$ 4\% \mbox{\citep{Takahashi2004, Johannsen2010}} between a non-spinning and maximally spinning black hole.

The GRMHD simulation is post-processed with the general relativistic ray tracing code {\tt RAPTOR} \citep{bronzwaer2018}. A major and relatively unconstrained free parameter in ray-traced GRMHD model images is the shape of the electron distribution function. Therefore, we consider two models: firstly a thermal-jet model which is based on the work by \cite{Moscibrodzka2016}, and secondly a $\kappa$-jet model which is an improved version of the model introduced in \cite{davelaar2018}.  The thermal-jet model uses a thermal distribution function in the full simulation domain. The $\kappa$-jet model deviates from this by adding electron acceleration. This is done by using a relativistic $\kappa$-distribution function \citep{xiao2006,pierrard2010,pandya2016}, where the power-law index is set by kinetic plasma simulations of trans-relativistic reconnection of an electron-ion plasma \citep{ball2018}. Both models have their best fits to the radio emission when the electrons are hot in the jet and cold in the disk. The $\kappa$-jet also recovers the near infrared part of the observed M87 spectrum. Both models were ray-traced from the same GRMHD frame at the EHT central frequency of 228 GHz, assuming a black hole mass of $6.6\times10^9 M_{\odot}$ and a distance of 16.7 Mpc. The resulting images are shown in Figure \ref{fig:models}, with different levels of blurring indicating the details that can in principle be uncovered with different array resolutions.

The different electron distribution functions result in model images where different parts of the accretion flow light up. The thermal-jet model has a relatively bright jet footprint appearing in front of the shadow. The $\kappa$-jet model shows more extended jet emission, and a bright knot at the point in the image plane, where the jet sheath crosses the photon ring in projection. It becomes difficult to visually distinguish between the models when they are blurred by a 20 $\mu$as beam. The models are described in more detail by \mbox{\citet{Davelaar2019}}.

   \begin{figure}[t]
   \centering
   \includegraphics[width=.49\textwidth]{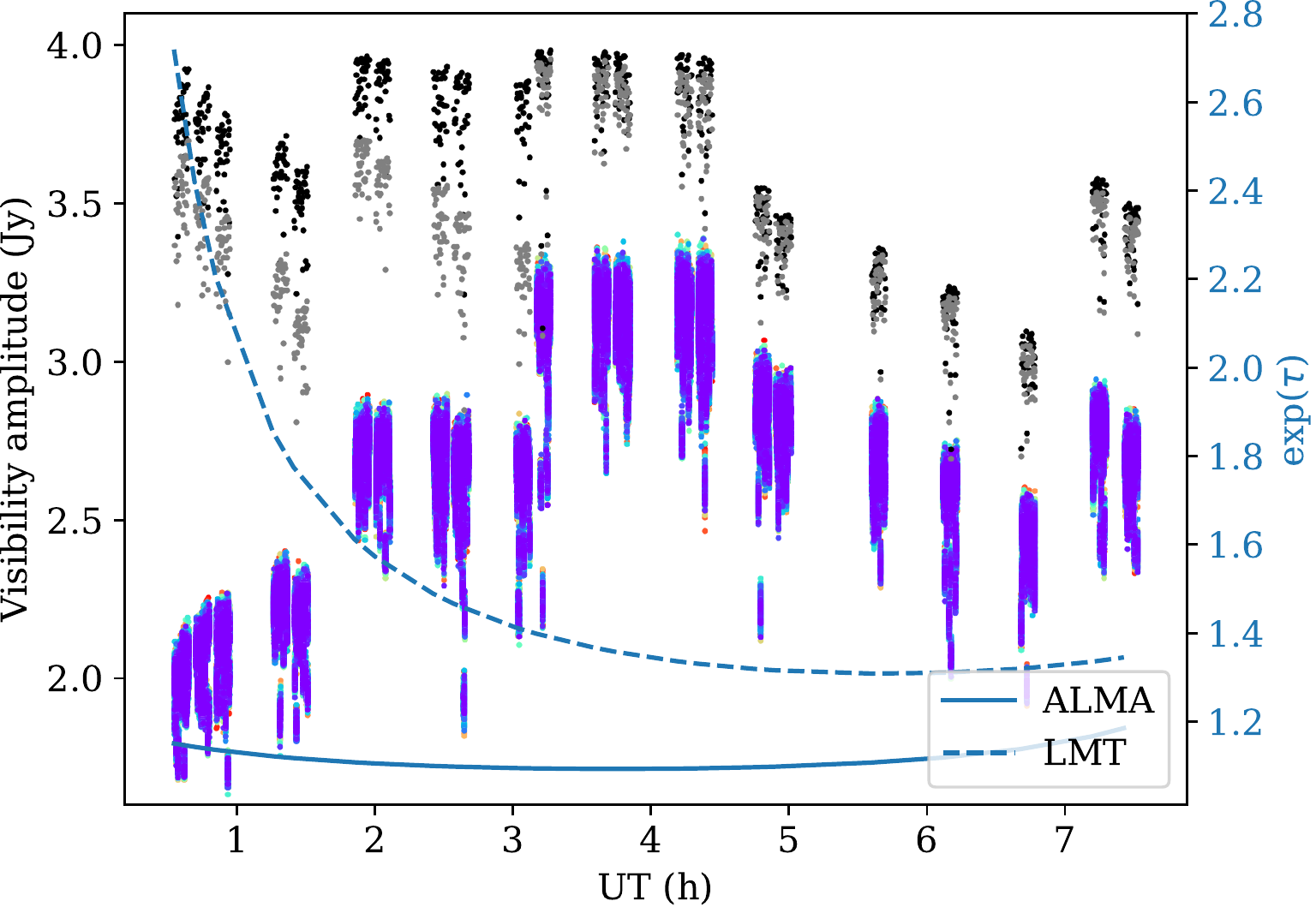}
      \caption{Visibility amplitude versus time at different calibration stages. The amplitudes on the ALMA-LMT baseline observing a 4 Jy point source are shown. The coloured data points represent the 64 channels spanning 2 GHz before calibration, with a time resolution of 1 second. After amplitude calibration, the visibilities are averaged in frequency and down to a time scale of 10 seconds (grey points). Network calibration is then applied to the averaged data, with a solution interval of 10 seconds (black points). The amplitude attenuation factors $\mathrm{exp}(\tau)$ at the centre of the band for the two stations are overplotted as blue lines.
      }
         \label{fig:ptsrc_amp}
   \end{figure}

\subsubsection{Pre-EHT2017 models}
\label{sec:models_pre2017}
An important motivation for synthetic data pipelines is to have the ability to directly compare predictions of theoretical source models to observations. As an illustration, we use \pipe{} to simulate observations of GRMHD model images by \mbox{\citet{Dexter2012}} and \mbox{\citet{Moscibrodzka2016}}. In contrast to the models from \mbox{\citet{Davelaar2019}}, these models were developed before the EHT2017 observations took place.

These models were rotated and scaled in flux and angular size to obtain the best fit the EHT2017 data (11 April, low band) using the GRMHD scoring pipeline described in \citet{eht-paperV, eht-paperVI}. For the model from \citet{Moscibrodzka2016} we used the best-fit model with $R_{\mathrm{high}}=80$. The parameter $R_{\mathrm{high}}$ sets the electron-to-proton temperature ratio in this model. Based on the EHT2017 data alone, $R_{\mathrm{high}}=1$ produced a slightly better fit, but it has not been used here since it does not produce jet-dominated emission. The two models and their image reconstructions are shown in Section~\ref{sec:pre2017}.

\section{Corruption and calibration impacts}
\label{sec:study_corandcal}

In this section, we demonstrate the impact of various corruption and calibration effects included in \pipe{}. Using a point source model, we show the corruption and calibration effects on the synthetic visibility data. Using a crescent and GRMHD models, we demonstrate the impact of the full set of corruption and calibration effects as opposed to thermal noise only synthetic data generation, when reconstructing source models.

\subsection{Point source study}

    \begin{figure}[t]
   \centering
   \includegraphics[width=.49\textwidth]{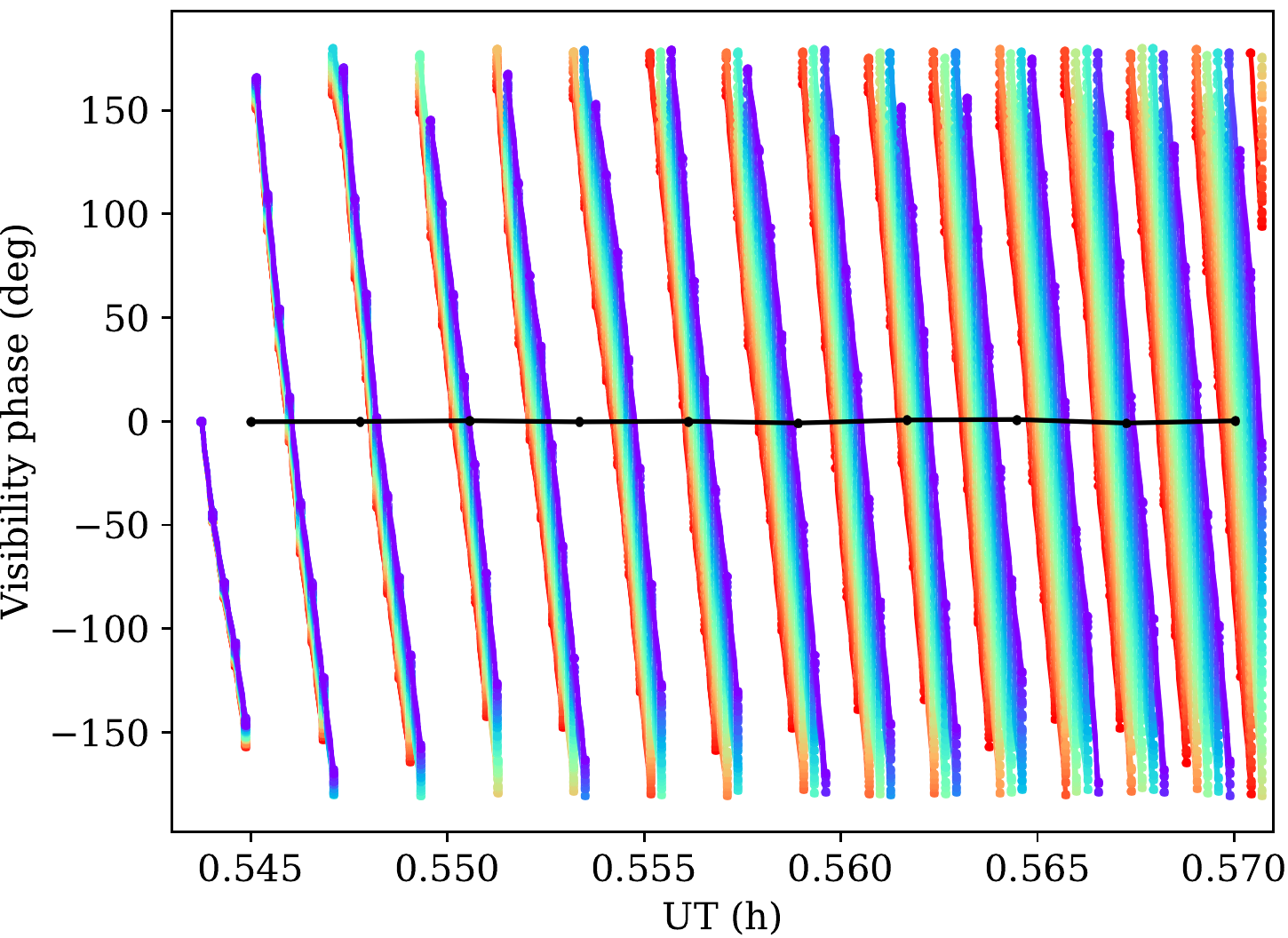}
      \caption{Visibility phase versus time at different calibration stages. A subset of the phases on the ALMA-LMT baseline observing a 4 Jy point source is shown. The colours represent the 64 channels, spanning 2 GHz before calibration, with a time resolution of 1 second. After fringe fitting, the visibilities are averaged in frequency and down to a time scale of 10 seconds (black points).
      }
         \label{fig:ptsrc_phase}
   \end{figure}

As a demonstration of the signal corruption and calibration effects, we observe a point source (Section~\ref{sec:model_pointsource}). In order to clearly show the effects of the individual corruptions on the data, the gain errors $\mathcal{G}_{\mathrm{err}}$ have not been included here. They have been included in our synthetic observations of GRMHD models in Section \ref{sec:grmhd_imagerecon}.
   
  \begin{figure*}[h]
   \centering
   \includegraphics[scale=0.44]{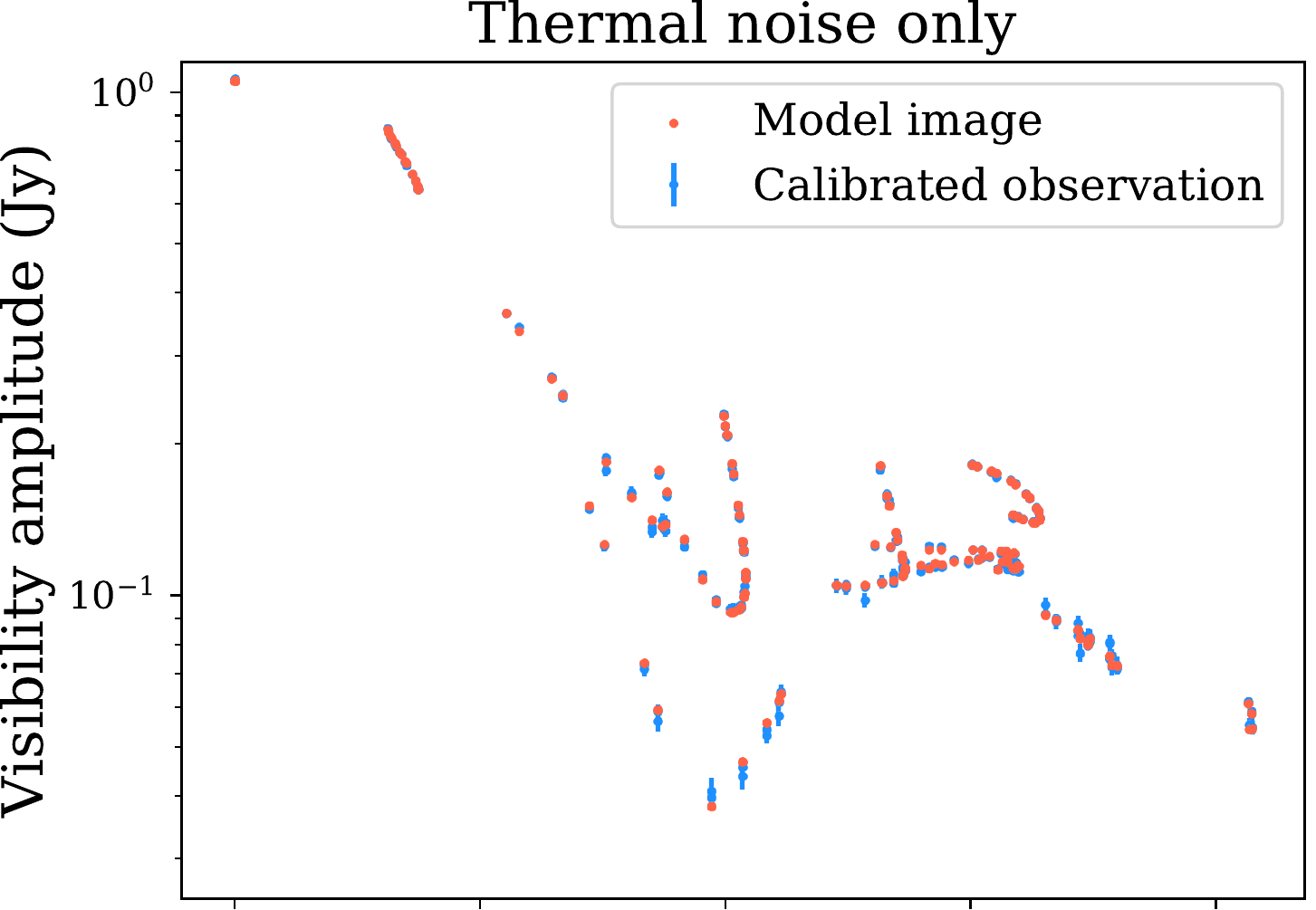}
   \includegraphics[scale=0.44]{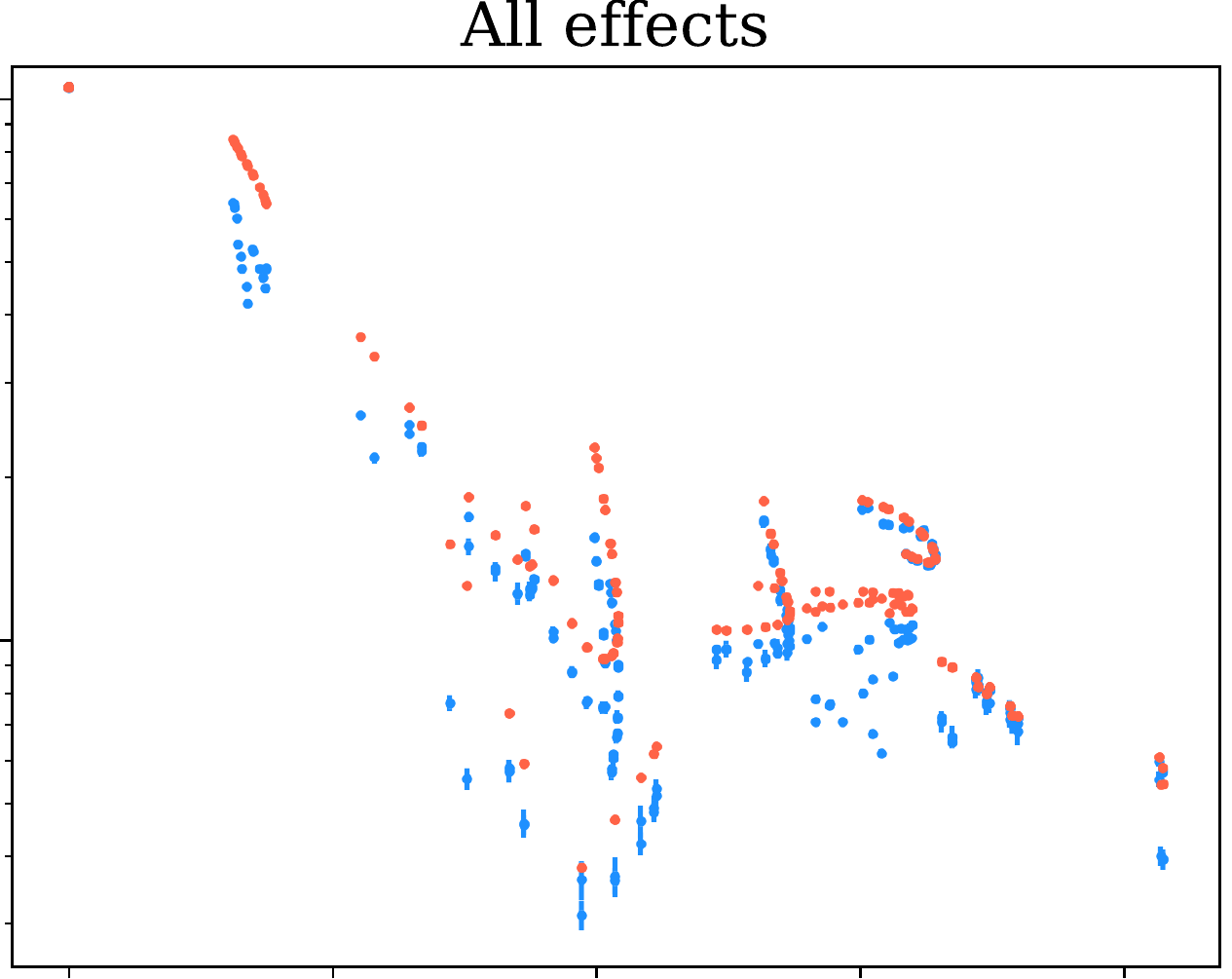}
   \includegraphics[scale=0.44]{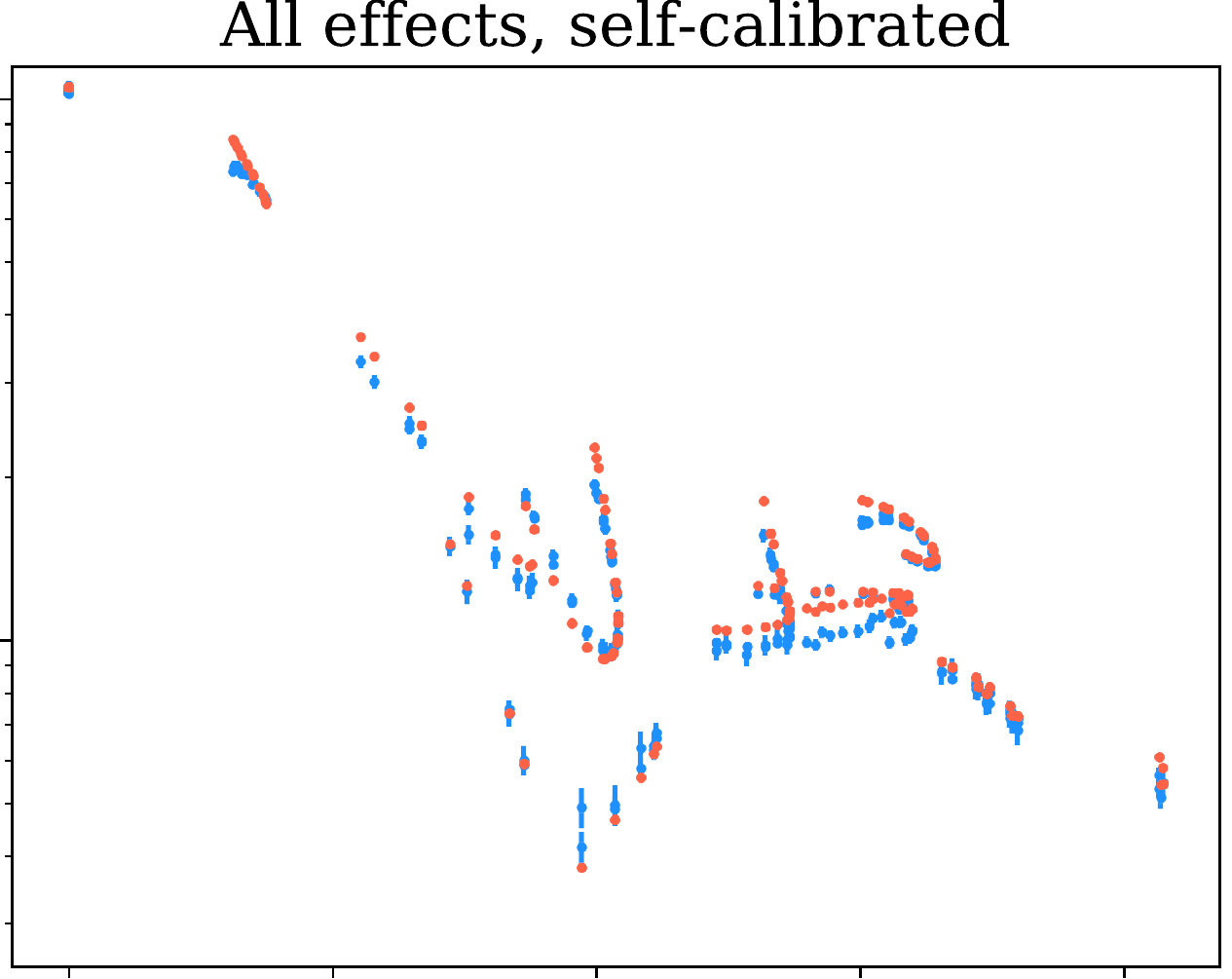} \\
   \includegraphics[scale=0.44]{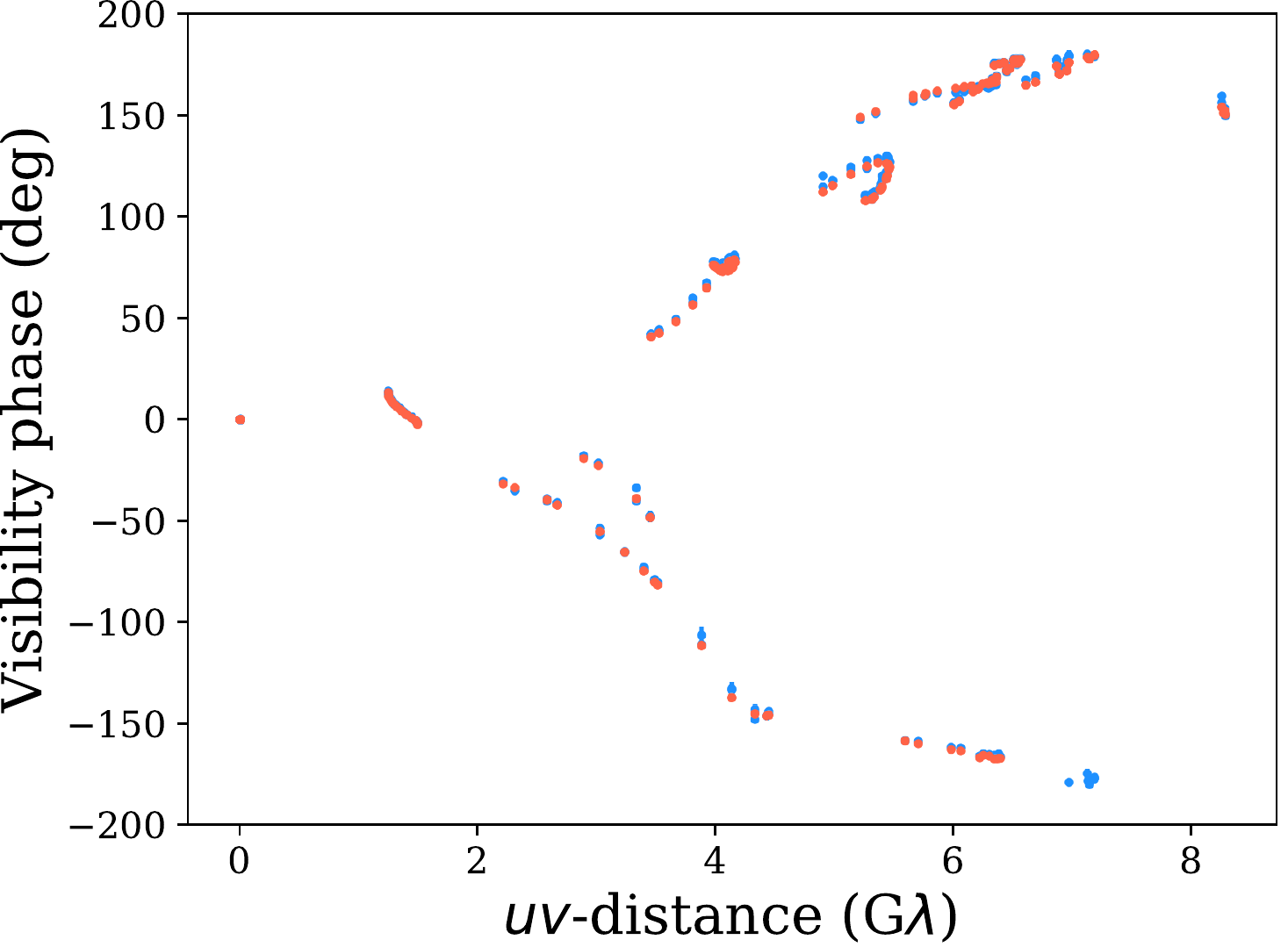}
   \includegraphics[scale=0.44]{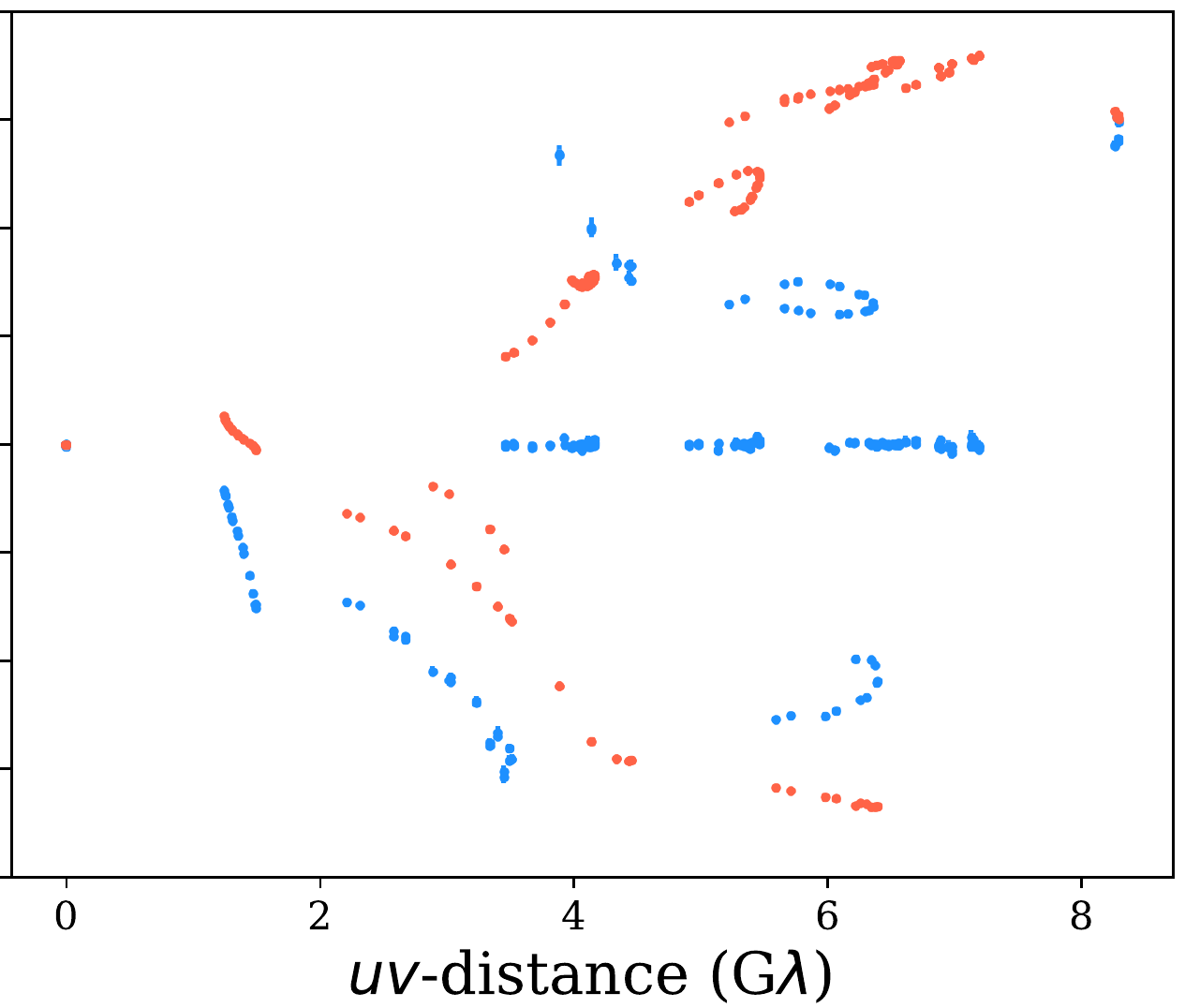}
   \includegraphics[scale=0.44]{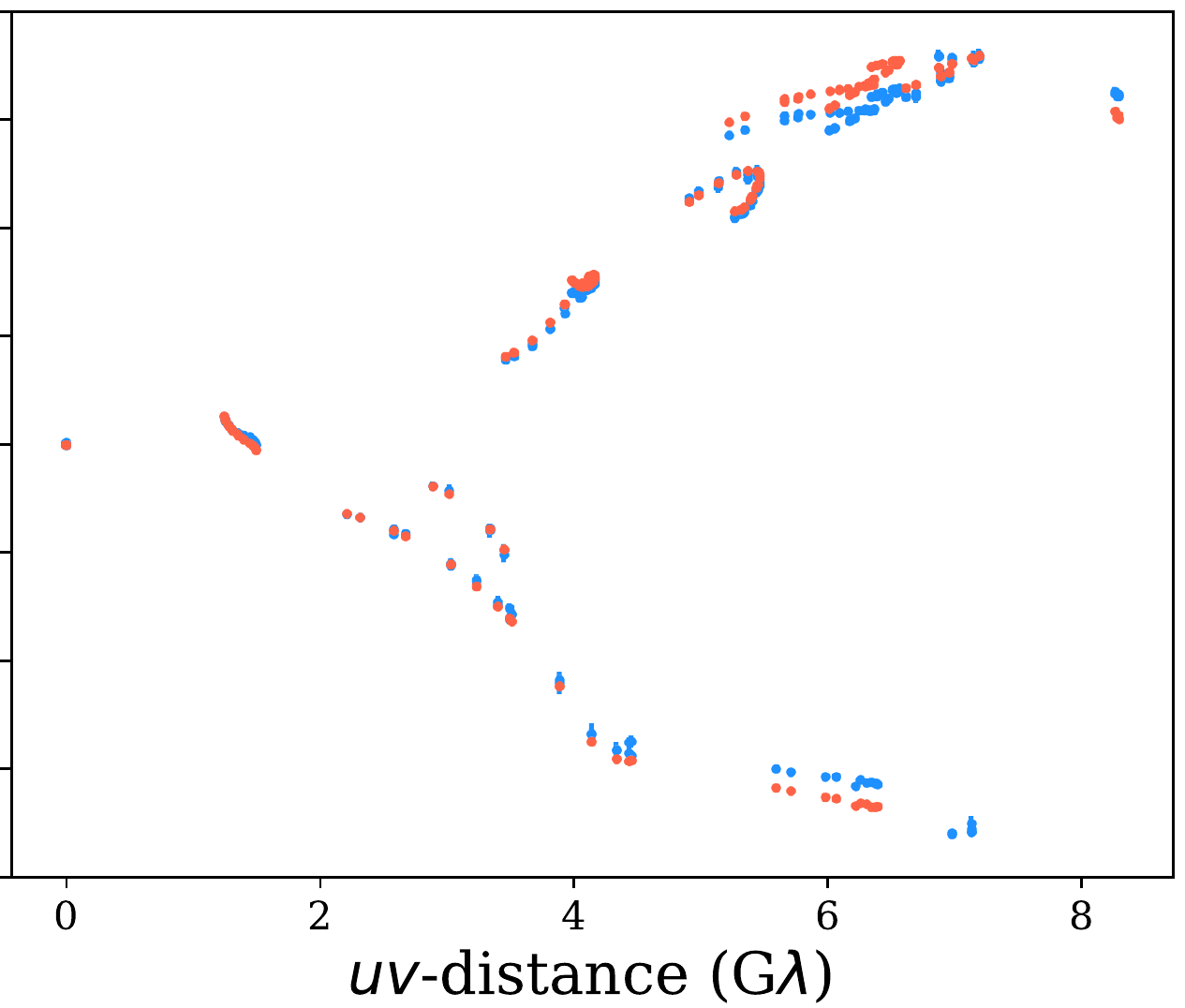} 
      \caption{Scan-averaged amplitude (upper panels) and phase (lower panels) versus baseline length of calibrated synthetic data. The $\kappa$-jet model (Fig. \ref{fig:models}) was used as input, applying either thermal noise only (left panels) or all corruption effects (middle and right panels). The right panels show visibilities that were self-calibrated to the reconstructed image of the source (Fig. \ref{fig:recs}). For the thermal noise only data, the calibration consists only of averaging in frequency (2 GHz across 64 channels) and time (scan-by-scan). Fringe fitting, amplitude calibration, and network calibration (on data averaged from the initial time resolution of 0.5 s down to 10 s) were applied to the synthetic data with all effects included. In order to make the phases of the self-calibrated reconstruction line up with the model image phases, the reconstruction was shifted in position to align with the thermal noise only reconstruction. The phases were then re-calibrated to this shifted reconstruction.} 
         \label{fig:radplots}
   \end{figure*}

Figure \ref{fig:ptsrc_amp} shows the visibility amplitudes on the LMT-ALMA baseline before and after calibration with \texttt{rPICARD}. Before calibration, the visibilities are split into 64 channels spanning a bandwidth of 2 GHz centred at 228 GHz, which is the central EHT observation frequency. There is a general rise and fall of the amplitudes as a function of time caused by atmospheric opacity attenuation (although part of the observed trend is also due to pointing offsets, see below). The attenuation factors $\mathrm{exp}(\tau)$ at the central frequency are overplotted in blue for both stations. Attenuation at the LMT is dominant in this case due to the higher precipitable water vapour column here (Table \ref{tab:weather}). As the source rises at the LMT, the attenuation decreases and the amplitudes increase. At the end of the track, the opposite trend occurs with a smaller slope when the source starts to set at ALMA. Apart from the general trend, the amplitudes show intra-scan variations due to mispointings caused by atmospheric seeing and wind, and inter-scan variations due to sub-optimal pointing solutions that deteriorate by 10\% for every scan and are renewed every 5 scans (see Section \ref{sec:pointing}).

After amplitude calibration (grey), the visibility amplitudes are close to the true 4 Jy point source flux. Some scatter remains due to the pointing offsets. These are partly corrected during network calibration (black), which solves for the gains assuming a fixed flux at the intra-site baselines (including ALMA-APEX). In cases where the pointing-induced amplitude attenuation is largely due to a mispointing at ALMA, network calibration thus corrects for it (e.g. in the second set of five scans). 
When a larger pointing offset occurs at the LMT (e.g. in the last set of five scans), network calibration does not correct for it since there is no intra-site baseline to the LMT. In this example, the amplitude drops due to pointing offsets are particularly large due to the small beam size of the LMT. At the beginning and end of the track, the telescopes observe at a low elevation and therefore through a large amount of airmass, resulting in significant atmospheric opacity effects. Since opacity measurements are only done between scans, while intra-scan trends are not corrected, visibility amplitudes are still exhibiting slopes within scans. At the end of the track, the opacity attenuation factor has a higher slope at ALMA. The intra-scan fall of the amplitudes is therefore partly corrected by network calibration here. The residual amplitude errors can typically be corrected with self-calibration methods (Section \ref{sec:grmhd_imagerecon}).

Figure \ref{fig:ptsrc_phase} shows the visibility phases before and after calibration on a short segment of the ALMA-LMT track. Before fringe fitting, the phase rotates fast due to tropospheric turbulence. The phases of the different frequency channels, shifted to start at the same value, drift apart as time progresses. After fringe fitting and averaging, the phase is close to zero.

\subsection{Crescent and GRMHD image reconstructions}
\label{sec:grmhd_imagerecon}
We use the geometric crescent model (Section~\ref{sec:model_crescent}) and the physically motivated $\kappa$-jet source model (Section~\ref{sec:models_davelaar}), to demonstrate the difference in visibility data and image reconstructions between simple synthetic observations where only thermal noise is included, and observations where all corruption and calibration effects are included.
We run these models through \pipe{} in two cases: one in which we apply only thermal noise, and one in which we apply all corruption and calibration effects described in Sections \ref{sec:meqsilhouette} -- \ref{sec:obsparameters}.
   
  \begin{figure*}[h]
   \centering
   \includegraphics[width=0.24\textwidth]{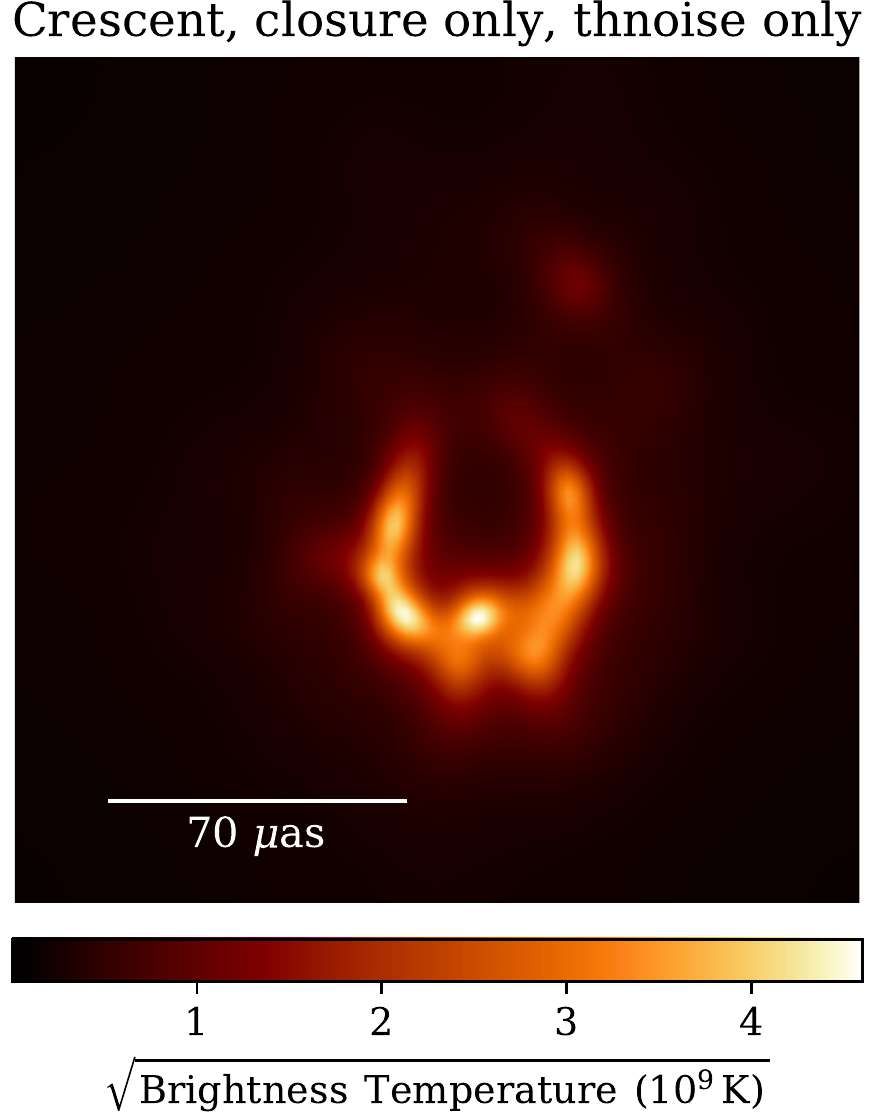}
   \includegraphics[width=0.24\textwidth]{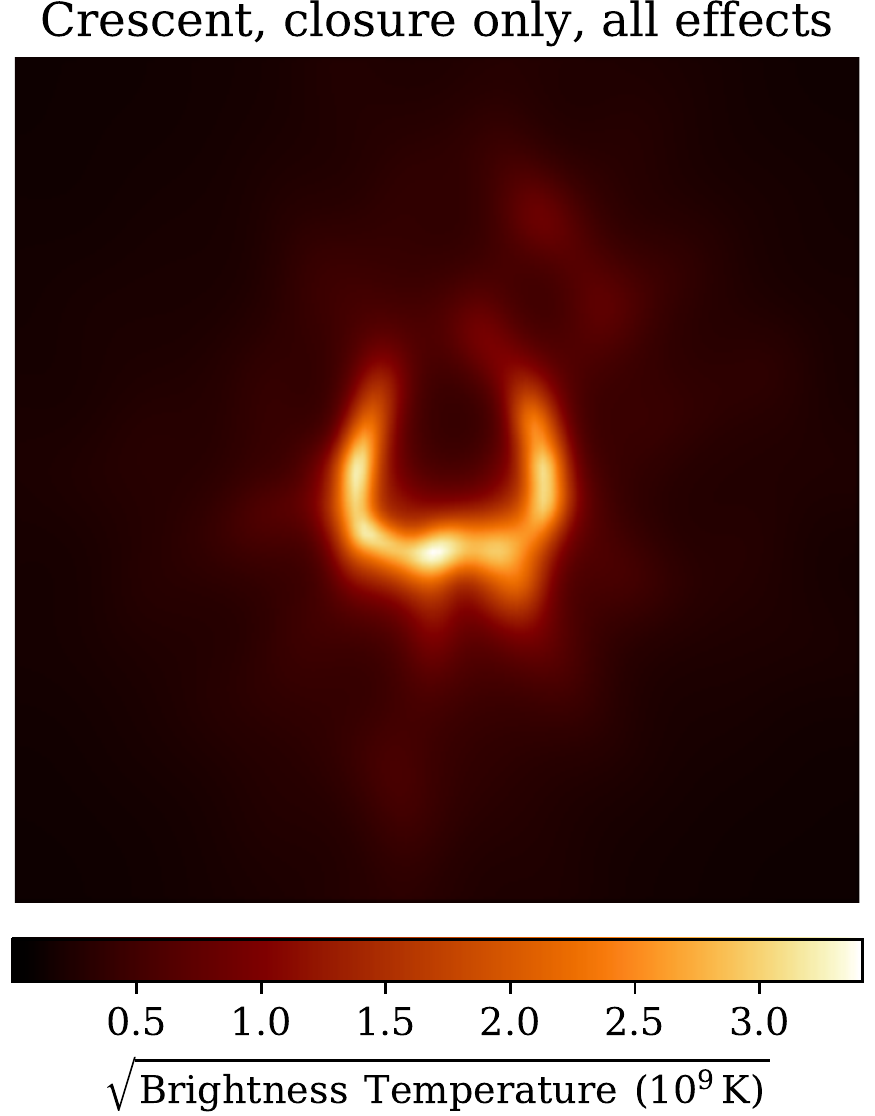}
   \includegraphics[width=0.24\textwidth]{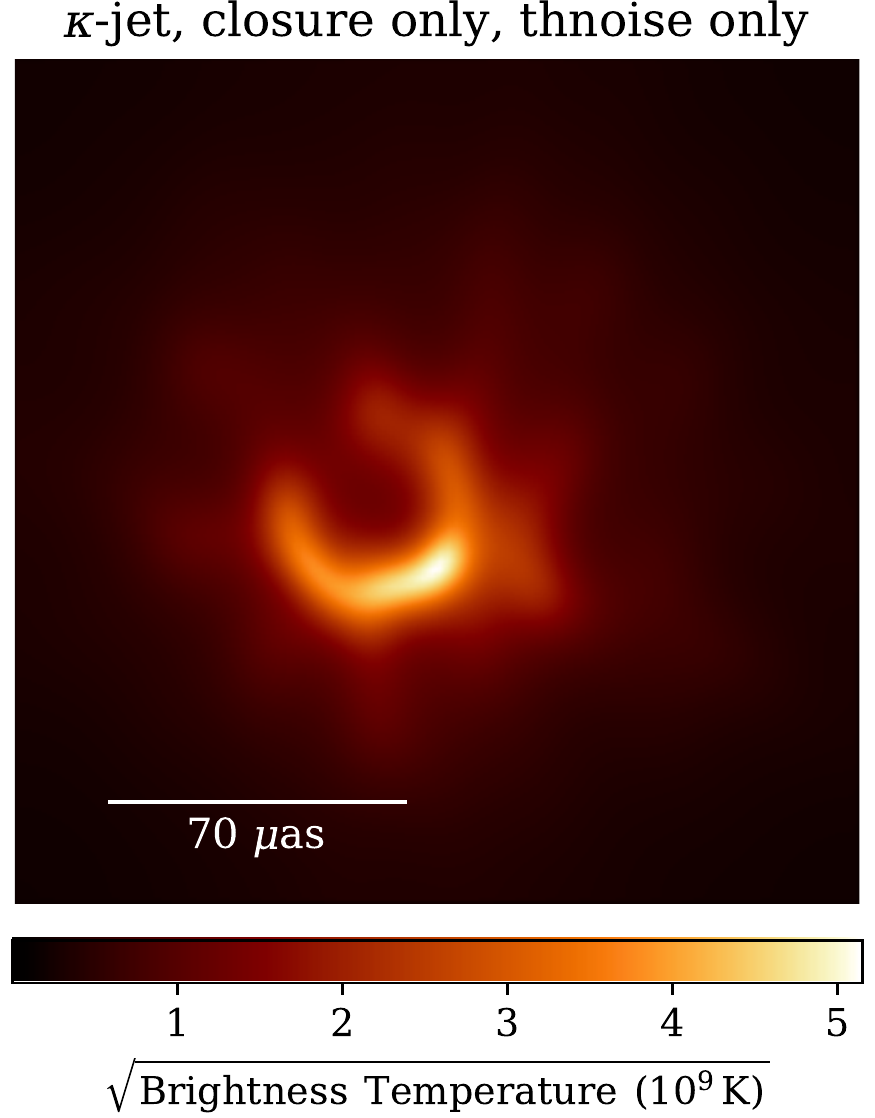}   \includegraphics[width=0.24\textwidth]{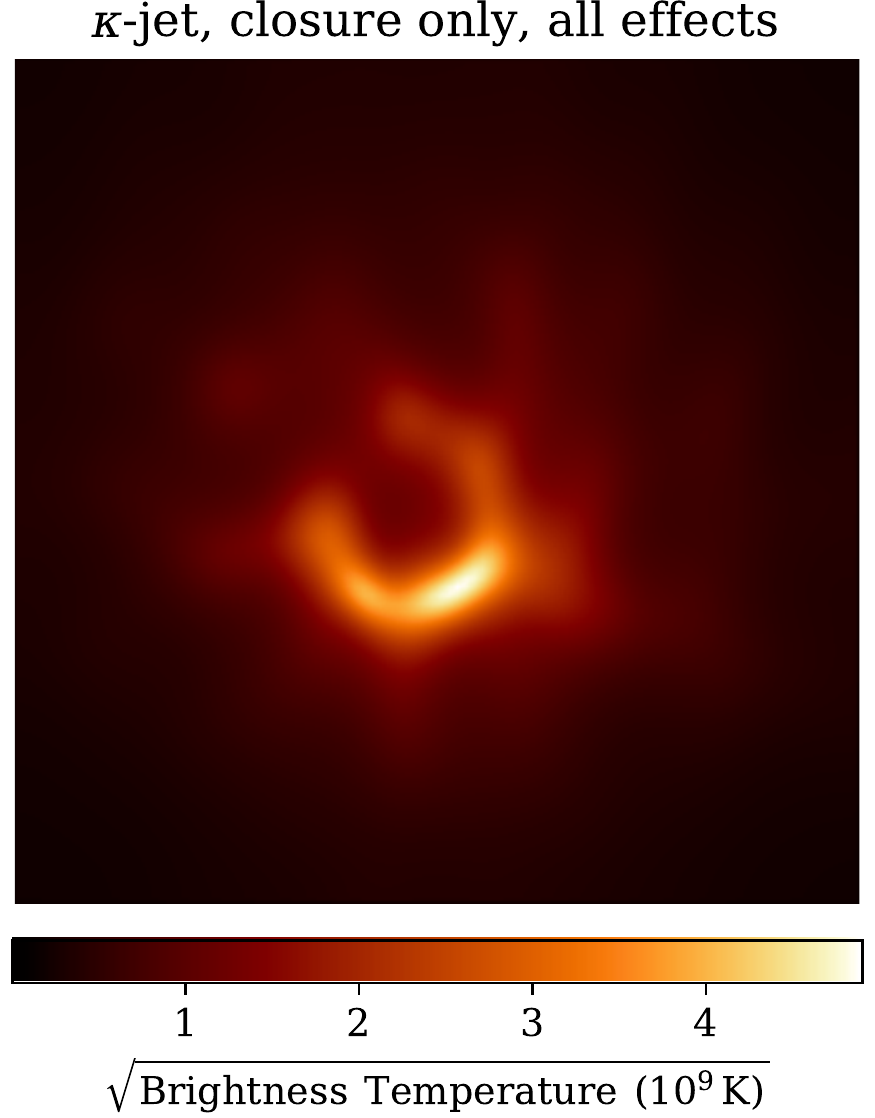} \\
   \includegraphics[width=0.24\textwidth]{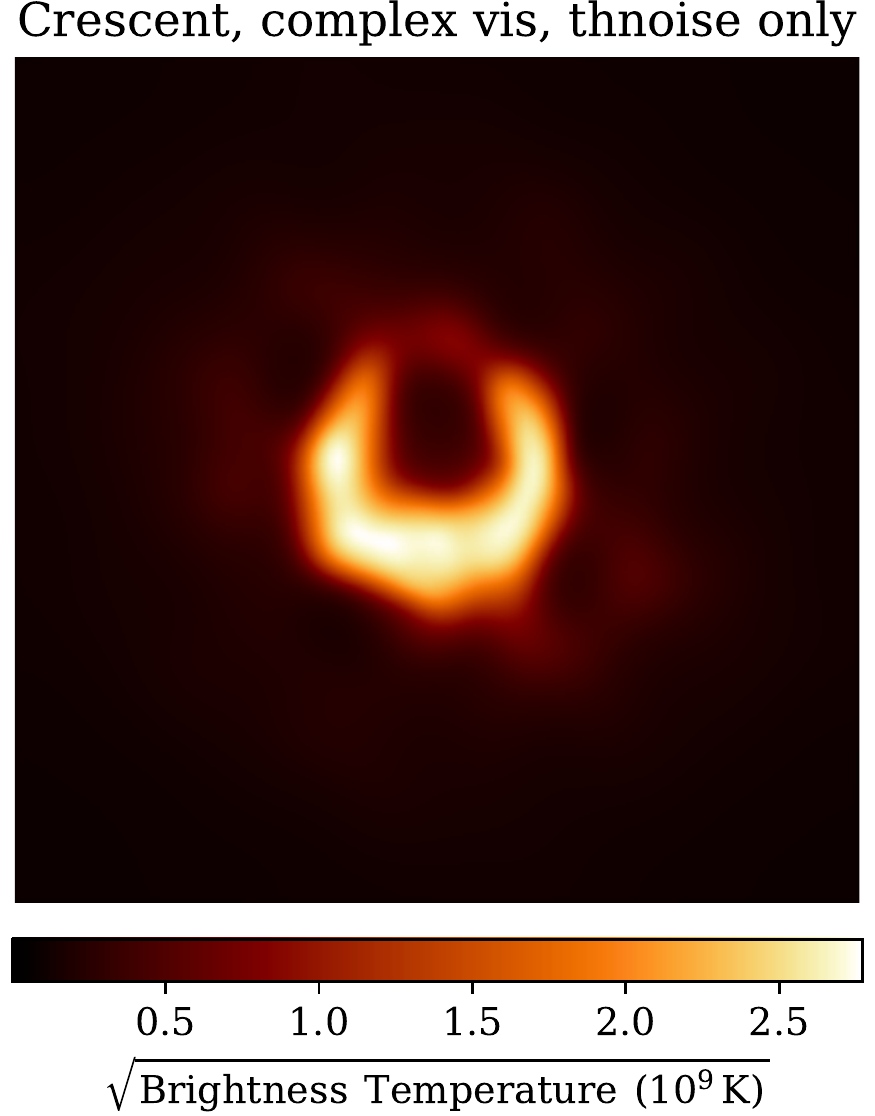}
   \includegraphics[width=0.24\textwidth]{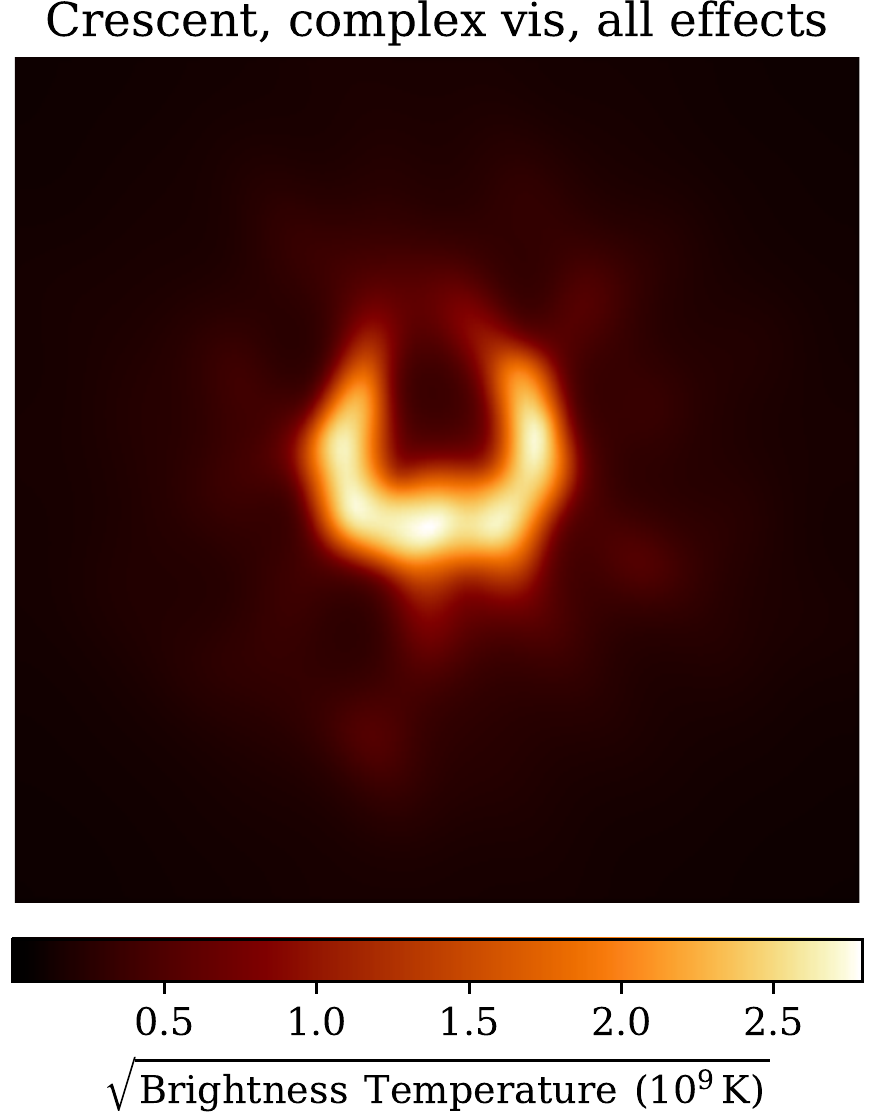}
   \includegraphics[width=0.24\textwidth]{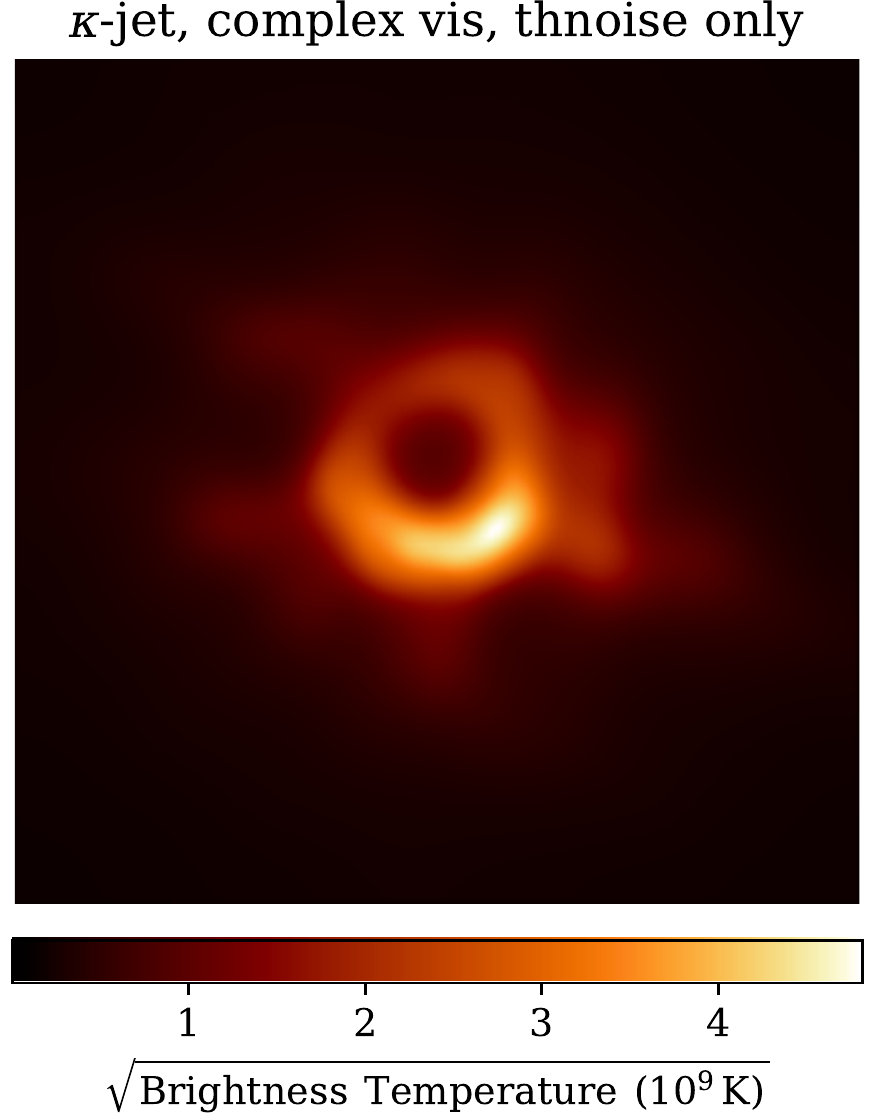}
   \includegraphics[width=0.24\textwidth]{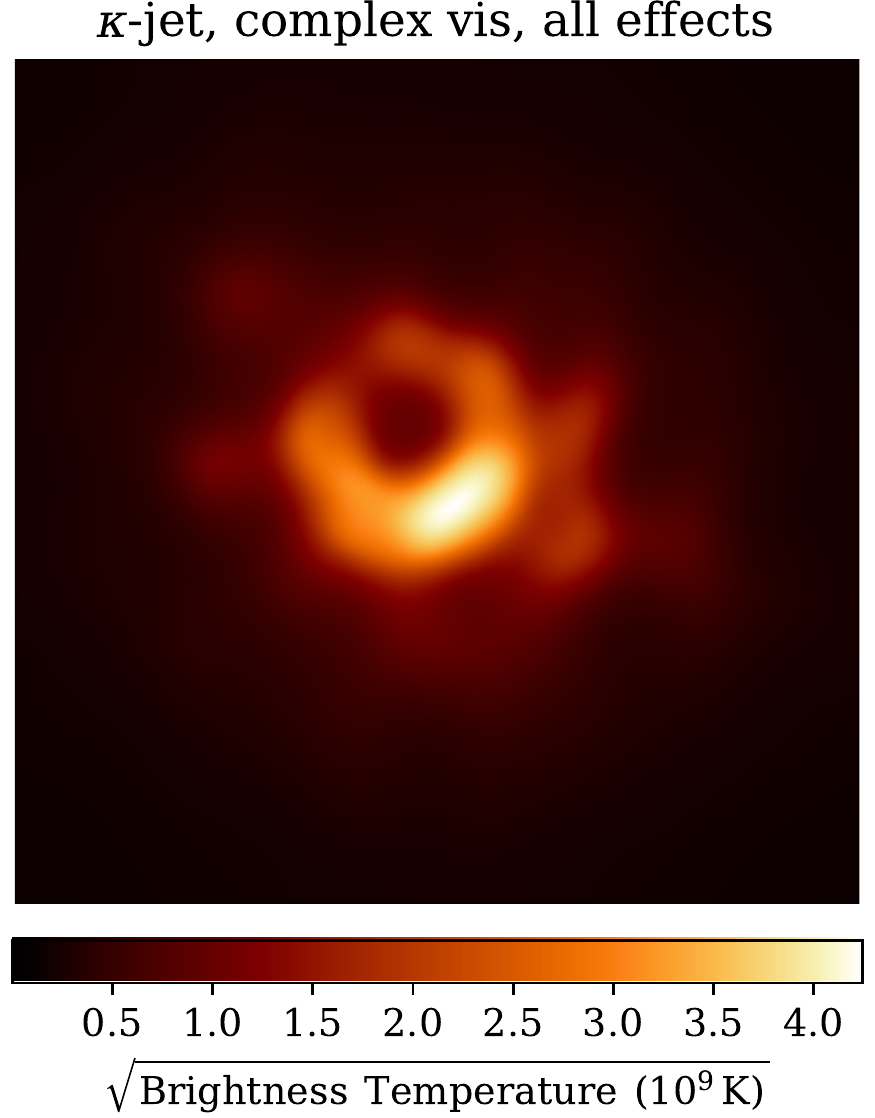}
      \caption{EHT2017 images reconstructed from calibrated synthetic data. The crescent model (Fig. \ref{fig:crescent}; columns 1 and 2) and the $\kappa$-jet model (Figure \ref{fig:radplots}; columns 3 and 4) were used as input. The images were reconstructed using closure quantities only (upper panels) or complex visibilities (lower panels), for synthetic data generated with only thermal noise (columns 1 and 3) and all corruption and calibration effects (columns 2 and 4) applied to the data. When all effects were included, the visibilities were self-calibrated in the imaging process. 
      }
     \label{fig:recs}
   \end{figure*}
  
  \begin{figure}[h]
   \centering
   \includegraphics[width=0.45\textwidth]{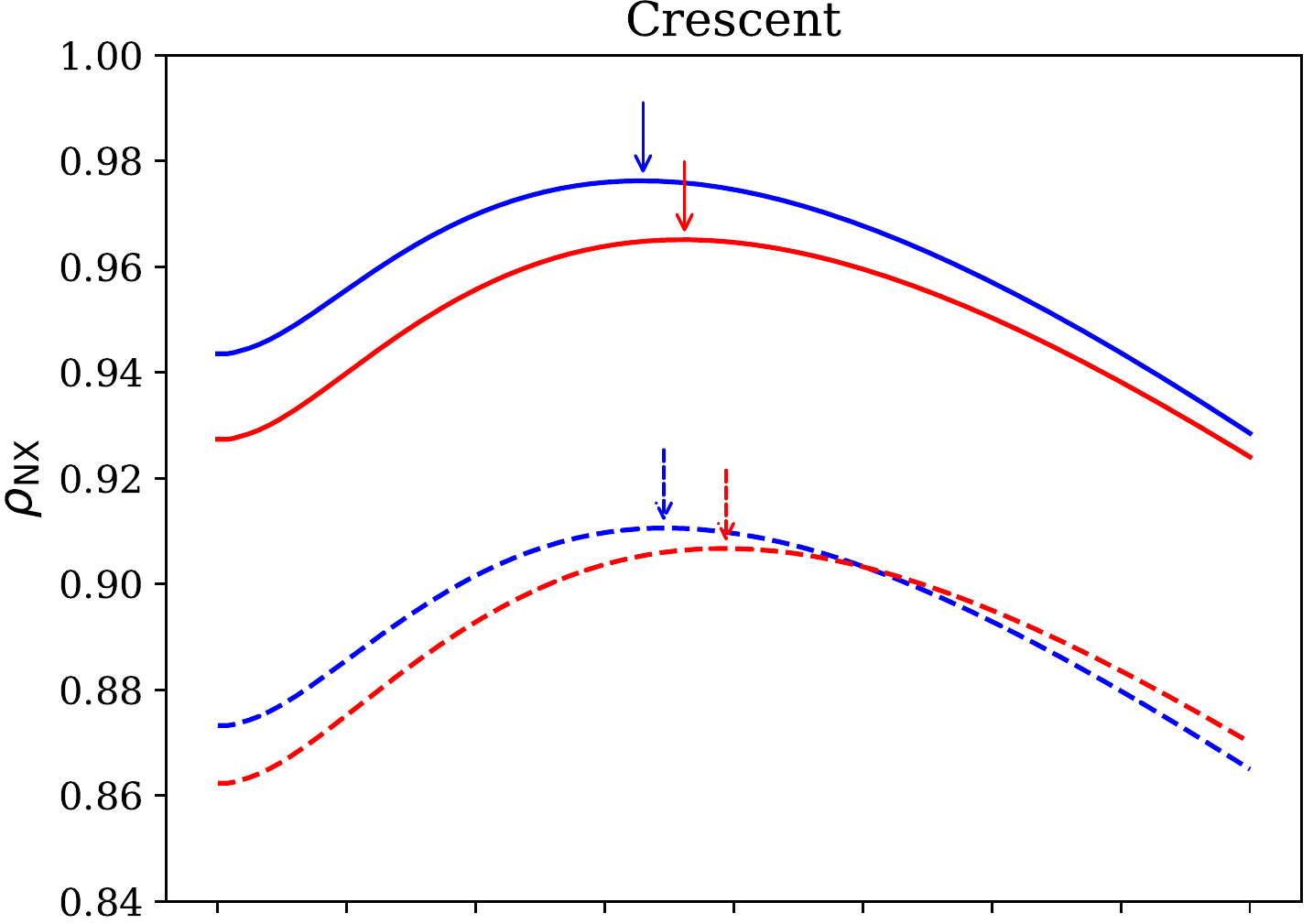}\\
   \includegraphics[width=0.45\textwidth]{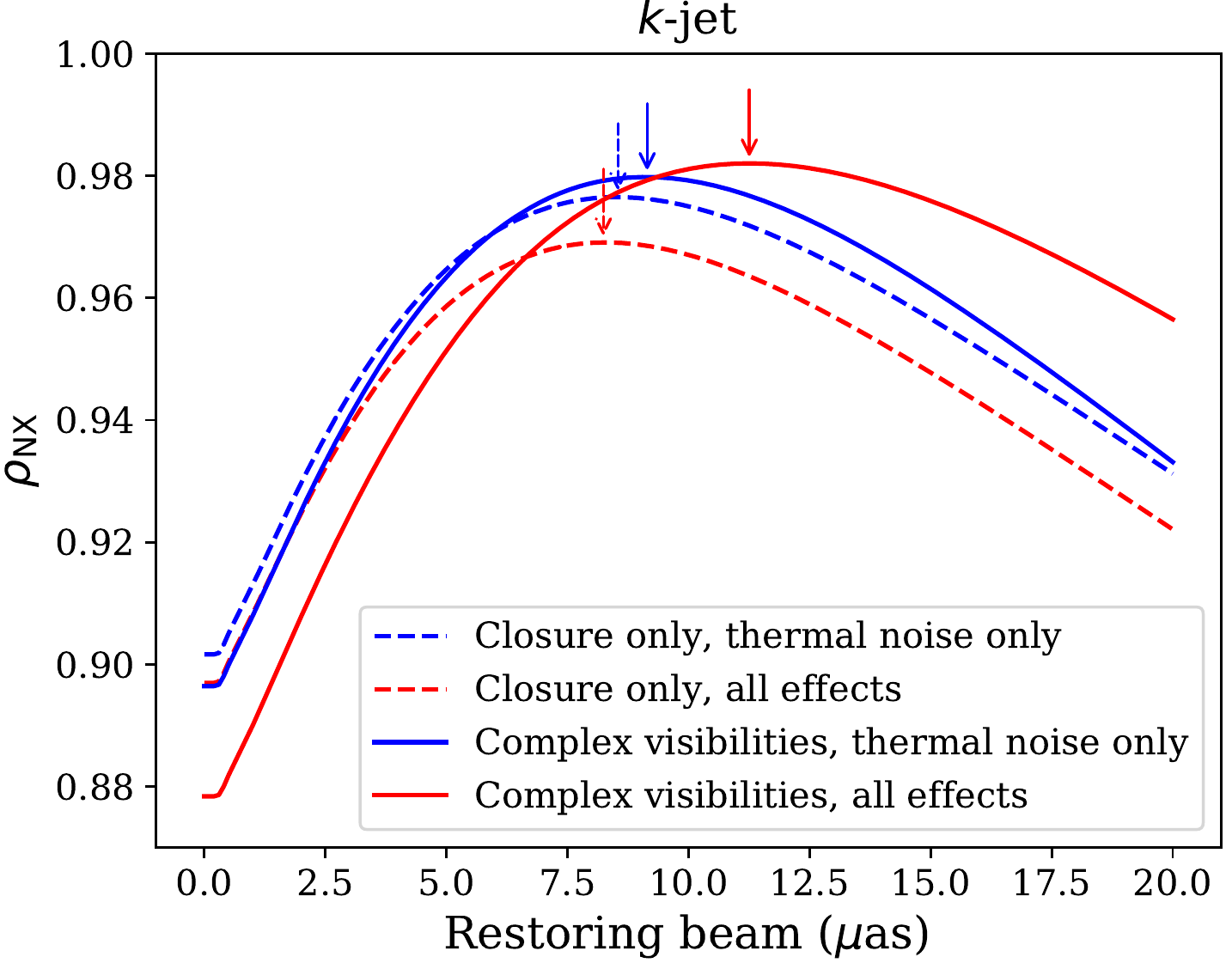}
      \caption{Normalized cross-correlation between image reconstructions in Figure \ref{fig:recs} and model images in Figures \ref{fig:crescent} and \ref{fig:models}. The crescent (top panel) and $\kappa$-jet (bottom panel) models were used, respectively, where the model images were convolved with a circular beam of varying size. The arrows indicate the peak positions.}
     \label{fig:nxcorr_conly_vis}
  \end{figure}

Figure \ref{fig:radplots} shows the scan-averaged synthetic visibility amplitudes and phases as a function of baseline length for both cases as compared to the direct Fourier Transform of the $\kappa$-jet model image. The amplitudes with only thermal noise (top row, left panel) line up with the model image, while there are systematic offsets for the amplitudes with all effects (top row, middle panel) due to pointing offsets and phase incoherence over the scan averaging time. The visibility phases with thermal noise only also line up with the model, while they are significantly different when all effects are included (bottom row, left and middle panel, respectively).

The offset between calibrated visibility phases and the phases computed directly from the model image is expected from the combination of rapid tropospheric phase fluctuations and station-based fringe fitting to a point source model, which causes the absolute phase information to be lost. The true source structure is nonetheless encoded in the closure quantities, which are robust against the station-based calibration errors, assuming there is no decorrelation when the complex visibilities are averaged to 10 seconds. After self-calibrating the data to the reconstructed source model (see below), the visibility phases match the model image more closely (bottom row, right panel). The remaining residual offsets are a result of uncertainties in the image reconstruction, introduced by the finite resolution and gaps in the $uv$-coverage.

Figure~\ref{fig:recs} shows reconstructed images for thermal noise only and full corruption plus calibration synthetic data sets generated from the crescent and $\kappa$-jet source models. The images are reconstructed with a regularized maximum likelihood (RML) method using the \ehtim{} software. The fiducial parameters and regularizers (Maximum Entropy, Total Variation, and Total Squared Variation) obtained from an extensive parameter survey by \citet{eht-paperIV} are adopted. Before imaging, the data are scan-averaged. The starting point for imaging is a circular Gaussian model with a FWHM of 40 $\mu$as. Images in the upper panels of Figure~\ref{fig:recs} are reconstructed using only closure quantities, that is log-closure amplitudes and closure phases. The images were reconstructed iteratively while increasing the weights of the data terms with respect to the weights of the regularizer terms. When imaging with the full set of complex visibilities (bottom row), we use the fiducial \ehtim{} script from \citet{eht-paperIV} to start imaging with closure phases, log-closure amplitudes, and visibility amplitudes, iteratively self-calibrating the visibility amplitudes to the reconstructed image to solve for the antenna gains due to e.g. the pointing offsets that were introduced. The amplitude self-calibration starts after a first round of imaging using closure quantities and a priori calibrated visibility amplitudes, and is performed within the a priori and systematic error tolerances used in \citet{eht-paperIV}. The visibility phases are then self-calibrated and used for imaging as well, while maintaining the closure quantity fits. The fiducial \ehtim{} script is included in \pipe{} as an optional final step (see also Sec. \ref{sec:computingworkflow}).

Because closure quantities are robust against station-based errors introduced in our synthetic observations, the reconstructed images (Figure~\ref{fig:recs}, top row) are similar when only thermal noise is taken into account compared to the inclusion of all effects. This is true for both models. Because the crescent model has no extended features, any emission outside of the outer crescent ring in the reconstructed images can be classified as an imaging artefact. The reconstructions including all corruption and calibration effects show more of this spurious structure than the reconstructions including thermal noise only. The difference between including only thermal noise and including all effects is more apparent when the data are self-calibrated and complex visibilities are used as described above (Figure~\ref{fig:recs}, bottom row). The crescent model reconstruction is more irregular and has more noise when all effects are included. The $\kappa$-jet model shows a smoother and thinner ring when only thermal noise is included. These comparisons highlight the importance of synthetic observations where all corruption and calibration effects are taken into account when exploring how well an observed source can be reconstructed.

The fidelity of the image reconstructions in Figure \ref{fig:recs} can be quantified using an image similarity metric. We compute the normalized cross-correlation \citep{eht-paperIV} between the reconstructed image $X$ and the input model image $Y$, which is defined as
\begin{equation}
\rho_{\mathrm{NX}}(X,Y) = \frac{1}{N}\sum_i\frac{(X_i-\langle X\rangle)(Y_i-\langle Y\rangle)}{\sigma_X\sigma_Y}.
\end{equation}
Here, $N$ is the number of pixels in the images, $X_i$ is the $i$th pixel value of image $X$, $\langle X\rangle$ is the average pixel value of image $X$, and $\sigma_X$ is the standard deviation of the pixel values of image $X$. The possible values of $\rho_{\mathrm{NX}}$ range between -1 and 1, where a value of -1 indicates perfect anti-correlation between the images, 0 indicates no correlation, and 1 indicates perfect correlation. The images are shifted against each other to maximize $\rho_{\mathrm{NX}}$.

Figure \ref{fig:nxcorr_conly_vis} shows the $\rho_{\mathrm{NX}}$ values of the reconstructions in Figure~\ref{fig:recs}, which were cross-correlated with the model images in Figure~\ref{fig:crescent} for the crescent model and in Figure~\ref{fig:models} for the $\kappa$-jet model. The model images were convolved with a circular Gaussian beam of varying size. The trends seen in $\rho_{\mathrm{NX}}$ generally agree with the image inspections by eye as described above. For the crescent model, the closure only $\rho_{\mathrm{NX}}$ are similar for the thermal noise only reconstructions and reconstructions including all effects (top panel, dotted lines), although the former has a slightly higher peak $\rho_{\mathrm{NX}}$ at a slightly smaller beam size. $\rho_{\mathrm{NX}}$ improves substantially when complex visibilities are used for the image reconstructions (top panel, solid lines). Here, the thermal noise only reconstruction also gives a higher $\rho_{\mathrm{NX}}$ as one would intuitively expect. For all images, the peak value of $\rho_{\mathrm{NX}}$ is obtained for a restoring beam substantially smaller than the nominal array resolution of $\sim$23 $\mu$as, indicating the ability of RML image reconstruction to superresolve image structures \citep[e.g.][]{Chael2016, Akiyama2017}. For the $\kappa$-jet reconstructions, the (peak) $\rho_{\mathrm{NX}}$ also increases when complex visibilities are used for imaging. The reconstruction including all effects has a slightly higher peak $\rho_{\mathrm{NX}}$ than the thermal noise only reconstruction, but the peak is obtained for a larger restoring beam. Comparing the two bottom right images in Figure \ref{fig:recs}, the thermal noise only reconstruction indeed shows a sharper ring with a clearer outline of the black hole shadow. 

   \begin{figure*}[h]
   \centering
   \includegraphics[scale=0.55]{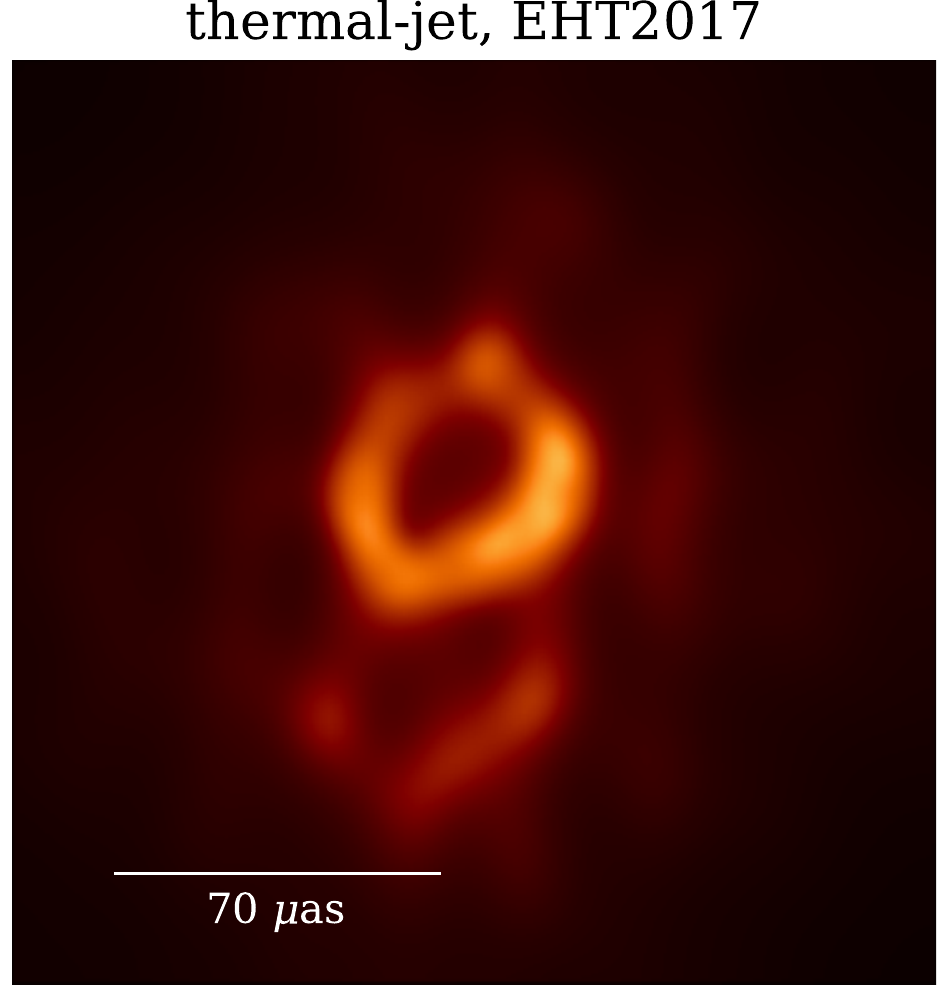}
   \includegraphics[scale=0.55]{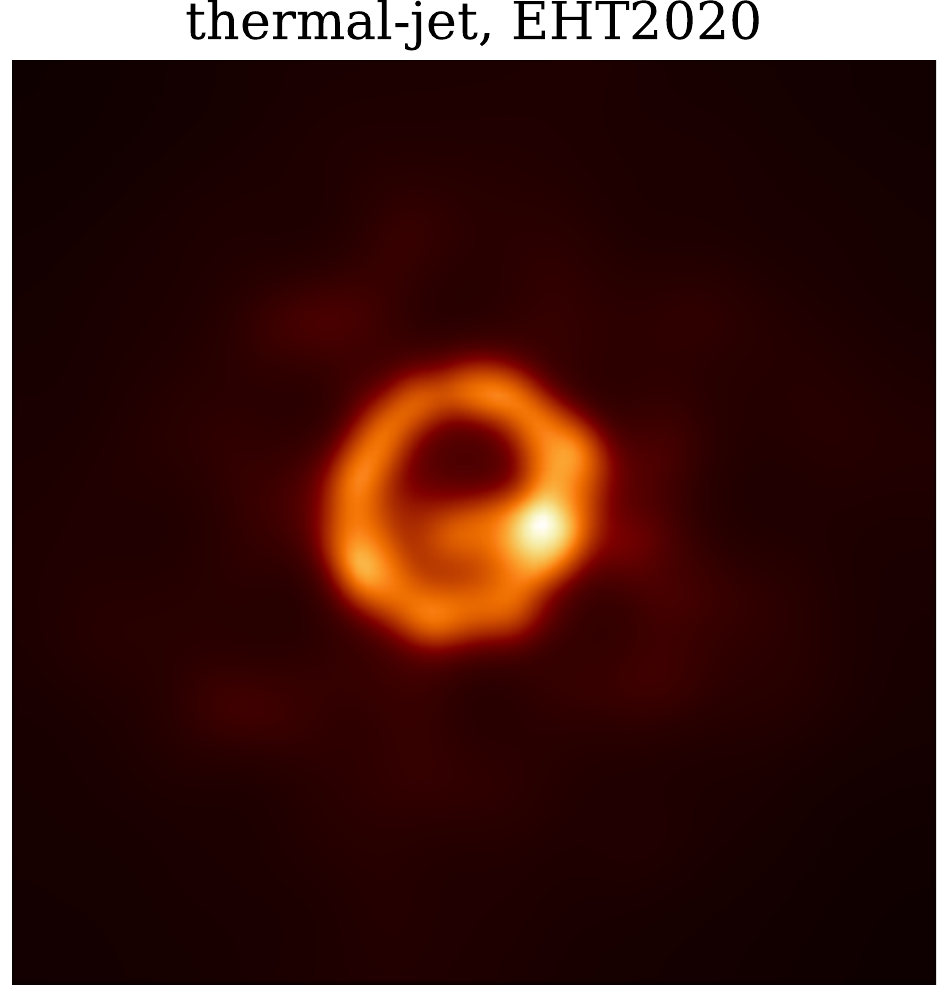}
   \includegraphics[scale=0.55]{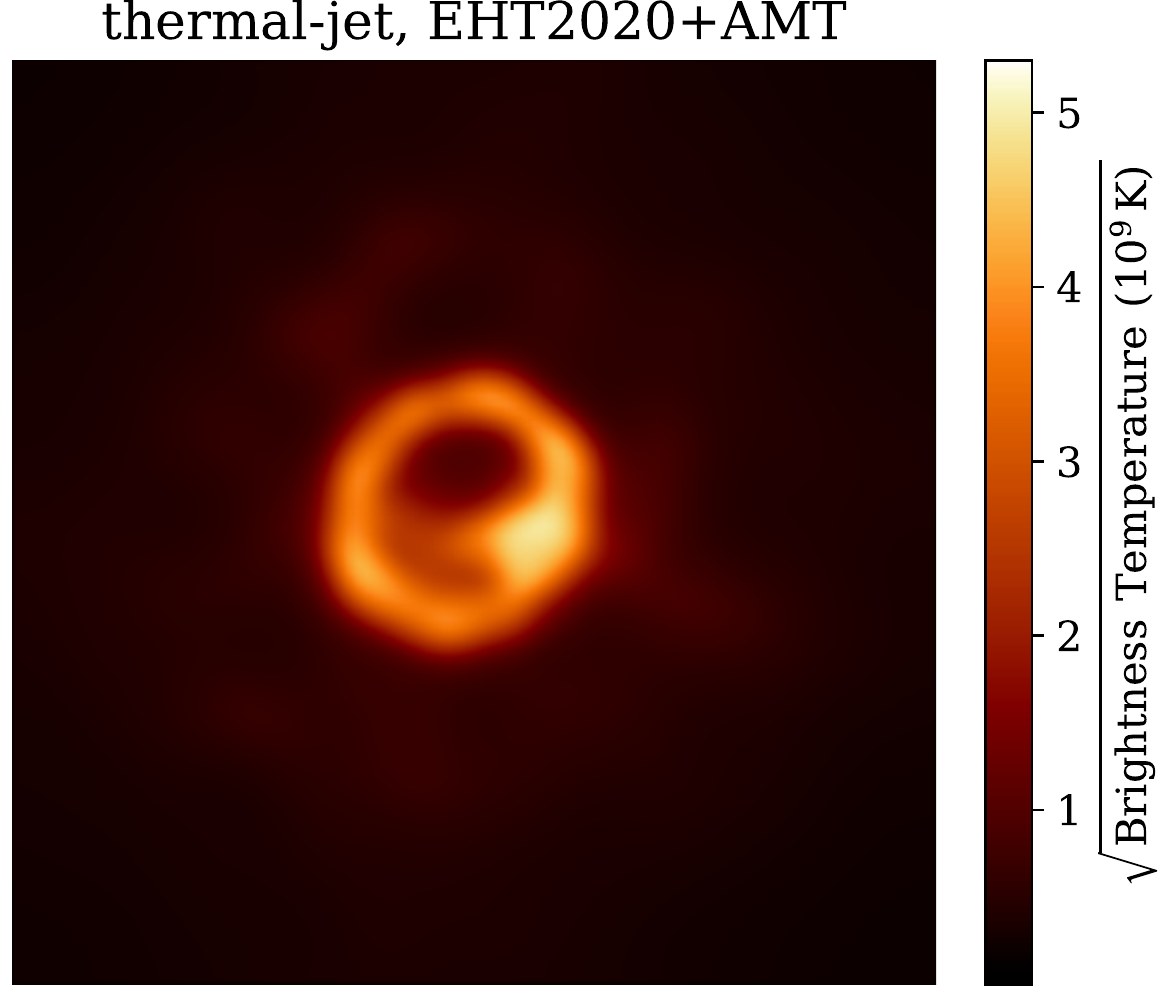} \\
   \includegraphics[scale=0.55]{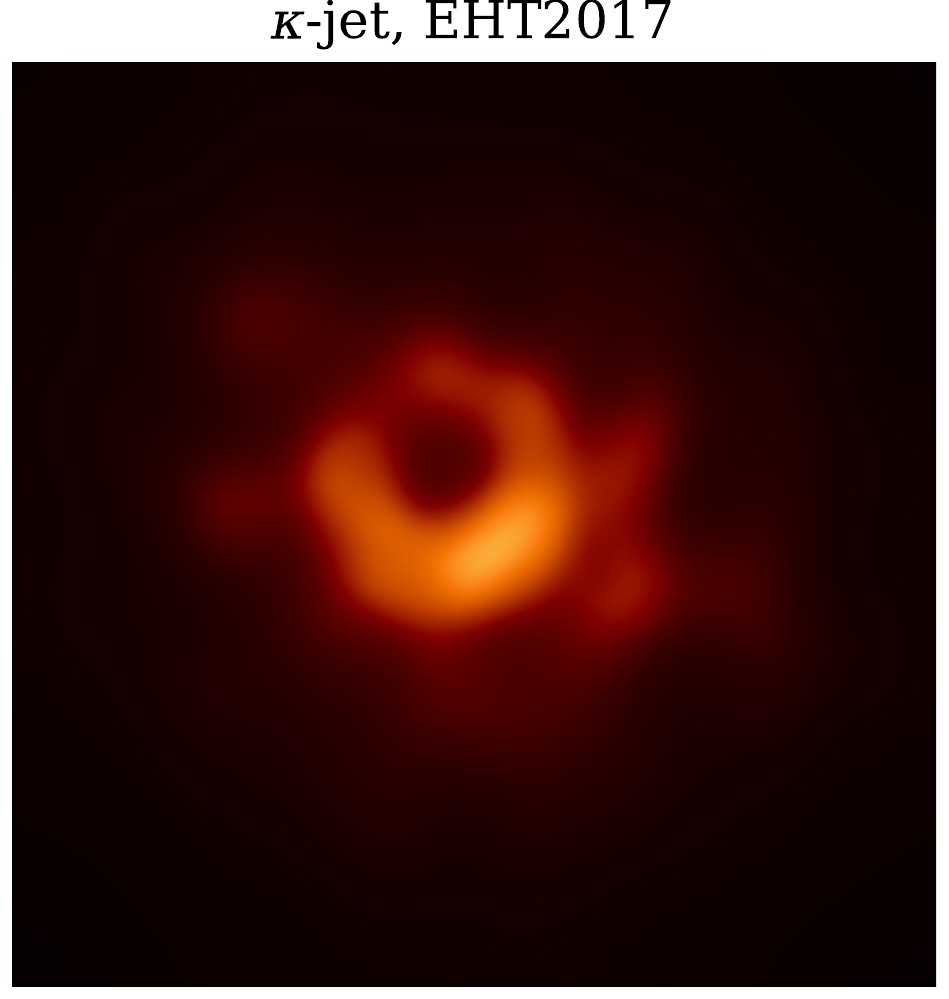}
   \includegraphics[scale=0.55]{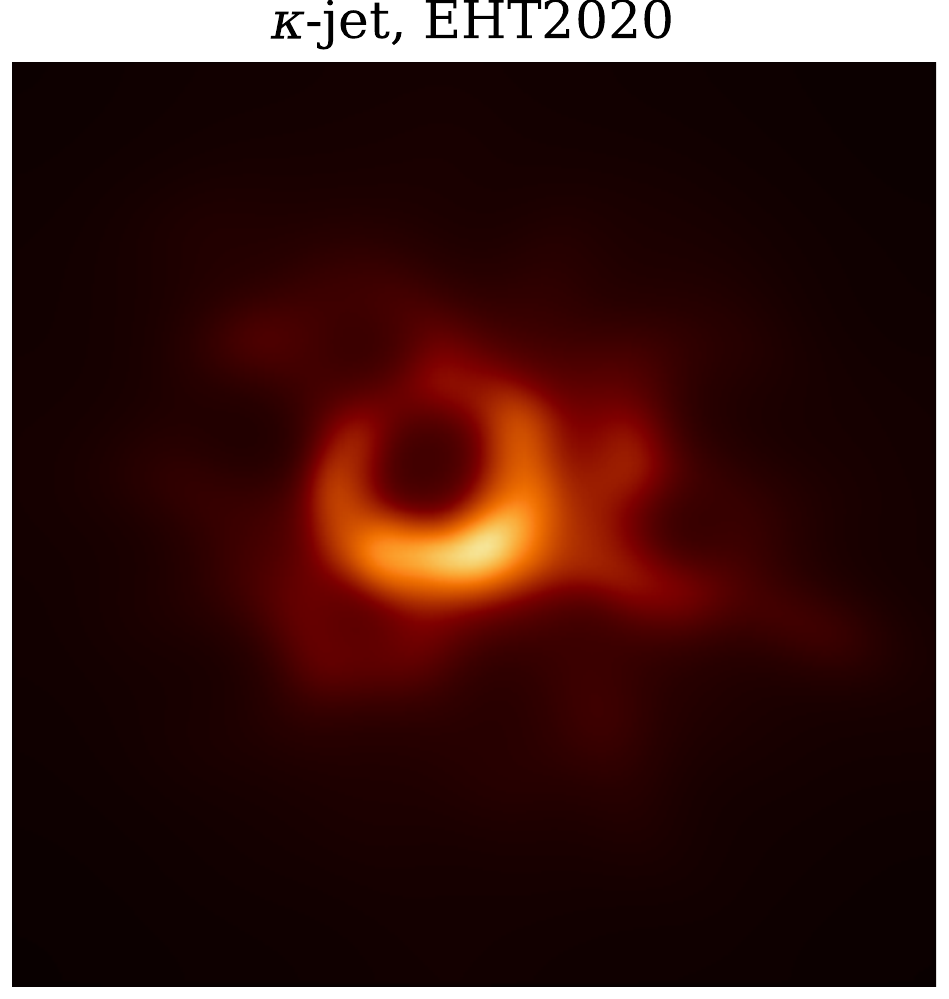}
   \includegraphics[scale=0.55]{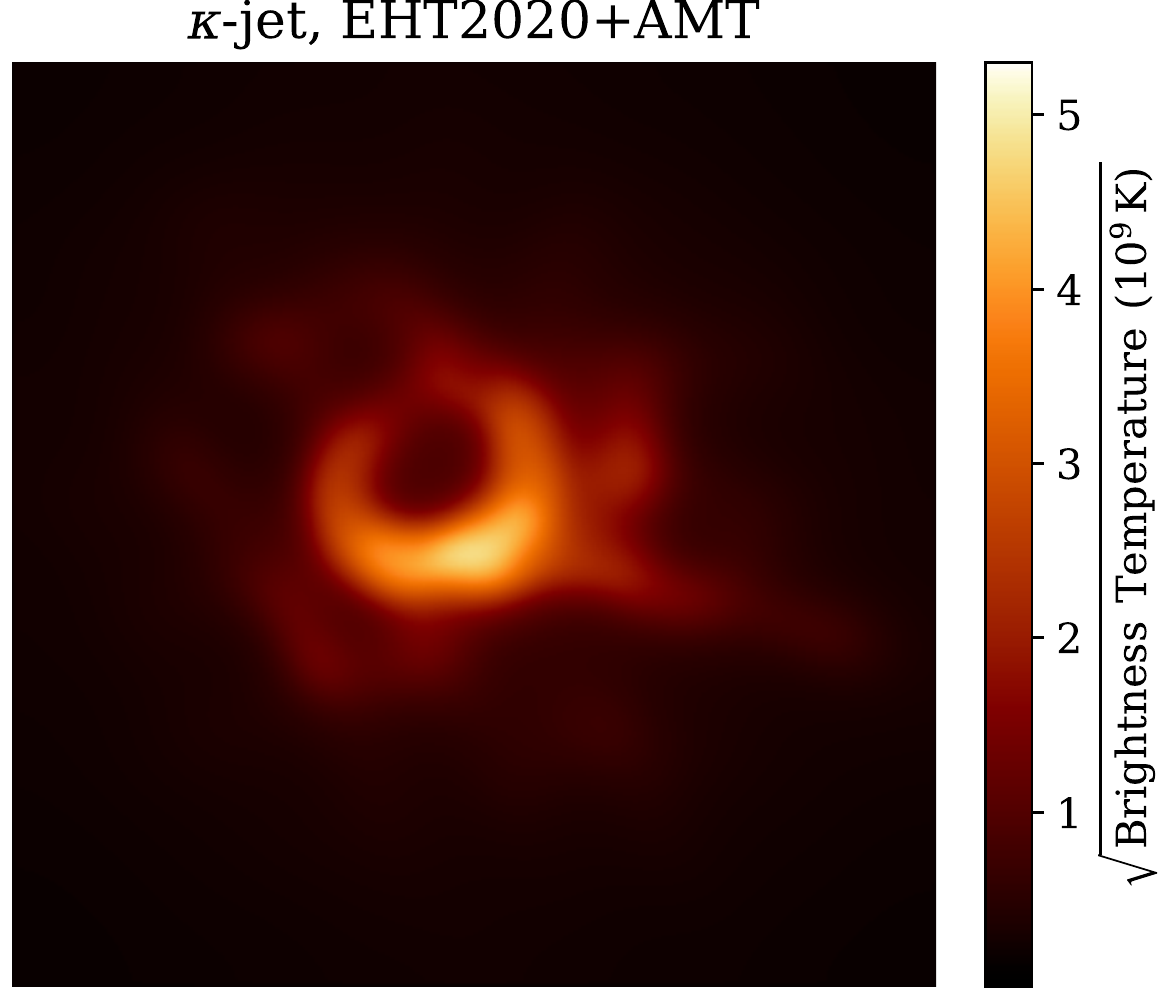} 
      \caption{Images reconstructed from synthetic observations with all effects included for different source models and arrays. The reconstructions are for the thermal (top) and $\kappa$-jet (bottom) models, using the EHT array in its 2017 configuration (left), with PDB, KP and GLT added as expected for 2020 (middle), and with the AMT, PDB, KP and GLT added (right). The image in the bottom left panel in this figure is the same as the image in the bottom right panel in Figure \ref{fig:recs}.}
         \label{fig:recs_future}
   \end{figure*}
   
      \begin{figure}[h]
   \centering
   \includegraphics[width=.45\textwidth]{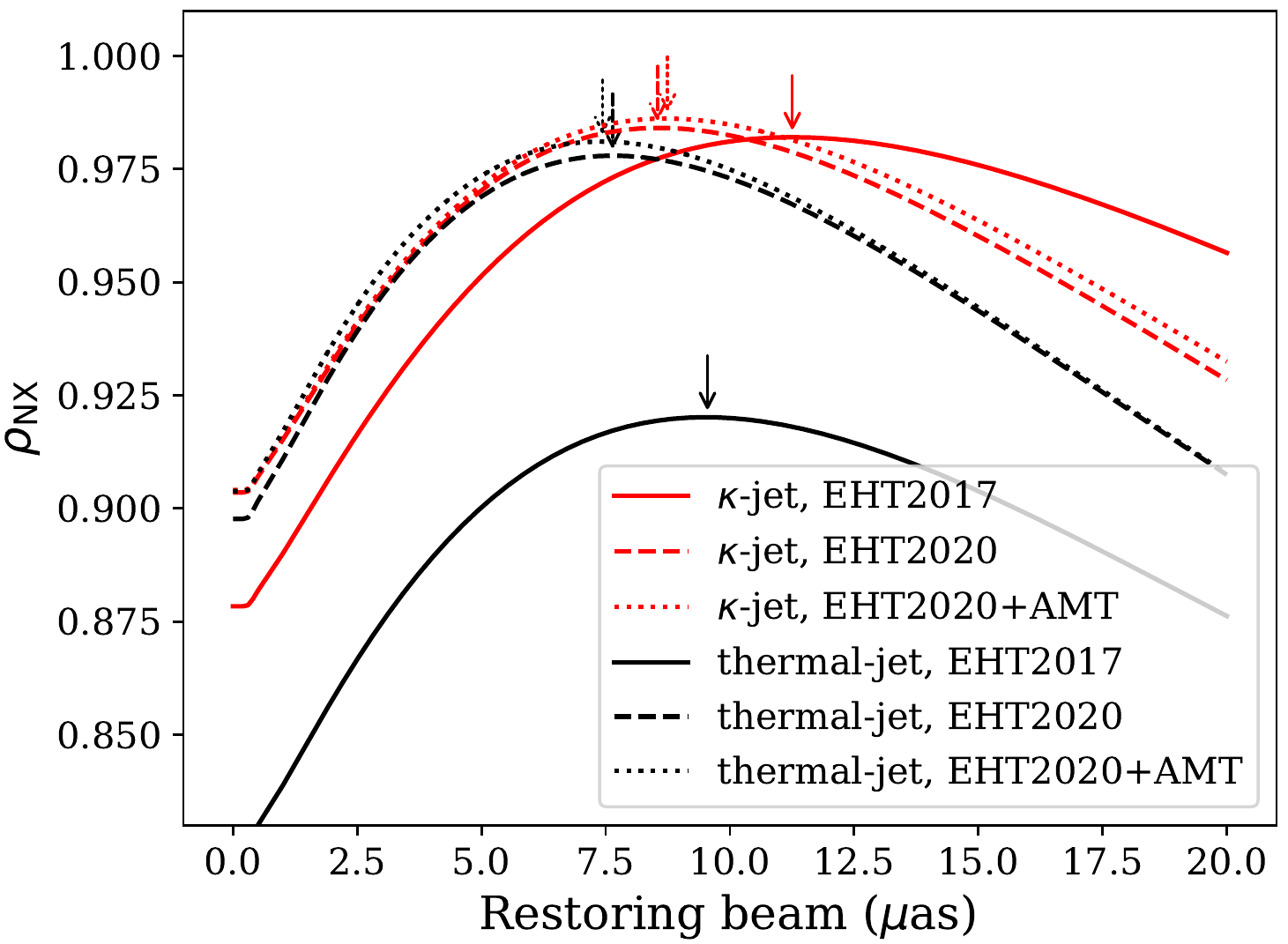}
      \caption{Normalized cross-correlation between image reconstructions in Figure \ref{fig:recs_future} and model images in Figure \ref{fig:models}. The model images were convolved with a circular beam of varying size. The arrows indicate the peak positions.}
         \label{fig:nxcorr}
   \end{figure}
   
      \begin{figure*}[h]
   \centering
   \includegraphics[scale=0.37]{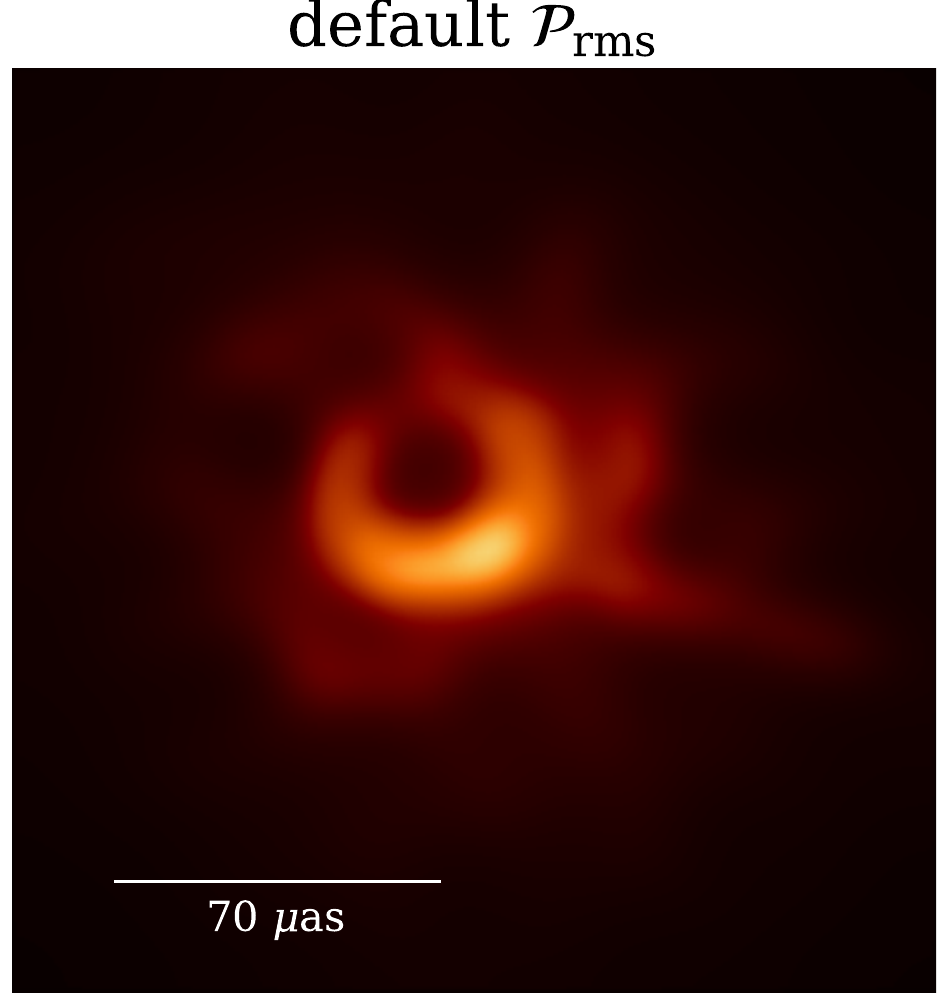}
   \includegraphics[scale=0.37]{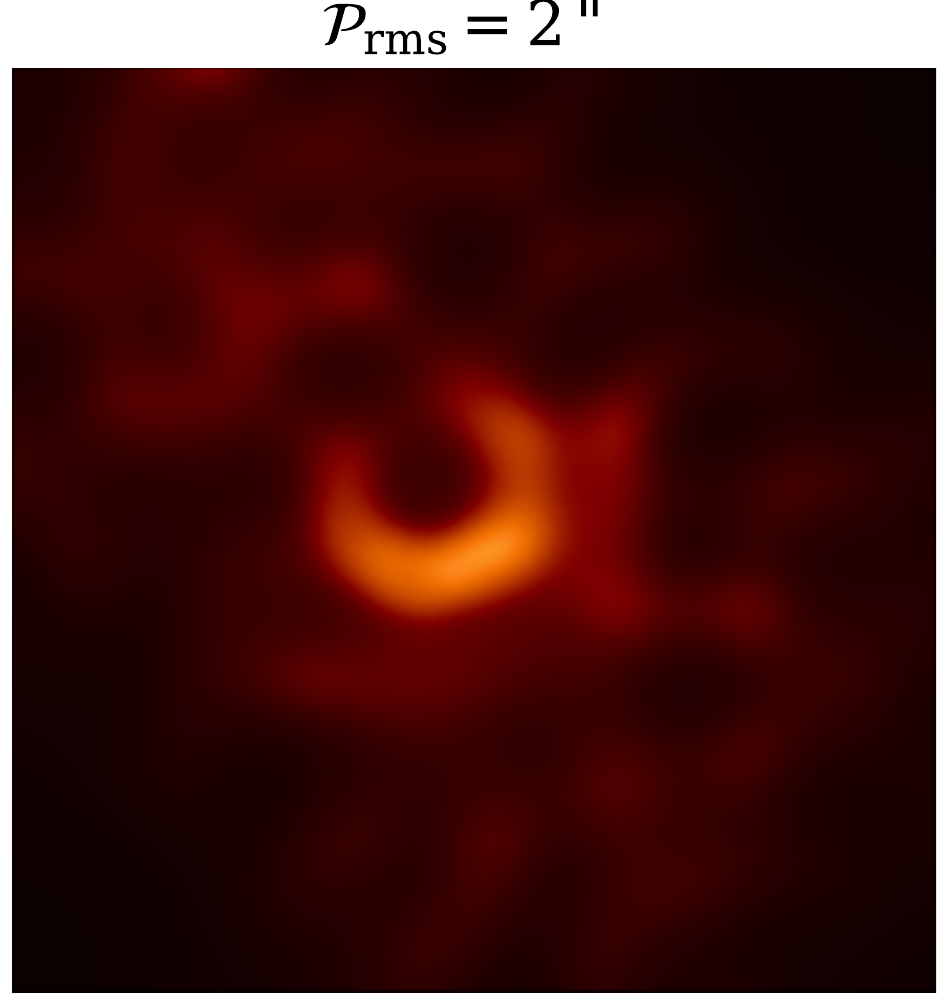}
   \includegraphics[scale=0.37]{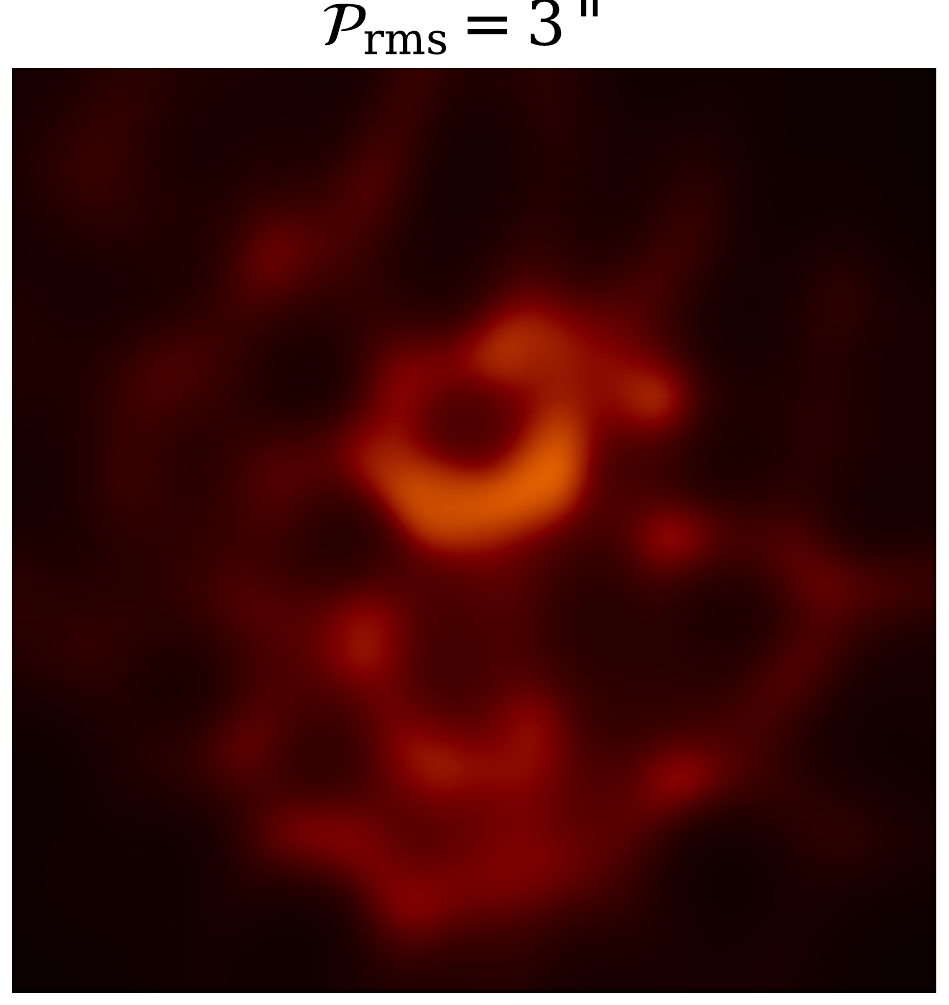}
   \includegraphics[scale=0.37]{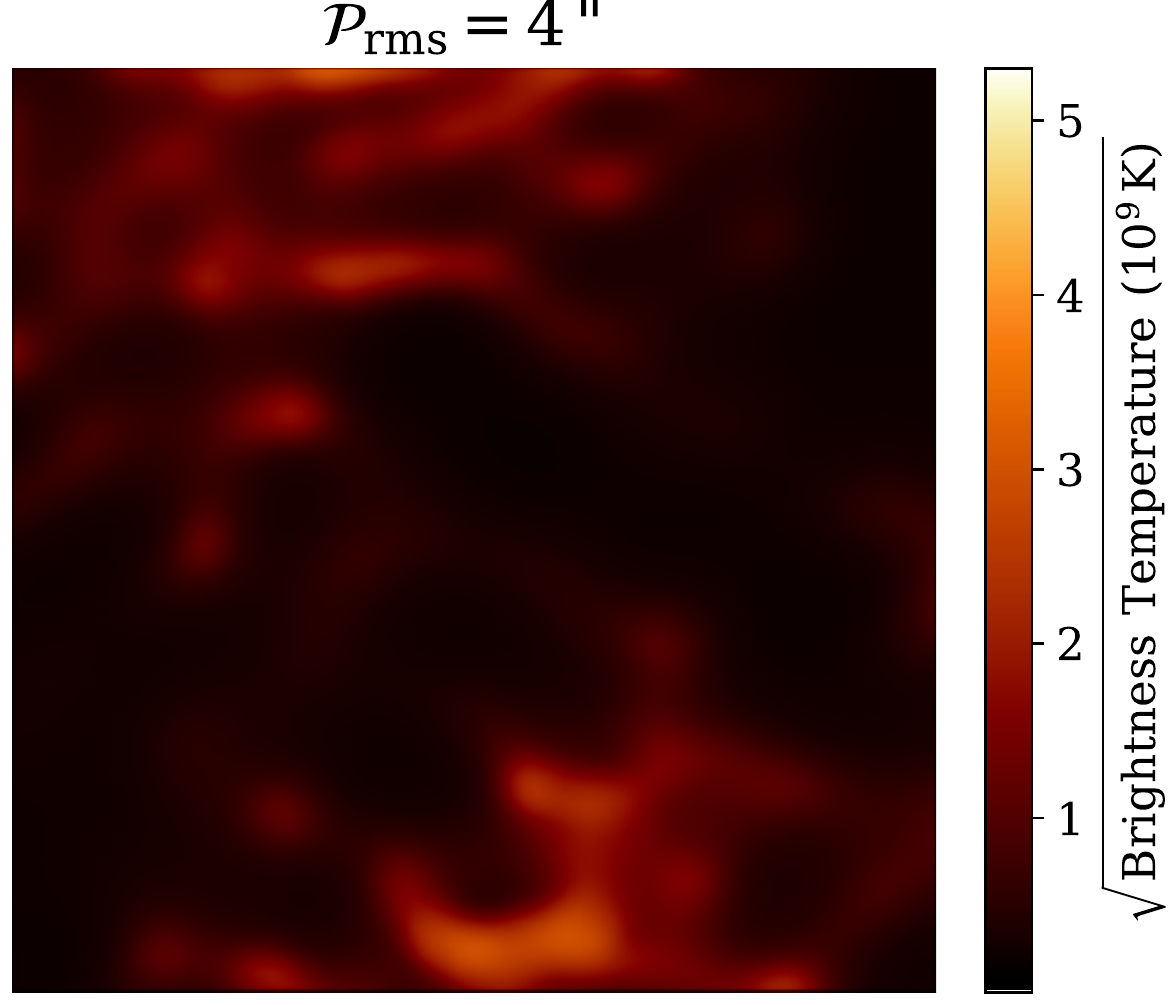}
      \caption{Images reconstructed from synthetic observations with all effects included under varying weather conditions. The $\kappa$-jet model was used, with the 2020 EHT array. These synthetic observations are run with 10\,\% gain errors, PWV = 5\,mm, and  $t_\mathrm{c}$ = 3\,s for all stations. The leftmost panel shows a reconstruction with the default pointing rms values listed in Table~\ref{tab:antennas}. Increasingly larger $\mathcal{P}_\mathrm{rms}$ values have been used in the other reconstructions: (from left to right) 2\,as, 3\,as, and 4\,as for all stations.
      }
         \label{fig:weather}
   \end{figure*}
   
      \begin{figure}[t!]
   \centering
   \includegraphics[scale=0.36]{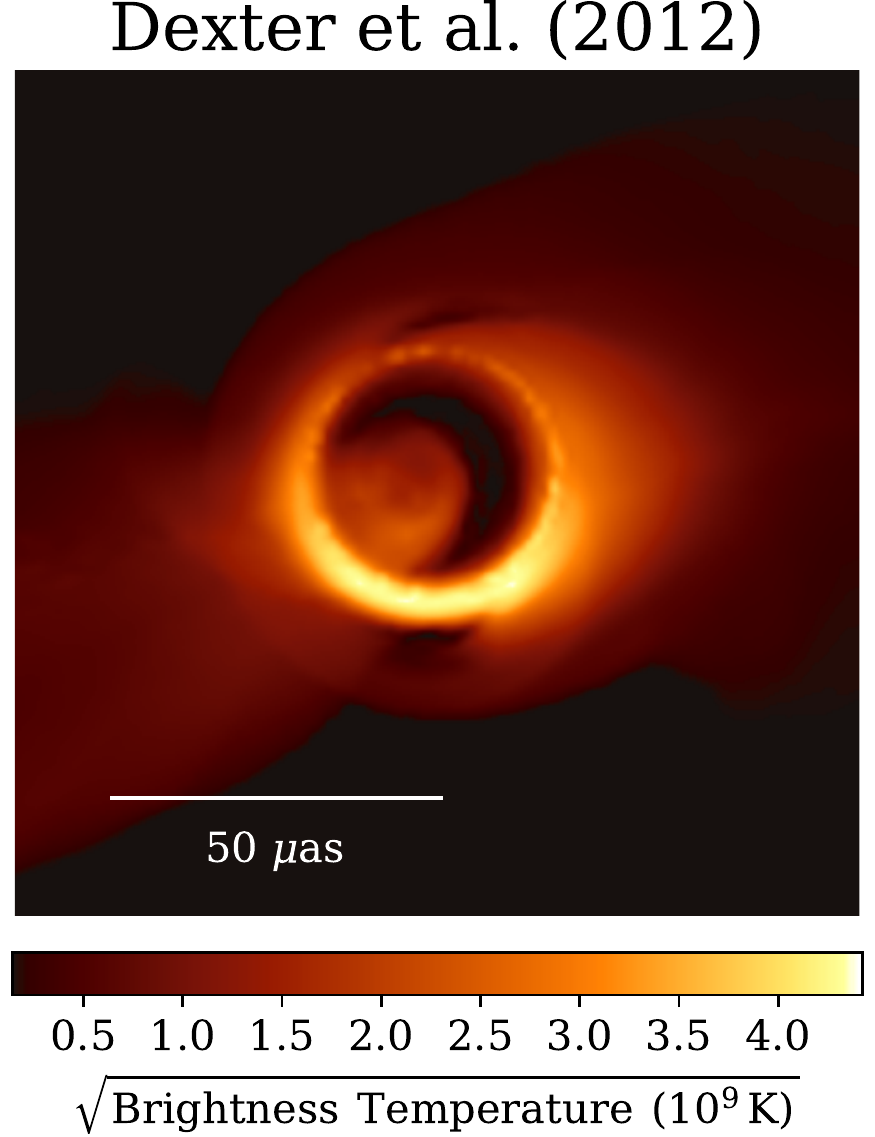}
   \includegraphics[scale=0.36]{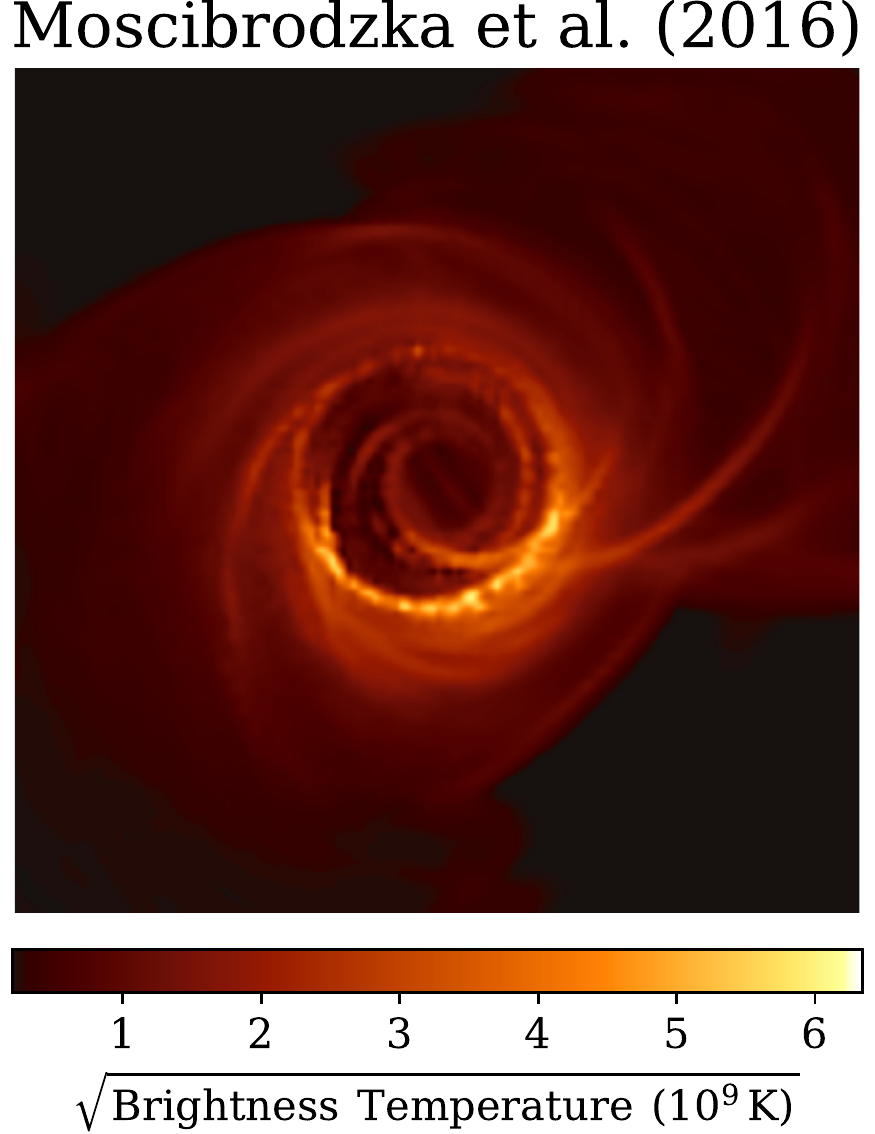} \\
   \includegraphics[scale=0.36]{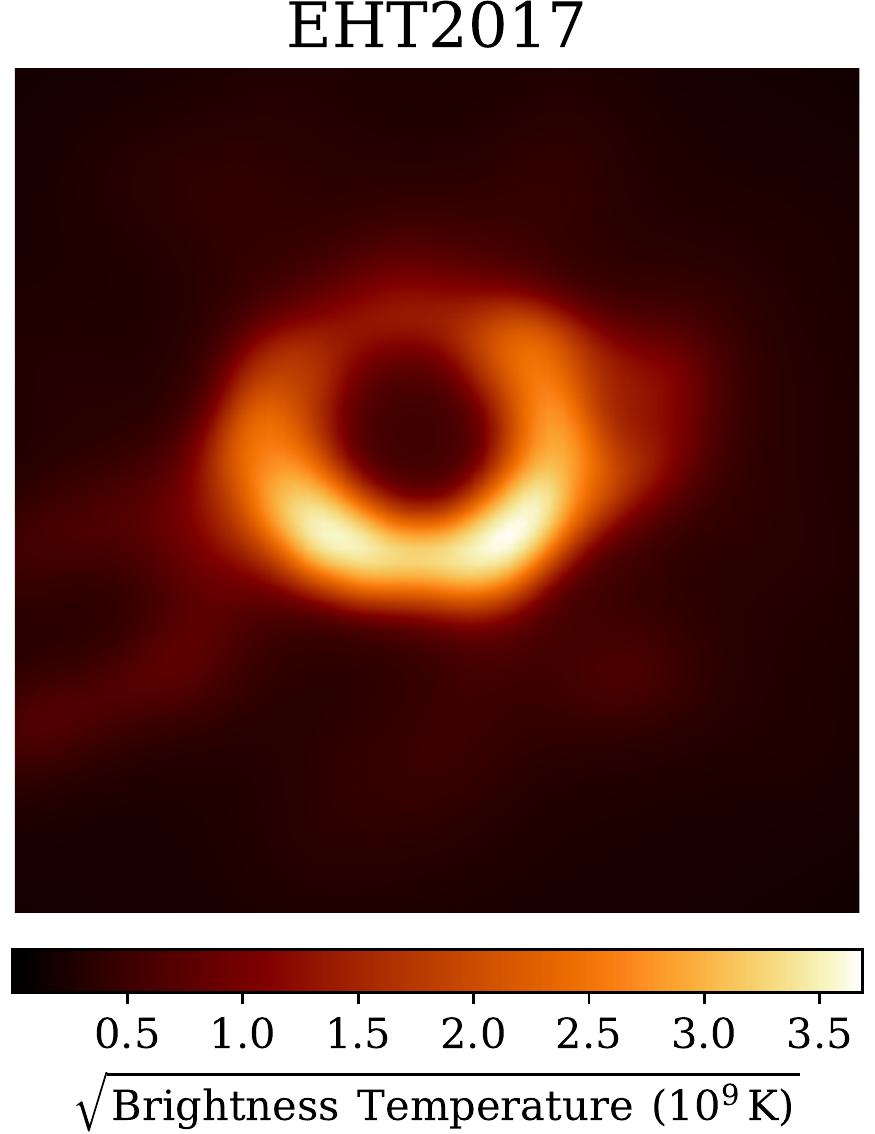}
   \includegraphics[scale=0.36]{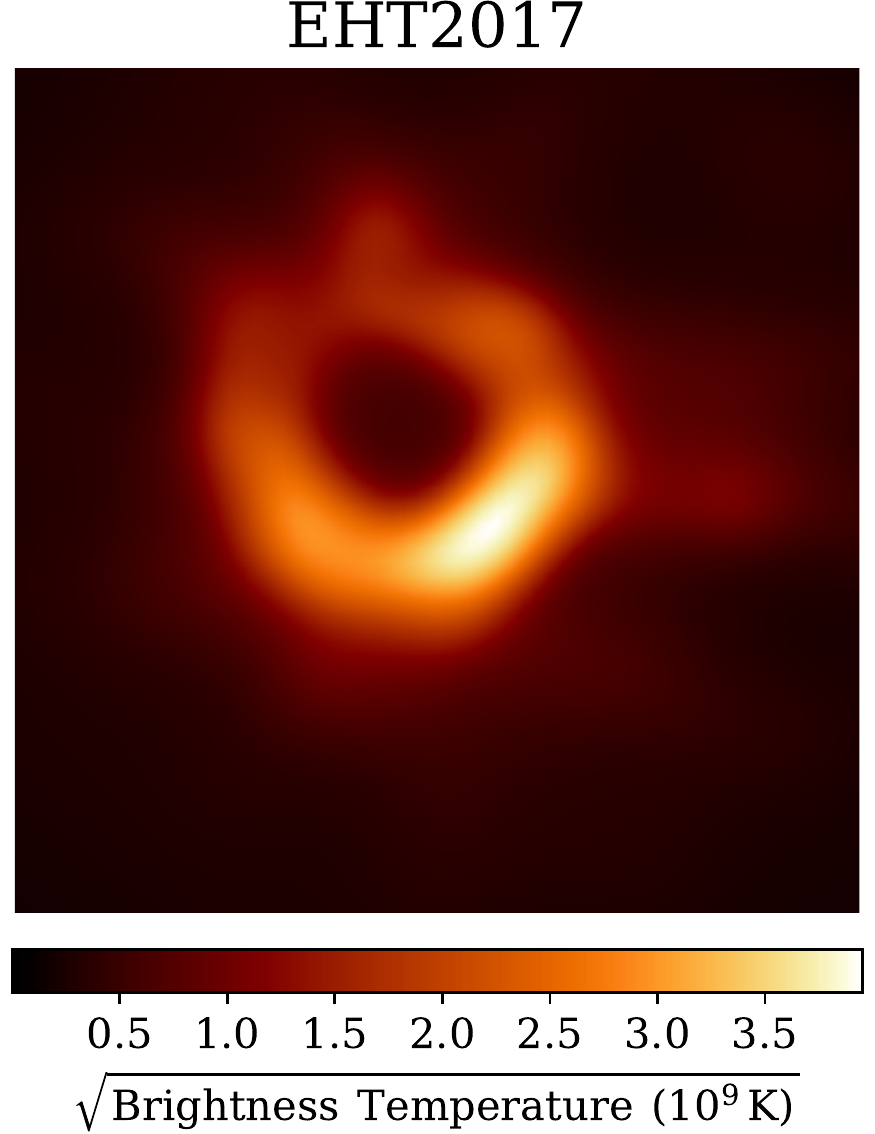} \\
   \includegraphics[scale=0.36]{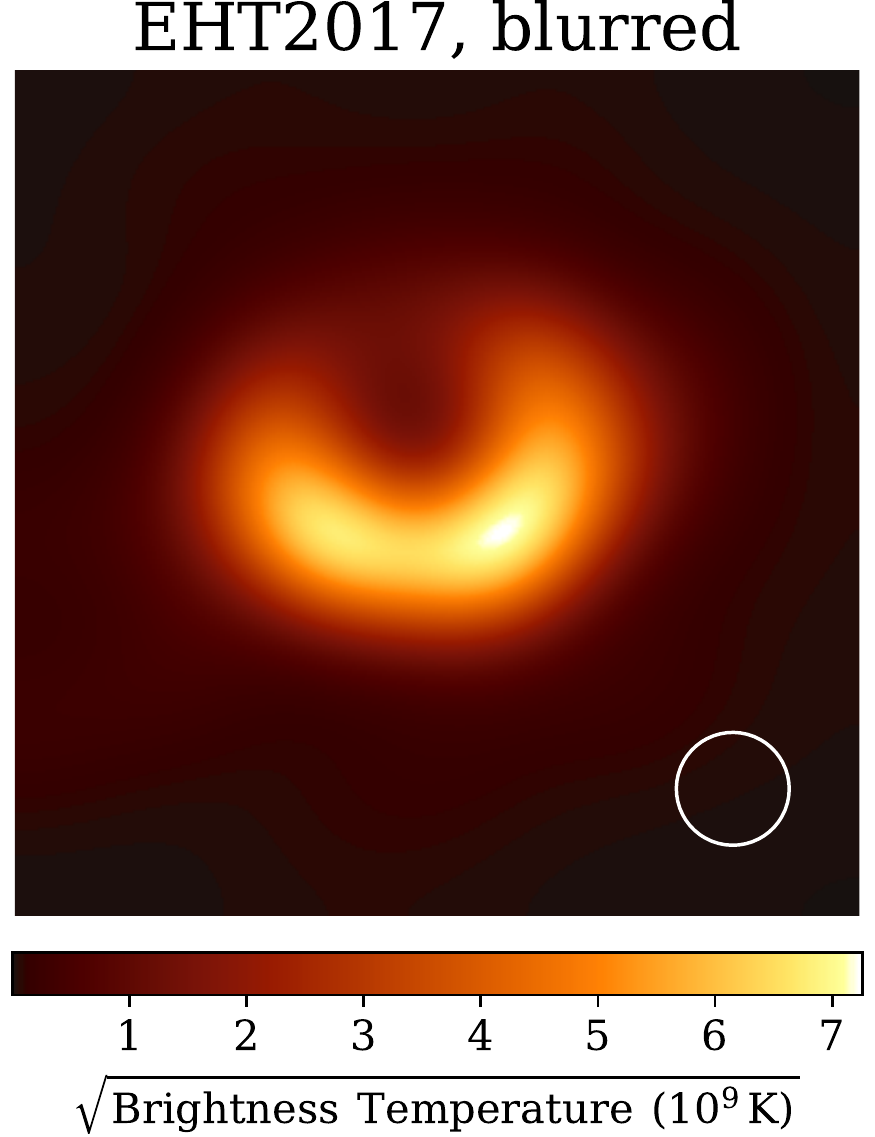}
   \includegraphics[scale=0.36]{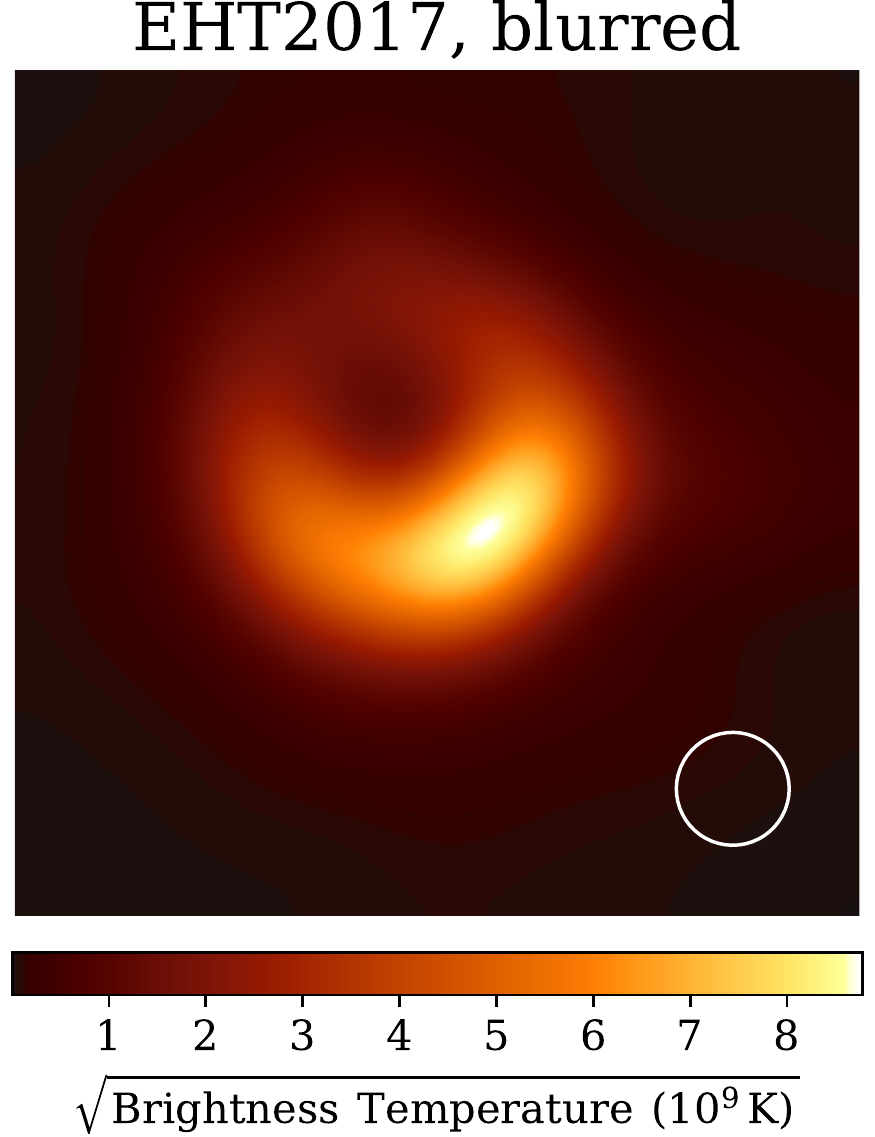} 
      \caption{EHT2017 reconstructions of pre-2017 source models. Top: GRMHD models of M87 from \citet{Dexter2012} (left) and \citet{Moscibrodzka2016} (right) described in Section~\ref{sec:models_pre2017}. Middle: images reconstructed with \pipe{} observing these images with the EHT2017 station and weather parameters, displayed on a square root scale like the other images in this paper. Bottom: the reconstructed images blurred with a circular Gaussian beam with FWHM of 17.1\,$\mu$as displayed on a linear scale as was done in \citet{eht-paperI, eht-paperIV}.}
         \label{fig:recs_oldmodels}
   \end{figure}

\section{Example case studies}
\label{sec:case_studies}
In this section, we give examples of studies that can be performed with \pipe{}. We illustrate possible image reconstruction improvements with future EHT observations (Sec. \ref{sec:newsites}), perform a case study of how well the EHT could perform under different observing conditions (Sec. \ref{sec:weather}), and compare models of M87 made before 2017 with the first results of the 2017 EHT observing campaign (Sec. \ref{sec:pre2017}). These synthetic observations are not meant as exhaustive studies, but could motivate more complete and quantitative parameter surveys.

\subsection{$\kappa$-jet versus thermal-jet model with different arrays}
\label{sec:newsites}
Figure \ref{fig:recs_future} shows images reconstructed from synthetic observations of both the thermal and $\kappa$-jet models (Fig. \ref{fig:models}), where all corruption and calibration effects were included. The images were reconstructed using self-calibrated complex visibilities as described above. The ring-like structure with the brightest spot in the south-west could be reconstructed with the EHT2017 array for both models (left panels). The thermal-jet model shows a more closed and thinner ring than the $\kappa$-jet model. The reconstructions, especially for the thermal-jet model, significantly improve with the addition of the GLT, PB, and KP (middle panels). With these stations included, the two models can be visually distinguished from each other more easily because of the appearance of model-specific features, such as the bright jet footprint for the thermal-jet model and the jet sheath extending towards the south-west for the $\kappa$-jet model. We note that we have only used one GRMHD snapshot with two specific electron temperature models. The EHT2020 array may not be able to let one visually distinguish between all possible source models with different electron distributions functions that exist in the literature. 

The M87 reconstruction shows only minor improvement when the AMT is added to the EHT2020 array. However, the AMT is useful for the purposes of array redundancy, and is expected to make a larger impact in observations of Sgr A*, which is a southern source with poorer east-west coverage than M87. Southern sites can observe Sgr A* during a large portion of the day. The AMT is planned to be built in southern Africa, providing east-west coverage to the American continent. New southern sites are particularly important given the short (minute-scale) time-variability of Sgr A*, requiring decent $uv$-coverage on short timescales as well. 

Figure \ref{fig:nxcorr} shows the $\rho_{\mathrm{NX}}$ values of the reconstructions in Figure~\ref{fig:recs_future}, which were cross-correlated with the model images in Figure~\ref{fig:models}. The model images were convolved with a circular Gaussian beam of varying size. For both models, $\rho_{\mathrm{NX}}$ between the non-convolved models and the reconstructions clearly increases as more stations are added to the array, with a small but noticeable effect when the AMT is added to the 2020 array. The peak moves towards smaller beam sizes as more stations are added and the nominal beam of the array becomes smaller due to e.g. the PDB-ALMA, GLT-ALMA, GLT-AMT, AMT-SMT, and AMT-LMT baselines. The thermal-jet model shows a stronger increase of the (peak) $\rho_{\mathrm{NX}}$ value than the $\kappa$-jet model as more stations are added. This is likely due to the fact that the jet footprint, which is a relatively dominant feature in the thermal-jet model, can be resolved with the EHT2020 array but not with the EHT2017 array. The sharp change of $\rho_{\mathrm{NX}}$ at $\sim0.4$ $\mu$as indicates the pixel size of the model images.  

\subsection{Varying observing conditions}
\label{sec:weather}
   
Here, we illustrate the use of \pipe{} for simulating observations under different observing conditions. Bad weather has several distinct effects on VLBI measurements; the most significant ones are an increase in precipitable water vapour, a decrease in coherence time, an increase in $\mathcal{P}_\mathrm{rms}$ due to worse atmospheric seeing conditions and poorer telescope pointing solutions, and a decrease in aperture efficiency together with an increase in gain errors due to poorer telescope focus solutions. We study these effects by reconstructing images based on the $\kappa$-jet model for the EHT2020 array under varying weather conditions.

Overall poor conditions are realized by setting PWV to 5\,mm and $t_\mathrm{c}$ to 3\,s for all stations. The source signal is attenuated by a factor of $\sim 1.3$ at zenith due to the PWV. The attenuation increases towards lower elevation by $\csc{(\mathrm{elevation})}$, which causes a significant loss of signal by a factor of about seven at an elevation of $10^\circ$. This is typically the lowest elevation at which telescopes can track a source while it is setting towards the horizon. The short coherence time results in more rapid phase wraps, which are difficult to fringe-fit. Moreover, amplitude variations occur beyond the thermal noise on short timescales due to the atmospheric seeing, since we have coupled small telescope mispointings to the atmospheric coherence time in \pipe{}. These amplitude variations necessitate a S/N limited self-calibration on short timescales. The telescopes with the smallest beams, PV and LMT, are most severely effected by the introduced pointing offsets. 

Additionally, we degrade $\eta_{\mathrm{ap}}$ by 10\,\%, add 10\,\% to $\mathcal{G_\mathrm{err}}$ and vary $\mathcal{P}_\mathrm{rms}$ between the default values of $\sim 1$" and $4$". The other weather parameters remain unchanged from the values given in Tables~\ref{tab:antennas} and \ref{tab:weather}. The results of this study are shown in Figure~\ref{fig:weather}. Due to the high PWV values across the array, a few scans of the LMT are already lost for the default pointing offsets, because fringes cannot be constrained. For $\mathcal{P}_\mathrm{rms} = 2$", a decent image can still be reconstructed. For $\mathcal{P}_\mathrm{rms} = 3$", the image quality is significantly reduced and the potential science return of an observation in these conditions is questionable. For $\mathcal{P}_\mathrm{rms} = 4$", the data quality is severely degraded by the weather conditions. The S/N is too low to calibrate for atmospheric phase variations on many baselines and every stations exhibits severe ($\mathcal{O}(2)$) scan-to-scan gain errors due to mispointings. The big LMT and PV dishes are affected most severely. The primary reason for the failed image reconstruction is that 50\% of the LMT and PV data is lost because of fringe non-detections, while the remaining data displays intra-scan gain errors $\mathcal{O}(10)$.

\subsection{Comparison of pre-EHT2017 models to EHT data}
\label{sec:pre2017}
   
We pass the predictive models from \citet{Dexter2012} and \citet{Moscibrodzka2016} introduced in Section~\ref{sec:models_pre2017} through the full \pipe{} pipeline, to test how they compare to the observed M87 black hole shadow image \citep{eht-paperI, eht-paperIV}.

For the synthetic observations with \pipe{}, all parameters and methods are the same as for other EHT2017 synthetic observations in this paper, except that a $uv$-flagging step was applied after network calibration. In this step, the scheduled EHT2017 $uv$-coverage as observed with \pipe{} was compared to the $uv$-coverage in the actual EHT data (11 April, low band), where some scans were not (fully) observed or no detections were obtained. Visibilities in the synthetic data for which there was no corresponding visibility in the real data within 1\% of the $uv$-coordinates were flagged.

Figure \ref{fig:recs_oldmodels} shows the fitted model images and reconstructions. The similarity of the reconstructions to the real M87 image of 11 April 2017 presented in \citet{eht-paperI, eht-paperIV}, showing the asymmetric ring structure, is striking given that these models were developed before the EHT2017 observations were done.

\section{Summary and outlook}
\label{sec:summary}
In this paper, we have presented \pipe{}, a new synthetic data pipeline for (mm-)VLBI observations, based on \meqsil{} and \rpicard{}. By introducing data corruptions from first principles and processing the data through a VLBI calibration pipeline, \pipe{} aims to mimic the full observation and calibration process as realistically as possible. Corruption effects that can be added include amplitude attenuation and phase corruptions by a mean and turbulent atmosphere, thermal noise with contributions from the receivers and atmosphere, antenna pointing offsets, polarization leakage, and complex gain errors. We have demonstrated the effects of these corruptions on synthetic EHT data and reconstructed images, taking point source, crescent, and GRMHD models of M87 as input, using an observed EHT schedule and including measured site conditions. Our synthetic observations show that the EHT2017 array is capable of reconstructing a black hole shadow from GRMHD model images, and that the image reconstruction quality could improve significantly with the addition of new sites in the future. In a comparison of reconstructed images from a thermal and non-thermal GRMHD model frame, these improvements allowed for a visual discrimination between these models.

In this work, we have focused on synthetic observations of a static total intensity model of M87. In future studies, observations of other sources, such as Sgr A*, could be simulated as well. \pipe{} also has the capability to simulate observations of time-variable, polarized source models and Faraday rotation. Synthetic observations using different (existing and future) VLBI arrays and different frequencies (e.g. 86 GHz GMVA, 345 GHz EHT, or cm VLBI observations) could also be done. In particular, we plan to extend the pipeline to handle wide-field ionospheric simulations. The elementary weather study shown in this work could be extended to a more in-depth study of the influence of various weather parameters across different sites, which is particularly useful for scheduling observations and commissioning new sites. Synthetic data from \pipe{} can also be used to test VLBI calibration (e.g., fringe-fitting) and self-calibration routines. Station's gain curves, which enter as an elevation dependent factor into the aperture efficiency, frequency dependent D-terms, and the simulation of inhomogeneous atmospheres will be added in future work. Furthermore, while this study has focused on investigating the effects of signal corruptions and the addition of new sites on the measured visibilities and reconstructed images, one could also investigate the precision with which model parameters, such as the black hole spin, electron temperature prescription, or inclination angle, can be fitted to the visibilities in different scenarios.

Finally, we believe that our open source end-to-end pipeline will have useful pedagogical applications. It could be used to teach students about a large variety of data corruption and calibration effects and their impact on the visibility data, and result in a rapid development of intuition and expertise in (mm-)VLBI calibration and imaging. 

\begin{acknowledgements}
This work is supported by the ERC Synergy Grant “BlackHoleCam: Imaging the Event Horizon of Black Holes” (Grant 610058).
I. Natarajan and R. Deane are grateful for the support from the New Scientific Frontiers with Precision Radio Interferometry Fellowship awarded by the South African Radio Astronomy Observatory (SARAO), which is a facility of the National Research Foundation (NRF), an agency of the Department of Science and Technology (DST) of South Africa.

The authors of the present paper further thank the following organizations and  programmes: 
the Academy of Finland (projects 274477, 284495, 312496);
the Advanced European Network of E-infrastructures for Astronomy with the SKA (AENEAS) project, supported by the European Commission Framework Programme Horizon 2020 Research and Innovation action under grant agreement 731016;
the Alexander von Humboldt Stiftung; 
the Black Hole Initiative at Harvard University, through a grant (60477) from the John Templeton Foundation; 
the China Scholarship Council;
Comisi\'{o}n Nacional de Investigaci\'{on} Cient\'{\i}fica y Tecnol\'{o}gica (CONICYT, Chile, via PIA ACT172033, Fondecyt 1171506, BASAL AFB-170002, ALMA-conicyt 31140007);
Consejo Nacional de Ciencia y Tecnolog\'{\i}a (CONACYT, Mexico, projects 104497, 275201, 279006, 281692);
the Delaney Family via the Delaney Family John A.\ Wheeler Chair at Perimeter Institute; 
Direcci\'{o}n General de Asuntos del Personal Acad\'{e}mico-—Universidad Nacional Aut\'{o}noma de M\'{e}xico (DGAPA-—UNAM, project IN112417); 
the Generalitat Valenciana postdoctoral grant APOSTD/2018/177; 
the GenT Program (Generalitat Valenciana) under project CIDEGENT/2018/021;
the Gordon and Betty Moore Foundation (grants GBMF-3561, GBMF-5278); the Istituto Nazionale di Fisica Nucleare (INFN) sezione di Napoli, iniziative specifiche TEONGRAV;
the International Max Planck Research School for Astronomy and Astrophysics at the Universities of Bonn and Cologne; 
the Jansky Fellowship program of the National Radio Astronomy Observatory (NRAO);
the Japanese Government (Monbukagakusho: MEXT) Scholarship; 
the Japan Society for the Promotion of Science (JSPS) Grant-in-Aid for JSPS Research Fellowship (JP17J08829);
the Key Research Program of Frontier Sciences, Chinese Academy of Sciences (CAS, grants QYZDJ-SSW-SLH057, QYZDJ-SSW-SYS008);
the Leverhulme Trust Early Career Research Fellowship;
the Max-Planck-Gesellschaft (MPG);
the Max Planck Partner Group of the MPG and the CAS;
the MEXT/JSPS KAKENHI (grants 18KK0090, JP18K13594, JP18K03656, JP18H03721, 18K03709, 18H01245, 25120007);
the MIT International Science and Technology Initiatives (MISTI) Funds; 
the Ministry of Science and Technology (MOST) of Taiwan (105-2112-M-001-025-MY3, 106-2112-M-001-011, 106-2119-M-001-027, 107-2119-M-001-017, 107-2119-M-001-020, and 107-2119-M-110-005);
the National Aeronautics and Space Administration (NASA, Fermi Guest Investigator grant 80NSSC17K0649); 
NASA through the NASA Hubble Fellowship grant HST-HF2-51431.001-A awarded by the Space Telescope Science Institute, which is operated by the Association of Universities for Research in Astronomy, Inc., for NASA, under contract
NAS5-26555;
the National Institute of Natural Sciences (NINS) of Japan;
the National Key Research and Development Program of China (grant 2016YFA0400704, 2016YFA0400702); 
the National Science Foundation (NSF, grants AST-0096454, AST-0352953, AST-0521233, AST-0705062, AST-0905844, AST-0922984, AST-1126433, AST-1140030, DGE-1144085, AST-1207704, AST-1207730, AST-1207752, MRI-1228509, OPP-1248097, AST-1310896, AST-1312651, AST-1337663, AST-1440254, AST-1555365, AST-1715061, AST-1615796, AST-1716327, OISE-1743747, AST-1816420); 
the Natural Science Foundation of China (grants 11573051, 11633006, 11650110427, 10625314, 11721303, 11725312, 11933007); 
the Natural Sciences and Engineering Research Council of Canada (NSERC, including a Discovery Grant and the NSERC Alexander Graham Bell Canada Graduate Scholarships-Doctoral Program);
the National Youth Thousand Talents Program of China;
the National Research Foundation of Korea (the Global PhD Fellowship Grant: grants NRF-2015H1A2A1033752, 2015-R1D1A1A01056807, the Korea Research Fellowship Program: NRF-2015H1D3A1066561); 
the Netherlands Organization for Scientific Research (NWO) VICI award (grant 639.043.513) and Spinoza Prize SPI 78-409; the New Scientific Frontiers with Precision Radio Interferometry Fellowship awarded by the South African Radio Astronomy Observatory (SARAO), which is a facility of the National Research Foundation (NRF), an agency of the Department of Science and Technology (DST) of South Africa;
the Onsala Space Observatory (OSO) national infrastructure, for the provisioning of its facilities/observational support (OSO receives funding through the Swedish Research Council under grant 2017-00648)
the Perimeter Institute for Theoretical Physics (research at Perimeter Institute is supported by the Government of Canada through the Department of Innovation, Science and Economic Development and by the Province of Ontario through the Ministry of Research, Innovation and Science);
the Princeton/Flatiron Postdoctoral Prize Fellowship;
the Russian Science Foundation (grant 17-12-01029); 
the Spanish Ministerio de Econom\'{\i}a y Competitividad (grants AYA2015-63939-C2-1-P,  AYA2016-80889-P); 
the State Agency for Research of the Spanish MCIU through the "Center of Excellence Severo Ochoa" award for the Instituto de Astrof\'{\i}sica de Andaluc\'{\i}a (SEV-2017-0709); 
the Toray Science Foundation; 
the US Department of Energy (USDOE) through the Los Alamos National Laboratory (operated by Triad National Security, LLC, for the National Nuclear Security Administration of the USDOE (Contract 89233218CNA000001));
the Italian Ministero dell'Istruzione Universit\`{a} e Ricerca through the grant Progetti Premiali 2012-iALMA (CUP C52I13000140001);
the European Union’s Horizon 2020 research and innovation programme under grant agreement No 730562 RadioNet;
ALMA North America Development Fund;
the Academia Sinica;
Chandra TM6-17006X.
This work used the Extreme Science and Engineering Discovery Environment (XSEDE), supported by NSF grant ACI-1548562, and CyVerse, supported by NSF grants DBI-0735191, DBI-1265383, and DBI-1743442.  XSEDE Stampede2 resource at TACC was allocated through TG-AST170024 and TG-AST080026N.  XSEDE JetStream resource at PTI and TACC was allocated through AST170028.  The simulations were performed in part on the SuperMUC cluster at the LRZ in Garching, on the LOEWE cluster in CSC in Frankfurt, and on the HazelHen cluster at the HLRS in Stuttgart.  This research was enabled in part by support provided by Compute Ontario (http://computeontario.ca), Calcul Quebec (http://www.calculquebec.ca) and Compute Canada (http://www.computecanada.ca).

We thank the staff at the participating observatories, correlation centers, and institutions for their enthusiastic support.
This paper makes use of the following ALMA data: ADS/JAO.ALMA\#2017.1.00841.V. ALMA is a partnership of the European Southern Observatory (ESO; Europe, representing its member states), NSF, and National Institutes of Natural Sciences of Japan, together with National Research Council (Canada), Ministry of Science and Technology (MOST; Taiwan), Academia Sinica Institute of Astronomy and Astrophysics (ASIAA; Taiwan), and Korea Astronomy and Space Science Institute (KASI; Republic of Korea), in cooperation with the Republic of Chile. The Joint ALMA Observatory is operated by ESO, Associated Universities, Inc.\  (AUI)/NRAO, and the National Astronomical Observatory of Japan (NAOJ).  The NRAO is a facility of the NSF operated under cooperative agreement by AUI. APEX is a collaboration between the Max-Planck-Institut f\"{u}r Radioastronomie (Germany), ESO, and the Onsala Space Observatory (Sweden). The SMA is a joint project between the SAO and ASIAA and is funded by the Smithsonian Institution and the Academia Sinica. The JCMT is operated by the East Asian Observatory on behalf of the NAOJ, ASIAA, and KASI, as well as the Ministry of Finance of China, Chinese Academy of Sciences, and the National Key R\&D Program (No.\ 2017YFA0402700) of China. Additional funding support for the JCMT is provided by the Science and Technologies Facility Council (UK) and participating universities in the UK and Canada. The LMT is a project operated by the Instituto Nacional de Astr\'{o}fisica, \'{O}ptica, y Electr\'{o}nica (Mexico) and the University of Massachusetts at Amherst (USA). The IRAM~30-m telescope on Pico Veleta, Spain is operated by IRAM and supported by CNRS (Centre National de la Recherche Scientifique, France), MPG (Max-Planck-Gesellschaft, Germany) and IGN (Instituto Geogr\'afico Nacional, Spain).
The SMT is operated by the Arizona Radio Observatory, a part of the Steward Observatory of the University of Arizona, with financial support of operations from the State of Arizona and financial support for instrumentation development from the NSF. The SPT is supported by the National Science Foundation through grant PLR- 1248097. Partial support is also provided by the NSF Physics Frontier Center grant PHY-1125897 to the Kavli Institute of Cosmological Physics at the University of Chicago, the Kavli Foundation and the Gordon and Betty Moore Foundation grant GBMF 947. The SPT hydrogen maser was provided on loan from the GLT, courtesy of ASIAA.
The EHTC has received generous donations of FPGA chips from Xilinx Inc., under the Xilinx University Program. The EHTC has benefited from technology shared under open-source license by the Collaboration for Astronomy Signal Processing and Electronics Research (CASPER). The EHT project is grateful to T4Science and Microsemi for their assistance with Hydrogen Masers. This research has made use of NASA's Astrophysics Data System.  We gratefully acknowledge the support provided by the extended staff of the ALMA, both from the inception of the ALMA Phasing Project through the observational campaigns of 2017 and 2018. We would like to thank A. Deller and W. Brisken for EHT-specific support with the use of DiFX.  We acknowledge the significance that Maunakea, where the SMA and JCMT EHT stations are located, has for the indigenous Hawaiian people.

The software presented in this work makes use of the Numpy \citep{numpy}, Scipy \citep{scipy}, Astropy \citep{astropy1,astropy2} libraries and the KERN software bundle \citep{molenaar2018kern}.
\end{acknowledgements}

\bibliographystyle{aa} 
\bibliography{bibliography}

\newcommand{\noop}[1]{}
\begin{thebibliography}{97}
\expandafter\ifx\csname natexlab\endcsname\relax\def\natexlab#1{#1}\fi

\bibitem[{{Akiyama} {et~al.}(2017){Akiyama}, {Ikeda}, {Pleau}, {Fish},
  {Tazaki}, {Kuramochi}, {Broderick}, {Dexter}, {Mo{\'s}cibrodzka},
  {Gowanlock}, {Honma}, \& {Doeleman}}]{Akiyama2017}
{Akiyama}, K., {Ikeda}, S., {Pleau}, M., {et~al.} 2017, \aj, 153, 159

\bibitem[{{Backes} {et~al.}(2016){Backes}, {M{\"u}ller}, {Conway}, {Deane},
  {Evans}, {Falcke}, {Fraga-Encinas}, {Goddi}, {Klein Wolt}, {Krichbaum},
  {MacLeod}, {Ribeiro}, {Roelofs}, {Shen}, \& {van Langevelde}}]{Backes2016}
{Backes}, M., {M{\"u}ller}, C., {Conway}, J.~E., {et~al.} 2016, in Proceedings
  of the 4th Annual Conference on High Energy Astrophysics in Southern Africa
  (HEASA 2016). 25-26 August, 2016. South African Astronomical Observatory
  (SAAO), Cape Town, South Africa, 29

\bibitem[{{Ball} {et~al.}(2018){Ball}, {Sironi}, \& {{\"O}zel}}]{ball2018}
{Ball}, D., {Sironi}, L., \& {{\"O}zel}, F. 2018, \apj, 862, 80

\bibitem[{Bambi \& Freese(2009)}]{Bambi2009}
Bambi, C. \& Freese, K. 2009, Phys. Rev. D, 79, 043002

\bibitem[{{Bardeen}(1973)}]{Bardeen1973}
{Bardeen}, J.~M. 1973, Timelike and Null Geodesies in the Kerr Metric, Houches
  Lecture Series (Gordon and Breach), 215--239

\bibitem[{{Beasley} \& {Conway}(1995)}]{Beasley1995}
{Beasley}, A.~J. \& {Conway}, J.~E. 1995, in Astronomical Society of the
  Pacific Conference Series, Vol.~82, Very Long Baseline Interferometry and the
  VLBA, ed. J.~A. {Zensus}, P.~J. {Diamond}, \& P.~J. {Napier}, 327

\bibitem[{{Bird} {et~al.}(2010){Bird}, {Harris}, {Blakeslee}, \&
  {Flynn}}]{Bird2010}
{Bird}, S., {Harris}, W.~E., {Blakeslee}, J.~P., \& {Flynn}, C. 2010, \aap,
  524, A71

\bibitem[{{Blackburn} {et~al.}(2019){Blackburn}, {Chan}, {Crew}, {Fish},
  {Issaoun}, {Johnson}, {Wielgus}, {Akiyama}, {Barrett}, \&
  {Bouman}}]{Blackburn2019}
{Blackburn}, L., {Chan}, C.-k., {Crew}, G.~B., {et~al.} 2019, arXiv e-prints,
  arXiv:1903.08832

\bibitem[{{Blakeslee} {et~al.}(2009){Blakeslee}, {Jord{\'a}n}, {Mei},
  {C{\^o}t{\'e}}, {Ferrarese}, {Infante}, {Peng}, {Tonry}, \&
  {West}}]{Blakeslee2009}
{Blakeslee}, J.~P., {Jord{\'a}n}, A., {Mei}, S., {et~al.} 2009, \apj, 694, 556

\bibitem[{{Blandford} \& {K{\"o}nigl}(1979)}]{Blandford1979}
{Blandford}, R.~D. \& {K{\"o}nigl}, A. 1979, \apj, 232, 34

\bibitem[{{Blecher} {et~al.}(2017){Blecher}, {Deane}, {Bernardi}, \&
  {Smirnov}}]{Blecher2017}
{Blecher}, T., {Deane}, R., {Bernardi}, G., \& {Smirnov}, O. 2017, \mnras, 464,
  143

\bibitem[{{Booth} {et~al.}(1989){Booth}, {Delgado}, {Hagstrom}, {Johansson},
  {Murphy}, {Olberg}, {Whyborn}, {Greve}, {Hansson}, {Lindstrom}, \&
  {Rydberg}}]{Booth1989}
{Booth}, R.~S., {Delgado}, G., {Hagstrom}, M., {et~al.} 1989, \aap, 216, 315

\bibitem[{{Bouman} {et~al.}(2017){Bouman}, {Johnson}, {Dalca}, {Chael},
  {Roelofs}, {Doeleman}, \& {Freeman}}]{Bouman2017}
{Bouman}, K.~L., {Johnson}, M.~D., {Dalca}, A.~V., {et~al.} 2017, ArXiv
  e-prints [\eprint[arXiv]{1711.01357}]

\bibitem[{{Broderick} {et~al.}(2016){Broderick}, {Fish}, {Johnson},
  {Rosenfeld}, {Wang}, {Doeleman}, {Akiyama}, {Johannsen}, \&
  {Roy}}]{Broderick2016}
{Broderick}, A.~E., {Fish}, V.~L., {Johnson}, M.~D., {et~al.} 2016, \apj, 820,
  137

\bibitem[{{Bronzwaer} {et~al.}(2018){Bronzwaer}, {Davelaar}, {Younsi},
  {Mo{\'s}cibrodzka}, {Falcke}, {Kramer}, \& {Rezzolla}}]{bronzwaer2018}
{Bronzwaer}, T., {Davelaar}, J., {Younsi}, Z., {et~al.} 2018, \aap, 613, A2

\bibitem[{{Cantiello} {et~al.}(2018){Cantiello}, {Blakeslee}, {Ferrarese},
  {C{\^o}t{\'e}}, {Roediger}, {Raimondo}, {Peng}, {Gwyn}, {Durrell}, \&
  {Cuillandre}}]{Cantiello2018}
{Cantiello}, M., {Blakeslee}, J.~P., {Ferrarese}, L., {et~al.} 2018, \apj, 856,
  126

\bibitem[{{Carilli} \& {Holdaway}(1999)}]{Carilli1999}
{Carilli}, C.~L. \& {Holdaway}, M.~A. 1999, Radio Science, 34, 817

\bibitem[{{Chael} {et~al.}(2018){Chael}, {Johnson}, {Bouman}, {Blackburn},
  {Akiyama}, \& {Narayan}}]{Chael2018}
{Chael}, A.~A., {Johnson}, M.~D., {Bouman}, K.~L., {et~al.} 2018, \apj, 857, 23

\bibitem[{{Chael} {et~al.}(2016){Chael}, {Johnson}, {Narayan}, {Doeleman},
  {Wardle}, \& {Bouman}}]{Chael2016}
{Chael}, A.~A., {Johnson}, M.~D., {Narayan}, R., {et~al.} 2016, \apj, 829, 11

\bibitem[{{Chan} {et~al.}(2015){Chan}, {Psaltis}, {{\"O}zel}, {Narayan}, \&
  {Sa{\c d}owski}}]{Chan2015}
{Chan}, C.-K., {Psaltis}, D., {{\"O}zel}, F., {Narayan}, R., \& {Sa{\c
  d}owski}, A. 2015, \apj, 799, 1

\bibitem[{{Conway} \& {Kronberg}(1969)}]{Conway1969}
{Conway}, R.~G. \& {Kronberg}, P.~P. 1969, \mnras, 142, 11

\bibitem[{{Davelaar} {et~al.}(2018){Davelaar}, {Mo{\'s}cibrodzka}, {Bronzwaer},
  \& {Falcke}}]{davelaar2018}
{Davelaar}, J., {Mo{\'s}cibrodzka}, M., {Bronzwaer}, T., \& {Falcke}, H. 2018,
  \aap, 612, A34

\bibitem[{{Davelaar} {et~al.}(2019){Davelaar}, {Olivares}, {Porth},
  {Bronzwaer}, {Janssen}, {Roelofs}, {Mizuno}, {Fromm}, {Falcke}, \&
  {Rezzolla}}]{Davelaar2019}
{Davelaar}, J., {Olivares}, H., {Porth}, O., {et~al.} 2019, arXiv e-prints,
  arXiv:1906.10065

\bibitem[{{Dexter} {et~al.}(2010){Dexter}, {Agol}, {Fragile}, \&
  {McKinney}}]{Dexter2010}
{Dexter}, J., {Agol}, E., {Fragile}, P.~C., \& {McKinney}, J.~C. 2010, \apj,
  717, 1092

\bibitem[{{Dexter} {et~al.}(2017){Dexter}, {Deller}, {Bower}, {Demorest},
  {Kramer}, {Stappers}, {Lyne}, {Kerr}, {Spitler}, {Psaltis}, {Johnson}, \&
  {Narayan}}]{Dexter2017}
{Dexter}, J., {Deller}, A., {Bower}, G.~C., {et~al.} 2017, \mnras, 471, 3563

\bibitem[{{Dexter} {et~al.}(2012){Dexter}, {McKinney}, \& {Agol}}]{Dexter2012}
{Dexter}, J., {McKinney}, J.~C., \& {Agol}, E. 2012, \mnras, 421, 1517

\bibitem[{{Doeleman} {et~al.}(2009){Doeleman}, {Fish}, {Broderick}, {Loeb}, \&
  {Rogers}}]{Doeleman2009cl}
{Doeleman}, S.~S., {Fish}, V.~L., {Broderick}, A.~E., {Loeb}, A., \& {Rogers},
  A.~E.~E. 2009, \apj, 695, 59

\bibitem[{{Event Horizon Telescope Collaboration}
  {et~al.}(2019{\natexlab{a}}){Event Horizon Telescope Collaboration},
  {Akiyama}, {Alberdi}, {Alef}, {Asada}, {Azulay}, {Baczko}, {Ball},
  {Balokovi{\'c}}, {Barrett}, \& et~al.}]{eht-paperI}
{Event Horizon Telescope Collaboration}, {Akiyama}, K., {Alberdi}, A., {et~al.}
  2019{\natexlab{a}}, \apjl, 875, L1

\bibitem[{{Event Horizon Telescope Collaboration}
  {et~al.}(2019{\natexlab{b}}){Event Horizon Telescope Collaboration},
  {Akiyama}, {Alberdi}, {Alef}, {Asada}, {Azulay}, {Baczko}, {Ball},
  {Balokovi{\'c}}, {Barrett}, \& et~al.}]{eht-paperII}
{Event Horizon Telescope Collaboration}, {Akiyama}, K., {Alberdi}, A., {et~al.}
  2019{\natexlab{b}}, \apjl, 875, L2

\bibitem[{{Event Horizon Telescope Collaboration}
  {et~al.}(2019{\natexlab{c}}){Event Horizon Telescope Collaboration},
  {Akiyama}, {Alberdi}, {Alef}, {Asada}, {Azulay}, {Baczko}, {Ball},
  {Balokovi{\'c}}, {Barrett}, \& et~al.}]{eht-paperIII}
{Event Horizon Telescope Collaboration}, {Akiyama}, K., {Alberdi}, A., {et~al.}
  2019{\natexlab{c}}, \apjl, 875, L3

\bibitem[{{Event Horizon Telescope Collaboration}
  {et~al.}(2019{\natexlab{d}}){Event Horizon Telescope Collaboration},
  {Akiyama}, {Alberdi}, {Alef}, {Asada}, {Azulay}, {Baczko}, {Ball},
  {Balokovi{\'c}}, {Barrett}, \& et~al.}]{eht-paperIV}
{Event Horizon Telescope Collaboration}, {Akiyama}, K., {Alberdi}, A., {et~al.}
  2019{\natexlab{d}}, \apjl, 875, L4

\bibitem[{{Event Horizon Telescope Collaboration}
  {et~al.}(2019{\natexlab{e}}){Event Horizon Telescope Collaboration},
  {Akiyama}, {Alberdi}, {Alef}, {Asada}, {Azulay}, {Baczko}, {Ball},
  {Balokovi{\'c}}, {Barrett}, \& et~al.}]{eht-paperV}
{Event Horizon Telescope Collaboration}, {Akiyama}, K., {Alberdi}, A., {et~al.}
  2019{\natexlab{e}}, \apjl, 875, L5

\bibitem[{{Event Horizon Telescope Collaboration}
  {et~al.}(2019{\natexlab{f}}){Event Horizon Telescope Collaboration},
  {Akiyama}, {Alberdi}, {Alef}, {Asada}, {Azulay}, {Baczko}, {Ball},
  {Balokovi{\'c}}, {Barrett}, \& et~al.}]{eht-paperVI}
{Event Horizon Telescope Collaboration}, {Akiyama}, K., {Alberdi}, A., {et~al.}
  2019{\natexlab{f}}, \apjl, 875, L6

\bibitem[{{Falcke} \& {Biermann}(1995)}]{Falcke1995}
{Falcke}, H. \& {Biermann}, P.~L. 1995, \aap, 293, 665

\bibitem[{{Falcke} \& {Markoff}(2000)}]{Falcke2000b}
{Falcke}, H. \& {Markoff}, S. 2000, \aap, 362, 113

\bibitem[{{Falcke} {et~al.}(2000){Falcke}, {Melia}, \& {Agol}}]{Falcke2000}
{Falcke}, H., {Melia}, F., \& {Agol}, E. 2000, \apjl, 528, L13

\bibitem[{{Fish} {et~al.}(2011){Fish}, {Doeleman}, {Beaudoin}, {Blundell},
  {Bolin}, {Bower}, {Chamberlin}, {Freund}, {Friberg}, {Gurwell}, {Honma},
  {Inoue}, {Krichbaum}, {Lamb}, {Marrone}, {Moran}, {Oyama}, {Plambeck},
  {Primiani}, {Rogers}, {Smythe}, {SooHoo}, {Strittmatter}, {Tilanus}, {Titus},
  {Weintroub}, {Wright}, {Woody}, {Young}, \& {Ziurys}}]{Fish2011}
{Fish}, V.~L., {Doeleman}, S.~S., {Beaudoin}, C., {et~al.} 2011, \apjl, 727,
  L36

\bibitem[{{Fish} {et~al.}(2009){Fish}, {Doeleman}, {Broderick}, {Loeb}, \&
  {Rogers}}]{Fish2009}
{Fish}, V.~L., {Doeleman}, S.~S., {Broderick}, A.~E., {Loeb}, A., \& {Rogers},
  A.~E.~E. 2009, \apj, 706, 1353

\bibitem[{{Fish} {et~al.}(2014){Fish}, {Johnson}, {Lu}, {Doeleman}, {Bouman},
  {Zoran}, {Freeman}, {Psaltis}, {Narayan}, {Pankratius}, {Broderick}, {Gwinn},
  \& {Vertatschitsch}}]{Fish2014}
{Fish}, V.~L., {Johnson}, M.~D., {Lu}, R.-S., {et~al.} 2014, \apj, 795, 134

\bibitem[{{Freund} {et~al.}(2014){Freund}, {Ziurys}, {Lauria}, \&
  {Reiland}}]{Freund2014}
{Freund}, R.~W., {Ziurys}, L.~M., {Lauria}, E.~F., \& {Reiland}, G.~P. 2014, in
  2014 XXXIth URSI General Assembly and Scientific Symposium (URSI GASS), 1--4

\bibitem[{{Gebhardt} {et~al.}(2011){Gebhardt}, {Adams}, {Richstone}, {Lauer},
  {Faber}, {G{\"u}ltekin}, {Murphy}, \& {Tremaine}}]{Gebhardt2011}
{Gebhardt}, K., {Adams}, J., {Richstone}, D., {et~al.} 2011, \apj, 729, 119

\bibitem[{Gelaro {et~al.}(2017)Gelaro, McCarty, Suárez, Todling, Molod,
  Takacs, Randles, Darmenov, Bosilovich, Reichle, Wargan, Coy, Cullather,
  Draper, Akella, Buchard, Conaty, da~Silva, Gu, Kim, Koster, Lucchesi,
  Merkova, Nielsen, Partyka, Pawson, Putman, Rienecker, Schubert, Sienkiewicz,
  \& Zhao}]{Gelaro2017}
Gelaro, R., McCarty, W., Suárez, M.~J., {et~al.} 2017, Journal of Climate, 30,
  5419

\bibitem[{{Goddi} {et~al.}(2017){Goddi}, {Falcke}, {Kramer}, {Rezzolla},
  {Brinkerink}, {Bronzwaer}, {Davelaar}, {Deane}, {de Laurentis}, {Desvignes},
  {Eatough}, {Eisenhauer}, {Fraga-Encinas}, {Fromm}, {Gillessen}, {Grenzebach},
  {Issaoun}, {Jan{\ss}en}, {Konoplya}, {Krichbaum}, {Laing}, {Liu}, {Lu},
  {Mizuno}, {Moscibrodzka}, {M{\"u}ller}, {Olivares}, {Pfuhl}, {Porth},
  {Roelofs}, {Ros}, {Schuster}, {Tilanus}, {Torne}, {van Bemmel}, {van
  Langevelde}, {Wex}, {Younsi}, \& {Zhidenko}}]{Goddi2016}
{Goddi}, C., {Falcke}, H., {Kramer}, M., {et~al.} 2017, International Journal
  of Modern Physics D, 26, 1730001

\bibitem[{{Gold} {et~al.}(2017){Gold}, {McKinney}, {Johnson}, \&
  {Doeleman}}]{Gold2017}
{Gold}, R., {McKinney}, J.~C., {Johnson}, M.~D., \& {Doeleman}, S.~S. 2017,
  \apj, 837, 180

\bibitem[{{Guilloteau} {et~al.}(1992){Guilloteau}, {Delannoy}, {Downes},
  {Greve}, {Guelin}, {Lucas}, {Morris}, {Radford}, {Wink}, {Cernicharo},
  {Forveille}, {Garcia-Burillo}, {Neri}, {Blondel}, {Perrigourad}, {Plathner},
  \& {Torres}}]{Guilloteau1992}
{Guilloteau}, S., {Delannoy}, J., {Downes}, D., {et~al.} 1992, \aap, 262, 624

\bibitem[{{Hada} {et~al.}(2011){Hada}, {Doi}, {Kino}, {Nagai}, {Hagiwara}, \&
  {Kawaguchi}}]{Hada2011}
{Hada}, K., {Doi}, A., {Kino}, M., {et~al.} 2011, \nat, 477, 185

\bibitem[{{Hamaker} {et~al.}(1996){Hamaker}, {Bregman}, \&
  {Sault}}]{Hamaker1996}
{Hamaker}, J.~P., {Bregman}, J.~D., \& {Sault}, R.~J. 1996, \aaps, 117, 137

\bibitem[{{Janssen} {et~al.}(2019{\natexlab{a}}){Janssen}, {Blackburn},
  {Issaoun}, {Wielgus}, {Krichbaum}, \& {Falcke}}]{Janssen2019b}
{Janssen}, M., {Blackburn}, L., {Issaoun}, S., {et~al.} 2019{\natexlab{a}}, EHT
  Memo Series, 2019-CE-01
  (\url{https://eventhorizontelescope.org/for-astronomers/memos})

\bibitem[{{Janssen} {et~al.}(2019{\natexlab{b}}){Janssen}, {Goddi}, {van
  Bemmel}, {Kettenis}, {Small}, {Liuzzo}, {Rygl}, {Mart{\'\i}-Vidal},
  {Blackburn}, {Wielgus}, \& {Falcke}}]{Janssen2019}
{Janssen}, M., {Goddi}, C., {van Bemmel}, I.~M., {et~al.} 2019{\natexlab{b}},
  \aap, 626, A75

\bibitem[{{Jennison}(1958)}]{Jennison1958}
{Jennison}, R.~C. 1958, Monthly Notices of the Royal Astronomical Society, 118,
  276

\bibitem[{{Johannsen} \& {Psaltis}(2010)}]{Johannsen2010}
{Johannsen}, T. \& {Psaltis}, D. 2010, \apj, 718, 446

\bibitem[{{Johnson}(2016)}]{Johnson2016so}
{Johnson}, M.~D. 2016, \apj, 833, 74

\bibitem[{{Johnson} {et~al.}(2017){Johnson}, {Bouman}, {Blackburn}, {Chael},
  {Rosen}, {Shiokawa}, {Roelofs}, {Akiyama}, {Fish}, \&
  {Doeleman}}]{Johnson2017}
{Johnson}, M.~D., {Bouman}, K.~L., {Blackburn}, L., {et~al.} 2017, \apj, 850,
  172

\bibitem[{{Johnson} \& {Gwinn}(2015)}]{Johnson2015}
{Johnson}, M.~D. \& {Gwinn}, C.~R. 2015, \apj, 805, 180

\bibitem[{{Johnson} {et~al.}(2018){Johnson}, {Narayan}, {Psaltis}, {Blackburn},
  {Kovalev}, {Gwinn}, {Zhao}, {Bower}, {Moran}, {Kino}, {Kramer}, {Akiyama},
  {Dexter}, {Broderick}, \& {Sironi}}]{Johnson2018}
{Johnson}, M.~D., {Narayan}, R., {Psaltis}, D., {et~al.} 2018, \apj, 865, 104

\bibitem[{Jones {et~al.}(2001)Jones, Oliphant, Peterson, {et~al.}}]{scipy}
Jones, E., Oliphant, T., Peterson, P., {et~al.} 2001, {SciPy}: Open source
  scientific tools for {Python}

\bibitem[{{Jones}(1941)}]{Jones1941}
{Jones}, R.~C. 1941, Journal of the Optical Society of America (1917-1983), 31,
  488

\bibitem[{{Kamruddin} \& {Dexter}(2013)}]{Kamruddin2013}
{Kamruddin}, A.~B. \& {Dexter}, J. 2013, \mnras, 434, 765

\bibitem[{{Kerr}(1963)}]{Kerr1963}
{Kerr}, R.~P. 1963, Physical Review Letters, 11, 237

\bibitem[{{Lu} {et~al.}(2014){Lu}, {Broderick}, {Baron}, {Monnier}, {Fish},
  {Doeleman}, \& {Pankratius}}]{Lu2014}
{Lu}, R.-S., {Broderick}, A.~E., {Baron}, F., {et~al.} 2014, \apj, 788, 120

\bibitem[{{Lu} {et~al.}(2016){Lu}, {Roelofs}, {Fish}, {Shiokawa}, {Doeleman},
  {Gammie}, {Falcke}, {Krichbaum}, \& {Zensus}}]{Lu2016}
{Lu}, R.-S., {Roelofs}, F., {Fish}, V.~L., {et~al.} 2016, \apj, 817, 173

\bibitem[{{Masson}(1994)}]{Masson1994}
{Masson}, C.~R. 1994, in Astronomical Society of the Pacific Conference Series,
  Vol.~59, IAU Colloq. 140: Astronomy with Millimeter and Submillimeter Wave
  Interferometry, ed. M.~{Ishiguro} \& J.~{Welch}, 87--95

\bibitem[{{McMullin} {et~al.}(2007){McMullin}, {Waters}, {Schiebel}, {Young},
  \& {Golap}}]{McMullin2007}
{McMullin}, J.~P., {Waters}, B., {Schiebel}, D., {Young}, W., \& {Golap}, K.
  2007, in Astronomical Society of the Pacific Conference Series, Vol. 376,
  Astronomical Data Analysis Software and Systems XVI, ed. R.~A. {Shaw},
  F.~{Hill}, \& D.~J. {Bell}, 127

\bibitem[{{Medeiros} {et~al.}(2017){Medeiros}, {Chan}, {{\"O}zel}, {Psaltis},
  {Kim}, {Marrone}, \& {Sadowski}}]{Medeiros2016}
{Medeiros}, L., {Chan}, C.-k., {{\"O}zel}, F., {et~al.} 2017, \apj, 844, 35

\bibitem[{{Middelberg} {et~al.}(2013){Middelberg}, {Deller}, {Norris},
  {Fotopoulou}, {Salvato}, {Morgan}, {Brisken}, {Lutz}, \&
  {Rovilos}}]{Middelberg2013}
{Middelberg}, E., {Deller}, A.~T., {Norris}, R.~P., {et~al.} 2013, \aap, 551,
  A97

\bibitem[{Molenaar \& Smirnov(2018)}]{molenaar2018kern}
Molenaar, G. \& Smirnov, O. 2018, Astronomy and Computing

\bibitem[{{Mo{\'s}cibrodzka} {et~al.}(2016){Mo{\'s}cibrodzka}, {Falcke}, \&
  {Shiokawa}}]{Moscibrodzka2016}
{Mo{\'s}cibrodzka}, M., {Falcke}, H., \& {Shiokawa}, H. 2016, Astronomy and
  Astrophysics, 586, A38

\bibitem[{{Mo{\'s}cibrodzka} {et~al.}(2014){Mo{\'s}cibrodzka}, {Falcke},
  {Shiokawa}, \& {Gammie}}]{Moscibrodzka2014}
{Mo{\'s}cibrodzka}, M., {Falcke}, H., {Shiokawa}, H., \& {Gammie}, C.~F. 2014,
  \aap, 570, A7

\bibitem[{Natarajan {et~al.}(2019)}]{Natarajan2019}
Natarajan, I. {et~al.} 2019, in prep.

\bibitem[{{Noordam} \& {Smirnov}(2010)}]{Noordam2010}
{Noordam}, J.~E. \& {Smirnov}, O.~M. 2010, \aap, 524, A61

\bibitem[{{Offringa} {et~al.}(2014){Offringa}, {McKinley}, {Hurley-Walker},
  {Briggs}, {Wayth}, {Kaplan}, {Bell}, {Feng}, {Neben}, {Hughes}, {Rhee},
  {Murphy}, {Bhat}, {Bernardi}, {Bowman}, {Cappallo}, {Corey}, {Deshpande},
  {Emrich}, {Ewall-Wice}, {Gaensler}, {Goeke}, {Greenhill}, {Hazelton},
  {Hindson}, {Johnston-Hollitt}, {Jacobs}, {Kasper}, {Kratzenberg}, {Lenc},
  {Lonsdale}, {Lynch}, {McWhirter}, {Mitchell}, {Morales}, {Morgan},
  {Kudryavtseva}, {Oberoi}, {Ord}, {Pindor}, {Procopio}, {Prabu}, {Riding},
  {Roshi}, {Shankar}, {Srivani}, {Subrahmanyan}, {Tingay}, {Waterson},
  {Webster}, {Whitney}, {Williams}, \& {Williams}}]{Offringa2014}
{Offringa}, A.~R., {McKinley}, B., {Hurley-Walker}, N., {et~al.} 2014, \mnras,
  444, 606

\bibitem[{{Owen} {et~al.}(2000){Owen}, {Eilek}, \& {Kassim}}]{Owen2000}
{Owen}, F.~N., {Eilek}, J.~A., \& {Kassim}, N.~E. 2000, \apj, 543, 611

\bibitem[{Paine(2019)}]{Paine2019}
Paine, S. 2019, The am atmospheric model

\bibitem[{{Pandya} {et~al.}(2016){Pandya}, {Zhang}, {Chandra}, \&
  {Gammie}}]{pandya2016}
{Pandya}, A., {Zhang}, Z., {Chandra}, M., \& {Gammie}, C.~F. 2016, \apj, 822,
  34

\bibitem[{Pardo {et~al.}(2001)Pardo, Cernicharo, \& Serabyn}]{Pardo2001}
Pardo, J.~R., Cernicharo, J., \& Serabyn, E. 2001, IEEE Transactions on
  Antennas and Propagation, 49, 1683

\bibitem[{{Pierrard} \& {Lazar}(2010)}]{pierrard2010}
{Pierrard}, V. \& {Lazar}, M. 2010, \solphys, 267, 153

\bibitem[{{Porth} {et~al.}(2017){Porth}, {Olivares}, {Mizuno}, {Younsi},
  {Rezzolla}, {Moscibrodzka}, {Falcke}, \& {Kramer}}]{Porth2017}
{Porth}, O., {Olivares}, H., {Mizuno}, Y., {et~al.} 2017, Computational
  Astrophysics and Cosmology, 4, 1

\bibitem[{{Psaltis} {et~al.}(2015){Psaltis}, {{\"O}zel}, {Chan}, \&
  {Marrone}}]{Psaltis2015}
{Psaltis}, D., {{\"O}zel}, F., {Chan}, C.-K., \& {Marrone}, D.~P. 2015, \apj,
  814, 115

\bibitem[{{Radford} \& {Holdaway}(1998)}]{Radford1998}
{Radford}, S.~J. \& {Holdaway}, M.~A. 1998, in \procspie, Vol. 3357, Advanced
  Technology MMW, Radio, and Terahertz Telescopes, ed. T.~G. {Phillips},
  486--494

\bibitem[{{Raffin} {et~al.}(2014){Raffin}, {Algaba-Marcosa}, {Asada},
  {Blundell}, {Burgos}, {Chang}, {Chen}, {Christensen}, {Grimes}, {Han}, {Ho},
  {Huang}, {Inoue}, {Koch}, {Kubo}, {Leiker}, {Liu}, {Martin-Cocher},
  {Matsushita}, {Nakamura}, {Nishioka}, {Nystrom}, {Paine}, {Patel}, {Pradel},
  {Pu}, {Shen}, {Snow}, {Sridharan}, {Srinivasan}, {Tong}, \&
  {Wang}}]{Raffin2014}
{Raffin}, P., {Algaba-Marcosa}, J.~C., {Asada}, K., {et~al.} 2014, in Society
  of Photo-Optical Instrumentation Engineers (SPIE) Conference Series, Vol.
  9145, Ground-based and Airborne Telescopes V, 91450G

\bibitem[{{Roelofs} {et~al.}(2017){Roelofs}, {Johnson}, {Shiokawa}, {Doeleman},
  \& {Falcke}}]{Roelofs2017}
{Roelofs}, F., {Johnson}, M.~D., {Shiokawa}, H., {Doeleman}, S.~S., \&
  {Falcke}, H. 2017, \apj, 847, 55

\bibitem[{{Rogers} {et~al.}(1974){Rogers}, {Hinteregger}, {Whitney},
  {Counselman}, {Shapiro}, {Wittels}, {Klemperer}, {Warnock}, {Clark},
  {Hutton}, {Marandino}, {Ronnang}, {Rydbeck}, \& {Niell}}]{Rogers1974}
{Rogers}, A.~E.~E., {Hinteregger}, H.~F., {Whitney}, A.~R., {et~al.} 1974,
  \apj, 193, 293

\bibitem[{{Schwab} \& {Cotton}(1983)}]{Schwab1983}
{Schwab}, F.~R. \& {Cotton}, W.~D. 1983, \aj, 88, 688

\bibitem[{{Smirnov}(2011{\natexlab{a}})}]{Smirnov2011a}
{Smirnov}, O.~M. 2011{\natexlab{a}}, \aap, 527, A106

\bibitem[{{Smirnov}(2011{\natexlab{b}})}]{Smirnov2011b}
{Smirnov}, O.~M. 2011{\natexlab{b}}, \aap, 527, A107

\bibitem[{{Smirnov}(2011{\natexlab{c}})}]{Smirnov2011c}
{Smirnov}, O.~M. 2011{\natexlab{c}}, \aap, 527, A108

\bibitem[{Takahashi(2004)}]{Takahashi2004}
Takahashi, R. 2004, \apj, 611, 996

\bibitem[{{The Astropy Collaboration} {et~al.}(2018){The Astropy
  Collaboration}, {Price-Whelan}, {Sip{\H o}cz}, {G{\"u}nther}, {Lim},
  {Crawford}, {Conseil}, {Shupe}, {Craig}, {Dencheva}, {Ginsburg},
  {VanderPlas}, {Bradley}, {P{\'e}rez-Su{\'a}rez}, {de Val-Borro}, {Aldcroft},
  {Cruz}, {Robitaille}, {Tollerud}, {Ardelean}, {Babej}, {Bach}, {Bachetti},
  {Bakanov}, {Bamford}, {Barentsen}, {Barmby}, {Baumbach}, {Berry}, {Biscani},
  {Boquien}, {Bostroem}, {Bouma}, {Brammer}, {Bray}, {Breytenbach},
  {Buddelmeijer}, {Burke}, {Calderone}, {Cano Rodr{\'{\i}}guez}, {Cara},
  {Cardoso}, {Cheedella}, {Copin}, {Corrales}, {Crichton}, {D'Avella}, {Deil},
  {Depagne}, {Dietrich}, {Donath}, {Droettboom}, {Earl}, {Erben}, {Fabbro},
  {Ferreira}, {Finethy}, {Fox}, {Garrison}, {Gibbons}, {Goldstein}, {Gommers},
  {Greco}, {Greenfield}, {Groener}, {Grollier}, {Hagen}, {Hirst}, {Homeier},
  {Horton}, {Hosseinzadeh}, {Hu}, {Hunkeler}, {Ivezi{\'c}}, {Jain}, {Jenness},
  {Kanarek}, {Kendrew}, {Kern}, {Kerzendorf}, {Khvalko}, {King}, {Kirkby},
  {Kulkarni}, {Kumar}, {Lee}, {Lenz}, {Littlefair}, {Ma}, {Macleod},
  {Mastropietro}, {McCully}, {Montagnac}, {Morris}, {Mueller}, {Mumford},
  {Muna}, {Murphy}, {Nelson}, {Nguyen}, {Ninan}, {N{\"o}the}, {Ogaz}, {Oh},
  {Parejko}, {Parley}, {Pascual}, {Patil}, {Patil}, {Plunkett}, {Prochaska},
  {Rastogi}, {Reddy Janga}, {Sabater}, {Sakurikar}, {Seifert}, {Sherbert},
  {Sherwood-Taylor}, {Shih}, {Sick}, {Silbiger}, {Singanamalla}, {Singer},
  {Sladen}, {Sooley}, {Sornarajah}, {Streicher}, {Teuben}, {Thomas},
  {Tremblay}, {Turner}, {Terr{\'o}n}, {van Kerkwijk}, {de la Vega}, {Watkins},
  {Weaver}, {Whitmore}, {Woillez}, {Zabalza}, \& {Astropy
  Contributors}}]{astropy2}
{The Astropy Collaboration}, {Price-Whelan}, A.~M., {Sip{\H o}cz}, B.~M.,
  {et~al.} 2018, \aj, 156, 123

\bibitem[{{The Astropy Collaboration} {et~al.}(2013){The Astropy
  Collaboration}, {Robitaille}, {Tollerud}, {Greenfield}, {Droettboom}, {Bray},
  {Aldcroft}, {Davis}, {Ginsburg}, {Price-Whelan}, {Kerzendorf}, {Conley},
  {Crighton}, {Barbary}, {Muna}, {Ferguson}, {Grollier}, {Parikh}, {Nair},
  {Unther}, {Deil}, {Woillez}, {Conseil}, {Kramer}, {Turner}, {Singer}, {Fox},
  {Weaver}, {Zabalza}, {Edwards}, {Azalee Bostroem}, {Burke}, {Casey},
  {Crawford}, {Dencheva}, {Ely}, {Jenness}, {Labrie}, {Lim}, {Pierfederici},
  {Pontzen}, {Ptak}, {Refsdal}, {Servillat}, \& {Streicher}}]{astropy1}
{The Astropy Collaboration}, {Robitaille}, T.~P., {Tollerud}, E.~J., {et~al.}
  2013, \aap, 558, A33

\bibitem[{{Thomas-Osip} {et~al.}(2007){Thomas-Osip}, {McWilliam}, {Phillips},
  {Morrell}, {Thompson}, {Folkers}, {Adams}, \&
  {Lopez-Morales}}]{Thomas-Osip2007}
{Thomas-Osip}, J., {McWilliam}, A., {Phillips}, M.~M., {et~al.} 2007, \pasp,
  119, 697

\bibitem[{{Thompson} {et~al.}(2017){Thompson}, {Moran}, \& {Swenson}}]{TMS2017}
{Thompson}, A.~R., {Moran}, J.~M., \& {Swenson}, Jr., G.~W. 2017,
  {Interferometry and Synthesis in Radio Astronomy, 3rd Edition}

\bibitem[{{Treuhaft} \& {Lanyi}(1987)}]{Treuhaft1987}
{Treuhaft}, R.~N. \& {Lanyi}, G.~E. 1987, Radio Science, 22, 251

\bibitem[{{Ulich} \& {Haas}(1976)}]{Ulich1976}
{Ulich}, B.~L. \& {Haas}, R.~W. 1976, \apjs, 30, 247

\bibitem[{{van der Walt} {et~al.}(2011){van der Walt}, {Colbert}, \&
  {Varoquaux}}]{numpy}
{van der Walt}, S., {Colbert}, S.~C., \& {Varoquaux}, G. 2011, Computing in
  Science and Engineering, 13, 22

\bibitem[{{Walsh} {et~al.}(2013){Walsh}, {Barth}, {Ho}, \& {Sarzi}}]{Walsh2013}
{Walsh}, J.~L., {Barth}, A.~J., {Ho}, L.~C., \& {Sarzi}, M. 2013, \apj, 770, 86

\bibitem[{Xiao(2006)}]{xiao2006}
Xiao, F. 2006, Plasma Physics and Controlled Fusion, 48, 203

\bibitem[{{Yuan} {et~al.}(2003){Yuan}, {Quataert}, \& {Narayan}}]{Yuan2003}
{Yuan}, F., {Quataert}, E., \& {Narayan}, R. 2003, \apj, 598, 301

\end{thebibliography}

\end{document}